\documentclass[10pt,journal,final,twocolumn,comosc]{IEEEtran}

\IEEEoverridecommandlockouts
\usepackage{amsmath,amssymb,amsfonts}
\usepackage{cite}
\usepackage{subfigure}
\usepackage{graphicx}
\usepackage{textcomp}
\usepackage{xcolor}
\usepackage{times}
\usepackage{subfigure}         
\usepackage{amssymb,amsmath}

\usepackage{acronym}  % make an acronym
\usepackage{balance}

\usepackage{epstopdf}

\usepackage{bm}         
\usepackage{algorithm} 
\usepackage{algorithmic}

\usepackage{lipsum}
\usepackage{stfloats}
\usepackage{url}

\usepackage{setspace}

\allowdisplaybreaks[4]

\newtheorem{remark}{\bf Remark}

\newtheorem{lemma}{\bf Lemma}

\definecolor{BLUE}{rgb}{0,0,1}

\newcommand{\paperTitleMarkboth}{Active RIS vs. Passive RIS: Which Will Prevail in 6G?}

\acrodef{siso}[SISO]{single-input single-output}%

\acrodef{ris}[RIS]{reconfigurable intelligent surface}%

\acrodef{csi}[CSI]{channel state information}%
\acrodef{awgn}[AWGN]{additive white Gaussian noise}%

\acrodef{bs}[BS]{base station}%
\acrodef{snr}[SNR]{signal-to-noise ratio}%
\acrodef{mmwave}[mmWave]{millimeter-wave}%
\acrodef{snr}[SNR]{signal-to-noise ratio}%
\acrodef{rf}[RF]{radio frequency}%

\acrodef{swipt}[SWIPT]{simultaneous wireless information and power transfer}%

\acrodef{sinr}[SINR]{signal-to-interference-plus-noise ratio}%

\acrodef{rc}[RC]{reflection coefficient}%
\acrodef{mimo}[MIMO]{multiple-input multiple-output}%
\def\BibTeX{{\rm B\kern-.05em{\sc i\kern-.025em b}\kern-.08em
		T\kern-.1667em\lower.7ex\hbox{E}\kern-.125emX}}

\begin{document}

\title{Active RIS vs. Passive RIS: Which \\ Will Prevail in 6G?}
\author{
	{{Zijian Zhang},~\IEEEmembership{Student Member,~IEEE}, {Linglong Dai},~\IEEEmembership{Fellow,~IEEE}, {Xibi Chen},~\IEEEmembership{Student Member,~IEEE},\\ {Changhao Liu},~\IEEEmembership{Graduate Student Member,~IEEE}, {Fan Yang},~\IEEEmembership{Fellow,~IEEE}, {Robert Schober},~\IEEEmembership{Fellow,~IEEE},\\ and {H. Vincent Poor},~\IEEEmembership{Life Fellow,~IEEE}\vspace{-1em}}
	%\thanks{This paper first appeared on ArXiv in March 2021 \cite{Zhang'21}.}
	%This work was supported in part by the National Key Research and Development Program of China (Grant No. 2020YFB1807201), in part by the National Natural Science Foundation of China (Grant No. 62031019), 
	\thanks{Manuscript received 6 September, 2022; revised 19 November, 2022; accepted 19 December, 2022. Date of publication XXX XX, 2022; date of current version XXX XX, 2022. This work was supported in part by the National Key Research and Development Program of China (Grant No. 2020YFB1807201), in part by the National Natural Science Foundation of China (Grant No. 62031019), in part by the European Commission through the H2020-MSCA-ITN META WIRELESS Research Project under Grant 956256, and in part by the U.S National Science Foundation under Grants CCF-1908308 and CNS-2128448. R. Schober's work was partly supported by the Federal Ministry of Education and Research of Germany under the programme of “Souveran. Digital. Vernetzt.” joint project 6G-RIC (project identification number: PIN 16KISK023) and by the Deutsche Forschungsgemeinschaft (DFG, German Research Foundation) under grant SCHO 831/15-1. An earlier version of this paper was presented in part at the IEEE GLOBECOM’22, Rio de Janeiro, Brazil, December 2022 \cite{Zhan2212:Active}. The associate editor coordinating the review of this article and approving it for publication was A. García Armada. {\it (Corresponding author: Linglong Dai.)}
	}

	\thanks{Zijian Zhang, Linglong Dai, Xibi Chen, Changhao Liu, and Fan Yang are with the Department of Electronic Engineering as well as the Beijing National Research Center for Information Science and Technology (BNRist), Tsinghua University, Beijing 100084, China (e-mail: zhangzj20@mails.tsinghua.edu.cn, daill@tsinghua.edu.cn, cxb17@tsinghua.org.cn, liu-ch21@mails.tsinghua.edu.cu, fan\_yang@tsinghua.edu.cn).
	}
	\thanks{Robert Schober is with the Institute for Digital Communications at Friedrich-Alexander Universität Erlangen-Nürnberg (FAU), 91054 Erlangen, Germany (e-mail: robert.schober@fau.de). }
	\thanks{H. Vincent Poor is  with the Department   of  Electrical and Computer Engineering,  Princeton  University, Princeton, NJ 08544,  USA  (e-mail: poor@princeton.edu).
	}
	\thanks{Color versions of one or more figures in this article are available at https://doi.org/10.1109/TCOMM.2022.3231893.}
	\thanks{Digital Object Identifier 10.1109/TCOMM.2022.3231893}

	%\thanks{This work was supported by the National Science and Technology Major Project of China under Grant 2018ZX03001004-003 and the National Natural Science Foundation of China for Outstanding Young Scholars under Grant 61722109.}
}
%\markboth{Journal of \LaTeX\ Class Files,~Vol.~18, No.~9, September~2020}%
%{How to Use the IEEEtran \LaTeX \ Templates}

\markboth{ IEEE Transactions on Communications}{ Zhang {\it et al.}: \paperTitleMarkboth}

\maketitle
	%\tableofcontents
%	\vspace{-1cm}
	\begin{abstract}
	%\vspace{-0.25cm}
	As a revolutionary paradigm for controlling wireless channels, reconfigurable intelligent surfaces (RISs) have emerged as a candidate technology for future 6G networks. However, due to the “multiplicative fading” effect, the existing passive RISs only achieve limited capacity gains in many scenarios with strong direct links.  In this paper, the concept of active RISs is proposed to overcome this fundamental limitation. Unlike passive RISs that reflect signals without amplification, active RISs can amplify the reflected signals via amplifiers integrated into their elements. To characterize the signal amplification and incorporate the noise introduced by the active components, we develop and verify the signal model of active RISs through the experimental measurements based on a fabricated active RIS element. Based on the verified signal model, we further analyze the asymptotic performance of active RISs to reveal the substantial capacity gain they provide for wireless communications. Finally, we formulate the sum-rate maximization problem for an active RIS aided multi-user multiple-input single-output (MU-MISO) system and a joint transmit beamforming and reflect precoding scheme is proposed to solve this problem. Simulation results show that, in a typical wireless system, passive RISs can realize only a limited sum-rate gain of 22\%, while active RISs can achieve a significant sum-rate gain of 130\%, thus overcoming the “multiplicative fading” effect.
	\end{abstract}
%\vspace{-2em}

%\vspace{-1em}
	\section{Introduction}	
	
As wireless communications have advanced from the first generation (1G) to 5G, the system capacity has been significantly increased by improving the transceiver designs, while the wireless channel has been considered to be uncontrollable. Recently, due to the advances in meta-materials, \acp{ris} have been proposed \cite{Zhang'18'na,Ren'20,Venkatesh'20} for the purpose of intelligently controlling wireless channels to achieve improved communication performance. Specifically, an RIS is an array composed of a very large number of passive elements that reflects electromagnetic signals in a desired manner so as to reconfigure the propagation properties of wireless environment \cite{Renzo'20}. Thanks to their high array gain, low cost, low power, and negligible noise \cite{Basar'19,LinglongDai,Renzo'20}, RISs promise to improve channel capacity \cite{Huang'20'J}, extend coverage \cite{Wang'19}, and save power \cite{Huang'18'2} in future 6G networks. Additionally, RISs are also projected to have other applications such as in WiFi \cite{Zhao'20'na}, precision measurement \cite{Faraji-Dana'18}, and navigation \cite{Park'21}.
\par
As an important advantage of RISs, the negligible noise introduced by passive RISs enables a high array gain. Particularly, in a RIS aided single-user single-input single-output (SU-SISO) system, the achievable \ac{snr} gain enabled by an $N$-element RIS is proportional to $N^2$ \cite{Wu'19}. Benefiting from this advantage, \acp{ris} are expected to introduce significant capacity gains in wireless systems \cite{Huang'20'J}. However, in practice, these capacity gains are typically only observed in communication scenarios where the direct link between transmitter and receiver is completely blocked or very weak \cite{Huang'20'J,Wang'19,Huang'18'2,Zhao'20,Hou'20,Zijian'20}. By contrast, in many scenarios where the direct link is not weak, conventional \acp{ris} achieve limited capacity gains \cite{Najafi'20}. The reason behind this phenomenon is the “multiplicative fading” effect introduced by RISs, i.e., the equivalent path loss of the transmitter-RIS-receiver link is the product (instead of the sum) of the path losses of the transmitter-RIS and RIS-receiver links, which is usually thousands of times larger than that of the direct link \cite{Najafi'20}. As a result, the “multiplicative fading” effect makes it almost impossible for passive RISs to achieve noticeable capacity gains in many wireless environments. Many existing works on RISs have bypassed this effect by only considering scenarios with severely obstructed direct links \cite{Huang'20'J,Wang'19,Huang'18'2,Zhao'20,Hou'20,Zijian'20}. Therefore, to advance the practicability of RISs in future 6G wireless networks, a critical issue for RISs to be addressed is: \textit{How to break the fundamental performance bottleneck caused by the “multiplicative fading” effect?}
\par
To overcome the fundamental physical limitation, in this paper, a new \ac{ris} architecture called \textit{active} \acp{ris} is proposed for wireless communication systems. Specifically, different from \textit{passive} \acp{ris} that passively reflect signals without amplification, the key feature of active \acp{ris} is their ability to actively reflect signals with amplification, which can be realized by integrating reflection-type amplifiers into their reflecting elements. At the expense of additional power consumption, active RIS can compensate for the large path loss of reflected links, which is promising to overcome the “multiplicative fading” effect. Some parallel works\footnote{In October 2019, we started to design an active RIS element integrating a reflection-type amplifier \cite{Xibi'20}. The fabrication of this active RIS element was finished in August 2020. Subsequently, we set out to establish an experimental environment for signal measurements with this element, and all measurements were completed in February 2021. This paper first appeared on ArXiv in March 2021 [DOI: 10.48550/arXiv.2103.15154].} have revealed the potential benefits of active RISs for enhancing wireless systems. For example, in \cite{Ruizhe'21}, an active \ac{ris} was introduced into a single-user single-input multi-output (SU-SIMO) system to enhance the user's \ac{snr}. In \cite{Changsheng'21}, the authors considered an active RIS aided SU-SISO system with limited RIS elements, and the impact of RIS location placement on the communication performance is analyzed.
\par
%\footnote{This paper first appeared on ArXiv in March 2021 \cite{Zhang'21}. Then, we became aware of some parallel works on active RISs, such as \cite{Ruizhe'21} published in August 2021 and \cite{Changsheng'21} published in December 2021.}

%Apart from the reflection-type amplification, one important feature of active RISs different from passive RISs is their additionally introduced thermal noise. Due to the integrated active components of active RISs, different from passive RISs with negligible noise introduced, active RISs will inevitably introduce non-negligible thermal noise. However, it is not well studied that how the additionally introduced thermal noise influences the signal model of active RISs. Besides, most existing works focused on the performance analysis of active RISs with limited elements and the beamforming design for single-user systems. The asymptotic performance analysis for active RISs with a extremely large number of elements and more practical multi-user beamforming design for active RISs aided systems are still worth further investigating.
\par
In this paper\footnote{Simulation codes are provided to reproduce the results presented in this article: http://oa.ee.tsinghua.edu.cn/dailinglong/publications/publications.html.}, we propose the concept of active RISs and focuses on the signal model verification, asymptotic performance analysis, and multi-user beamforming design of active RISs aided communication systems. Specifically, our contributions are summarized as follows:
	\begin{itemize}
		\item We develop and verify the signal model of active \acp{ris}, which characterizes the amplification of the incident signal and accounts for the non-negligible thermal noise introduced by the active elements. Particularly, the verification is made via the experimental measurements based on a designed and fabricated active RIS element.
		
		\item Based on the verified signal model, we analyze the asymptotic performance of an active RIS with extremely large number of active elements, and we further compare it to that of the existing passive RISs, which reveals the notable capacity gain enabled by the use of active RISs.
		
		\item To evaluate the performance of active RISs in typical communication systems, we formulate a sum-rate maximization problem for an active RIS aided multi-user multiple-input single-output (MU-MISO) system. Then, by exploiting fractional programming (FP), a joint transmit beamforming and reflect precoding scheme is proposed to solve this problem.
		
		\item To account for the non-ideal factors in practical systems, we extend the studied beamforming design in the scenario with the self-interference of active RISs. We model the feedback-type self-interference of active RISs, which allows us to formulate an mean-squared error minimization problem to suppress the self-interference. Then, by utilizing alternating direction method of multipliers (ADMM) \cite{admm} and sequential unconstrained minimization techniques (SUMT) \cite{fiacco1990nonlinear}, an alternating optimization scheme is proposed to solve the formulated problem.
	\end{itemize}

\par
The rest of this paper is organized as follows.
In Section \ref{sec:model}, the architectures as well as the signal models of passive \acp{ris} and active \acp{ris} are introduced, respectively. In Section \ref{sec:PA}, the asymptotic performance of active RISs is analyzed and compared to that of the passive RISs. In Section \ref{sec:Precoding}, a sum-rate maximization problem is formulated for an active RIS aided MU-MISO system, and a joint beamforming and precoding design is proposed to solve the formulated problem. In Section \ref{sec:IV-SI}, we extend the studied joint beamforming and precoding design to the practical case with self-interference. In Section \ref{subsec:IV-C}, the convergence and complexity of the proposed schemes are analyzed. In Section \ref{sec:sim}, validation results are presented to validate the signal model of active RISs and evaluate the performance of active RISs in typical communication scenarios. Finally, conclusions are drawn and future works are discussed in Section \ref{sec:con}.
	\begin{figure*}[!t]
		\centering
		\includegraphics[width = 1\textwidth]{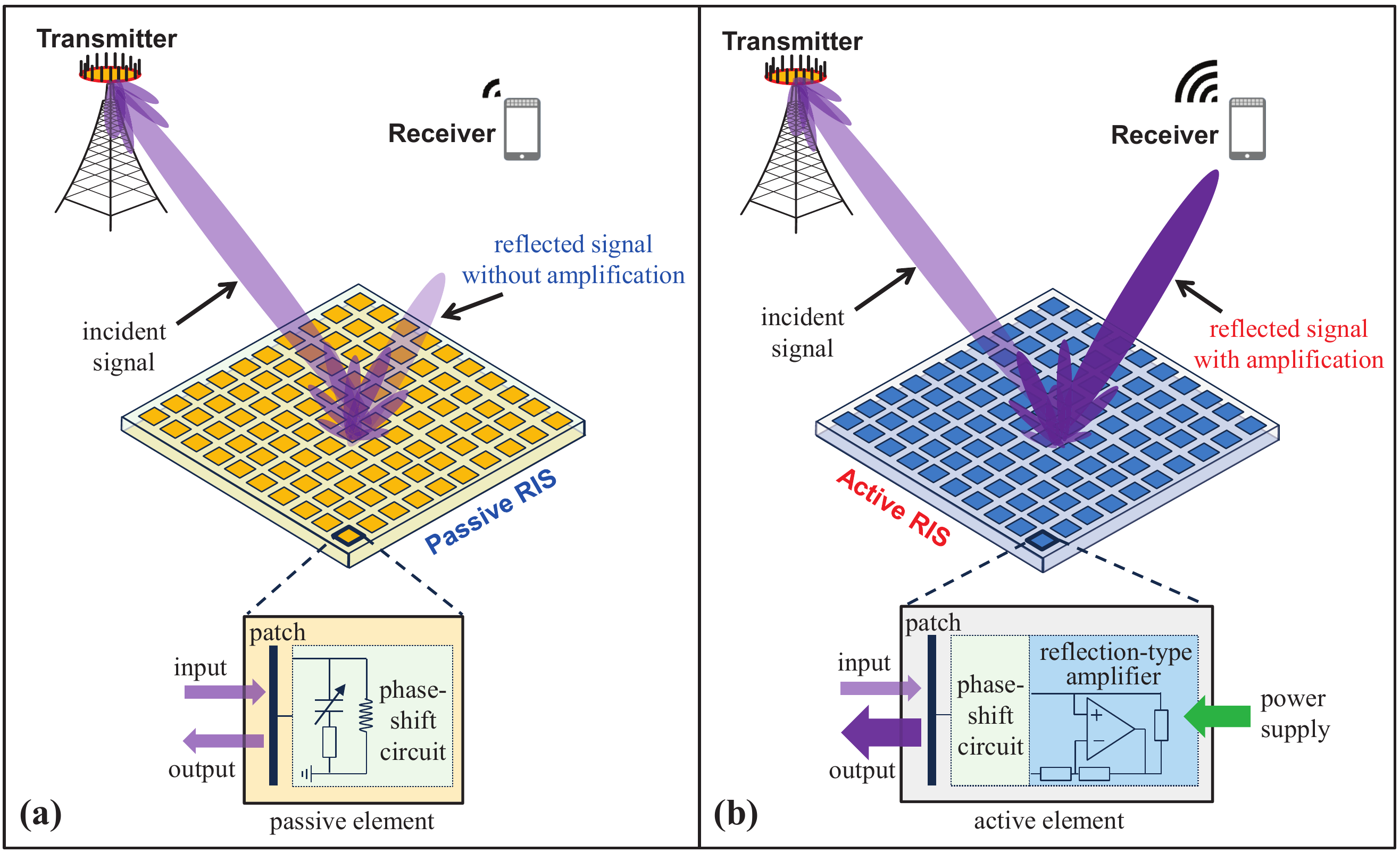}
		%\vspace*{-1em}
		\caption{\color{black}An illustration of the hardware architectures of (a) a passive RIS and (b) an active RIS.}
		\label{img:framework}
		%\vspace*{-1em}
	\end{figure*}

	\textit{Notations:} $\mathbb{C}$, $\mathbb{R}$, and $\mathbb{R}_{+}$ denote the sets of complex, real, and positive real numbers, respectively; ${[\cdot]^{-1}}$, ${[\cdot]^{*}}$, ${[\cdot]^{\rm T}}$, and ${[\cdot]^{\rm H}}$ denote the inverse, conjugate, transpose, and conjugate-transpose operations, respectively; $\|\cdot\|$ denotes the Euclidean norm of the argument; $\|\cdot\|_{\rm F}$ denotes the Frobenius norm of the argument; ${\rm diag}(\cdot)$ denotes the diagonalization operation; $\mathfrak{R}\{\cdot\}$ denotes the real part of the argument; $\otimes$ denotes the Kronecker product; $\angle[\cdot]$ denotes the angle of the complex argument; $\ln(\cdot)$ denotes the natural logarithm of its argument; $\mathcal{C} \mathcal{N}\!\left({\bm \mu}, {\bf \Sigma } \right)$ denotes the
	complex multivariate Gaussian distribution with mean ${\bm \mu}$ and variance ${\bf \Sigma }$; $\mathbf{I}_{L}$ is an $L\times L$ identity matrix, and $\mathbf{0}_{L}$ is an $L\times 1$ zero vector.

\section{Passive RISs and Active RISs}\label{sec:model}
In this section, we introduce the architectures of RISs. First, in Subsection \ref{subsec:model1}, we review passive RISs and point out their physical limitation imposed by the “multiplicative fading” effect. Then, in Subsection \ref{subsec:model2}, to overcome this limitation, we propose the novel concept of active RISs along with their hardware structure and signal model. Finally, in Subsection \ref{subsec:model3}, we present the transmission model for an active RIS aided MU-MISO system.
	\subsection{Passive RISs}\label{subsec:model1}
	The RISs widely studied in most existing works are passive RISs \cite{Zhang'18'na,Ren'20,Venkatesh'20,Renzo'20,Basar'19,LinglongDai,Huang'20'J,Huang'18'2,Wang'19}. Specifically, as shown in Fig. \ref{img:framework} (a), a passive RIS comprises a large number of passive elements each being able to reflect the incident signal with a controllable phase shift. In general, each passive RIS element consists of a reflective patch terminated with an impedance-adjustable circuit for phase shifting \cite{Yang'17}. Thanks to the passive mode of operation without active radio-frequency (RF) components, a passive RIS element practically consumes zero direct-current power \cite{Yang'17}, and the introduced thermal noise is usually negligible \cite{Renzo'20,Basar'19,LinglongDai,Huang'20'J,Huang'18'2,Wang'19}. Thereby, the signal model of an $N$-element passive RIS widely used in the literature is given as follows \cite{Basar'19}:
	\begin{equation}\label{eqn:passive_model}
		\begin{aligned}
			{\bf{y}} = {\bf \Theta}{\bf{x}},
		\end{aligned}
	\end{equation}
	where ${\bf x}\in{\mathbb C}^N$ denotes the incident signal, ${\bf \Theta}:={\rm diag}\left(e^{j\theta_1},\cdots,e^{j\theta_N}\right)\in{\mathbb C}^{N\times N}$ denotes the reflection coefficient matrix of the passive RIS with $\theta_n$ being the phase shift of the $n$-th passive element, and ${\bf y}\in{\mathbb C}^N$ denotes the signal reflected by the RIS. Note that the impact of noise is neglected in (\ref{eqn:passive_model}). As a consequence, by properly adjusting ${\bf \Theta}$ to manipulate the $N$ signals 
	reflected by the $N$ RIS elements to coherently add with the same phase at the receiver, a high array gain can be achieved. This is expected to significantly increase the receiver \ac{snr} \cite{Basar'19,LinglongDai,Renzo'20}, which is one of the key reasons for why RISs have attracted so much research interest recently \cite{Huang'20'J,Wang'19,Huang'18'2,Zhao'20'na,Faraji-Dana'18,Park'21,Zhao'20,Hou'20,Zijian'20}.
	\par
	Unfortunately, in practice, this expected high capacity gain often cannot be realized, especially in communication scenarios where the direct link between the transmitter and the receiver is not weak. The reason for this negative result is the “multiplicative fading” effect introduced by passive RISs. Specifically, the equivalent path loss of the transmitter-RIS-receiver reflected link is the product (instead of the sum) of the path losses of the transmitter-RIS and RIS-receiver links, and therefore, it is thousands of times larger than that of the unobstructed direct link. Thereby, for an RIS to realize a noticeable capacity gain, thousands of RIS elements are required to compensate for this extremely large path loss.
	\begin{remark}
		To illustrate the above fact, let us consider an SU-SISO system aided by a passive \ac{ris}. Assume that the transceiver antennas is omnidirectional and RIS elements are tightly deployed with half-wavelength spacing \cite{Najafi'20}. Let $d=200$ m, $d_{\rm t}=110$ m, and $d_{\rm r}=110$ m denote the distances between transmitter and receiver, transmitter and \ac{ris}, \ac{ris} and receiver, respectively. Assume that all channels are line-of-sight (LoS) and the RIS phase shift is optimally configured to maximize the channel gain of the transmitter-RIS-receiver reflected link. Then, for carrier frequencies of $5/10/28$ GHz, $N =\frac{4}{{d\lambda }}{d_{\rm{t}}}{d_{\rm{r}}}= 4034/8067/22587$ RIS elements are required to make the reflected link as strong as the direct link \cite{Najafi'20}. The high signaling overhead introduced by the $N$ pilots required for channel estimation \cite{Huchen} and the high complexity of ${\cal O}(N^2)$ for real-time beamforming \cite{Pan'19} make the application of such a large number of passive RIS elements in practical wireless networks very challenging \cite{Najafi'20}. Consequently, many existing works have bypassed the “multiplicative fading” effect by only considering the scenario where the direct link is completely blocked or very weak \cite{Basar'19,Wang'19,LinglongDai,Renzo'20,Huang'20'J,Huang'18'2,Zhao'20,Hou'20,Zijian'20}.
	\end{remark}

	\subsection{Active RISs}\label{subsec:model2}
To overcome the fundamental performance bottleneck caused by the “multiplicative fading” effect of RISs, in this paper, we propose active RISs as a promising solution. As shown in Fig. \ref{img:framework} (b), similar to passive RISs, active RISs can also reflect the incident signals with reconfigurable phase shifts. Different from passive \acp{ris} that just reflect the incident signals without amplification, active \acp{ris} can further amplify the reflected signals. To achieve this goal, the key component of an active RIS element is the additionally integrated active reflection-type amplifier, which can be realized by different existing active components, such current-inverting converters \cite{Loncar'19}, asymmetric current mirrors \cite{Bousquet'12}, or some integrated circuits \cite{Kishor'12}.%\footnote{In this paper, we focus on studying reflective active RISs, while the investigation of transmissive active RISs is left for future work \cite{Mu'21,Liu'21,Shuhao'21}.}
		
	With reflection-type amplifiers supported by a power supply, the reflected and amplified signal of an $N$-element active RIS can be modeled as follows:
	\begin{equation}\label{eqn:active_model}
		\begin{aligned}
			{\bf{y}} = \underbrace{{\bm\Psi}{\bf{x}}}_{\text{Desired signal}} + \underbrace {{\bm \Psi} {\bf{v}}}_{\text{Dynamic noise}} +\underbrace {{{\bf{n}}_{\text{s}}}}_{\text{Static noise}},
		\end{aligned}
	\end{equation}
	where ${\bm \Psi}:={\rm diag}\left(p_1e^{j\theta_1},\cdots,p_Ne^{j\theta_N}\right)\in{\mathbb C}^{N\times N}$ denotes the reflection coefficient matrix of the active RIS, wherein $p_n\in{\mathbb R}_+$ denotes the amplification factor of the $n$-th active element and $p_n$ can be larger than one thanks to the integrated reflection-type amplifier. Due to the use of active components, active RISs consume additional power for amplifying the reflected signals, and the thermal noise introduced by active RIS elements cannot be neglected as is done for passive RISs. Particularly, as shown in (\ref{eqn:active_model}), the noise introduced at active RISs can be classified into dynamic noise ${\bm \Psi} {\bf{v}}$ and static noise ${\bf{n}}_{\text{s}}$, where ${\bm \Psi} {\bf{v}}$ is the noise introduced and amplified by the reflection-type amplifier and ${\bf{n}}_{\text{s}}$ is generated by the patch and the phase-shift circuit \cite{Bousquet'12}. More specifically, ${\bf{v}}$ is related to the input noise and the inherent device noise of the active RIS elements \cite{Bousquet'12}, while the static noise ${\bf{n}}_{\text{s}}$ is unrelated to ${\bm \Psi}$ and is usually negligible compared to the dynamic noise ${\bm \Psi} {\bf{v}}$, as will be verified by experimental results in Section \ref{sub:vr:sm}. Thus, here we neglect  ${\bf{n}}_{\text{s}}$ and model ${\bf{v}}$ as ${\bf{v}} \sim \mathcal{C} \mathcal{N}\left(\mathbf{0}_N, \sigma _{v}^2{\bf{I}}_N\right)$.

	\begin{remark}
	Note that active \acp{ris} are fundamentally different from the relay-type \acp{ris} equipped with RF components \cite{He'21,Nguyen'21,Basar'19C} and relays \cite{Ntontin'19}. Specifically, in \cite{He'21,Nguyen'21,Basar'19C}, a subset of the passive RIS elements are connected to active RF chains, which are used for sending pilot signals and processing baseband signals. Thus, these relay-type RIS elements have signal processing capabilities \cite{He'21,Nguyen'21,Basar'19C}. On the contrary, active RISs do not have such capabilities but only reflect and amplify the incident signals to strengthen the reflected link. Besides, although active RISs can amplify the incident signals, similar to full-duplex amplify-and-forward (FD-AF) relays, their respective hardware architectures and transmission models are quite different. Specifically, an FD-AF relay is equipped with RF chains to receive the incident signal and then transmit it after amplification \cite{Ntontin'19}. Due to the long delay inherent to this process, two timeslots are needed to complete the transmission of one symbol, and the received signal at the receiver in a timeslot actually depends on two different symbols, which were transmitted by the transmitter and the FD-AF relay, respectively \cite{Ntontin'19}. As a consequence, in order to efficiently decode the symbols, the receiver in an FD-AF relay aided system has to combine the signals received in two successive timeslots to maximize the \ac{snr}. Thus, the transmission model for FD-AF relaying \cite[Eq. (22), Eq. (25)]{Ntontin'19} differs substantially from that for active RIS (\ref{eqn:signal}), which also leads to different achievable rates \cite[Table I]{Ntontin'19}.
	\end{remark}

\subsection{Active RIS Aided MU-MISO System}\label{subsec:model3}
	
	\begin{figure}[!t]
		\centering
		\includegraphics[width = 0.41\textwidth]{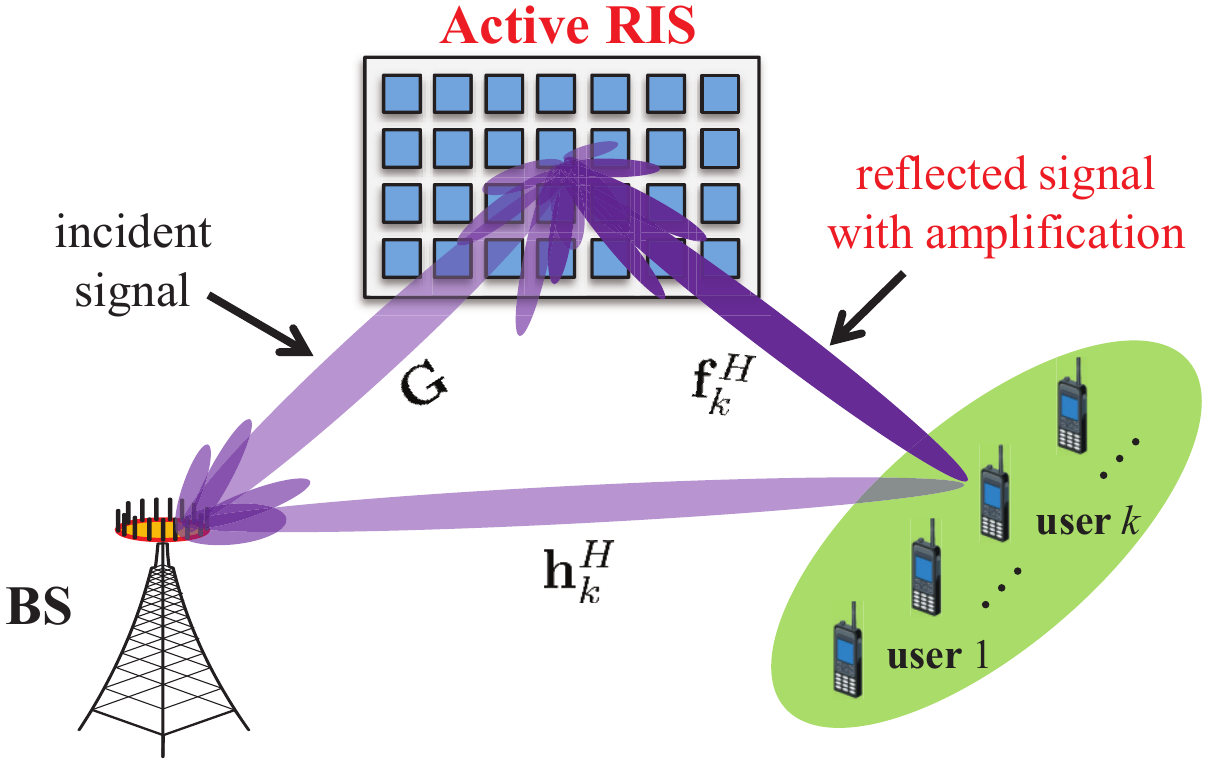}
		%\vspace*{-1em}
		\caption{An illustration of the downlink transmission in an active RIS aided MU-MISO system.
		}
		\label{img:model}
		%\vspace*{-1em}
	\end{figure}

	To characterize the performance gains enabled by our proposed active RISs in typical communication scenarios, we consider an active RIS aided downlink MU-MISO system as shown in Fig. \ref{img:model}, where an $M$-antenna \ac{bs} serves $K$ single-antenna users simultaneously with the aid of an $N$-element active \ac{ris}. 
	
	Let $\mathbf{s} :=\left[s_{1}, \cdots, s_{K}\right]^{\rm T} \in \mathbb{C}^{K}$ denote the transmitted symbol vector for the $K$ users with ${\mathbb E}\left\{\mathbf{s}\mathbf{s}^{\rm H}\right\}={\bf I}_K$. We assume that multi-user linear precoding is employed at the \ac{bs} for downlink transmission. Then, according to  (\ref{eqn:active_model}), signal $r_k\in\mathbb{C}$ received at user $k$ can be modeled as follows:
	\begin{align}
		\notag {r_k} = &(\underbrace {{\bf{h}}_k^{\rm H}}_{{\text{Direct link}}} + \underbrace {{\bf{f}}_k^{\rm H}{{\bm\Psi}}{\bf{G}}}_{{\text{Reflected link}}})\sum\limits_{j = 1}^K {{{\bf{w}}_j}{s_j}} +\\ &\!\!\!\!\!\! \underbrace{{\bf{f}}_k^{\rm H}{{\bm\Psi}}{\bf{v}}}_{\text{Noise introduced by active RIS}}  +\underbrace {z_k}_{\text{Noise introduced at user $k$}},\label{eqn:signal}
	\end{align}
	where ${\bf{G}}\in\mathbb{C}^{N\times M}$, ${\bf{h}}_k^{\rm H}\in\mathbb{C}^{1\times M}$ and ${\bf{f}}_k^{\rm H}\in\mathbb{C}^{1\times N}$ denote the channel vector between the \ac{bs} and the \ac{ris}, that between the \ac{bs} and user $k$, and that between the \ac{ris} and user $k$, respectively; ${\bf{w}}_k\in{\mathbb C}^{M\times 1}$ denotes the \ac{bs} beamforming vector for symbol $s_k$; and $z_k$ denotes the \ac{awgn} at user $k$ with zero mean and variance $\sigma^2$.
	\par
	To analytically illustrate how active RISs can overcome the “multiplicative fading” effect, based on the signal model in (\ref{eqn:active_model}), the performance gain enabled by the use of active RISs will be studied in the next section.
	
\section{Performance Analysis}\label{sec:PA}
In this section, we analyze the asymptotic performance of active RISs to reveal their notable capacity gains. To this end, in order to make the problem analytically tractable and get insightful results, similar to \cite{Wu'19}, we consider a SU-SISO system with $M=1$ \ac{bs} antenna and $K=1$ user, while the general MU-MISO case is studied in Section \ref{sec:Precoding}. %In Subsection \ref{subsec:PA1}, the asymptotic \acp{snr} of a passive RIS aided system and an active RIS aided system are provided. Then, in Subsection \ref{subsec:PA2}, to show the superiority of active RISs, the derived asymptotic \acp{snr} for the two systems are compared.
	
\subsection{Asymptotic \ac{snr} for Passive RISs and Active RISs}\label{subsec:PA1}
To illustrate the capacity gain provided by passive/active RIS aided reflected links, for the moment, we ignore the direct link by setting ${\bf h}_k$ to zero, as was done in, e.g., \cite{Wu'19}. Furthermore, to obtain analytical results and find more insights, we assume that each active RIS element has the same amplification factor (i.e., $p_n:= p$ for all $ n\in\{1,\cdots,N\}$), while the power allocation among active elements will be considered in Section \ref{sec:Precoding}. For a fair comparison with the asymptotic performance of passive RISs, similar to \cite{Wu'19}, we assume Rayleigh-fading channels. %This assumption has also been widely used for the performance analysis of massive \ac{mimo} systems, e.g., \cite{Marzetta'10} and \cite{Marzetta'99}.
	
For the above RIS aided SU-SISO system without direct link, we first redefine the \ac{bs}-RIS channel matrix and the RIS-user channel vector as ${\bf G}:={\bf g}\in{\mathbb C}^{N\times 1}$ and ${\bf f}_k:={\bf f}\in{\mathbb C}^{N\times 1}$, respectively, to simplify the notations. Then, we recall the following lemma from \cite{Wu'19} for the asymptotic \ac{snr} achieved by passive RISs.
\begin{lemma}[Asymptotic \ac{snr} for passive RISs]
	Assuming ${\bf{f}} \sim \mathcal{C} \mathcal{N}\left(\mathbf{0}_N, \varrho_{f}^{2} {\bf{I}}_N\right)$, ${\bf{g}} \sim \mathcal{C} \mathcal{N}\left(\mathbf{0}_N, \varrho_{g}^{2} {\bf{I}}_N\right)$ and letting $N \to \infty$, the asymptotic \ac{snr} $\gamma_{\text{passive}}$ of a passive RIS aided SU-SISO system is given by
	\begin{align}\label{eqn:ag_passive}
		\gamma_{\text{passive}}  \to {N^2}\frac{{P_{{\text{BS}}}^{{\max}}{\pi ^2}\varrho _f^2\varrho _g^2}}{{16{\sigma ^2}}},
	\end{align}
	where ${P^{\max}_{\text{BS}}}$ denotes the maximum transmit power at the \ac{bs}.
\end{lemma}
\begin{IEEEproof}
	The proof can be found in \cite[Proposition 2]{Wu'19}.
\end{IEEEproof}

For comparison, under the same transmission conditions, we provide the asymptotic \ac{snr} of an active RIS aided SU-SISO system in the following lemma.
	
\begin{lemma}[Asymptotic \ac{snr} for active RISs]
	Assuming ${\bf{f}} \sim \mathcal{C} \mathcal{N}\left(\mathbf{0}_N, \varrho_{f}^{2} {\bf{I}}_N\right)$, ${\bf{g}} \sim \mathcal{C} \mathcal{N}\left(\mathbf{0}_N, \varrho_{g}^{2} {\bf{I}}_N\right)$ and letting $N \to \infty$, the asymptotic \ac{snr} $\gamma_{\text{active}}$ of an active RIS aided SU-SISO system is given by
	 \begin{align}\label{eqn:ag_active}
	  \gamma_{\text{active}}  \to N \frac{{P_{{\text{BS}}}^{{\max}}P_{\text{A}}^{{\max}}{\pi ^2}\varrho _f^2\varrho _g^2}}{{16\left( {P_{\text{A}}^{{\max}}\sigma _v^2\varrho _f^2 + P_{{\text{BS}}}^{{\max}}{\sigma ^2}\varrho _g^2 + {\sigma ^2}\sigma _v^2} \right)}},
	 \end{align}
where ${P^{\max}_{\text{A}}}$ denotes the maximum reflect power of the active \ac{ris}.
\end{lemma}
\begin{IEEEproof}
Please see Appendix A.
\end{IEEEproof}
\begin{remark}
From (\ref{eqn:ag_active}) we observe that the asymptotic \ac{snr} of an active RIS aided SU-SISO system depends on both the \ac{bs} transmit power $P_{{\text{BS}}}^{{\max}}$ and the reflect power of the active \ac{ris} $P_{\text{A}}^{{\max}}$. When $P_{{\text{BS}}}^{{\max}} \to \infty$, it can be proved that the asymptotic \ac{snr} of the active RIS aided system will be upper-bounded by ${\gamma _{{\text{active}}}} \to N\frac{{P_{\rm{A}}^{\max }{\pi ^2}\varrho _f^2}}{{16{\sigma ^2}}}$, which only depends on the RIS-user channel gain $\varrho _f^2$ and the noise power at the user $\sigma^2$. This indicates that, when the \ac{bs} transmit power is high enough, the \ac{bs}-RIS channel $\bf g$ and the noise power at the active RIS have negligible impact on the user's \ac{snr}. Similarly, if $P_{\rm{A}}^{\max } \to \infty$, the asymptotic \ac{snr} ${\gamma _{{\text{active}}}}$ in (\ref{eqn:ag_active}) will be upper-bounded by ${\gamma _{{\text{active}}}} \to N\frac{{P_{{\text{BS}}}^{\max }{\pi ^2}\varrho _g^2}}{{16\sigma _v^2}}$. Compared with (\ref{eqn:ag_active}), this upper bound is independent of the RIS-user channel $\bf f$ and the noise power at the user ${\sigma^2}$. It indicates that, the negative impact of small $\bf f$ and large ${\sigma^2}$ can be reduced by increasing the reflect power of the active RIS $P_{\text{A}}^{{\max}}$, which may provide guidance for the design of practical active RIS-aided systems.
\end{remark}

Next, we compare the asymptotic \acp{snr} for passive RISs in {\it Lemma 1} and active RISs in {\it Lemma 2} to reveal the superiority of active \acp{ris} in wireless communications.

\subsection{Comparisons between Passive RISs and Active RISs}\label{subsec:PA2}
\begin{figure*}[!t]
	\centering
	\subfigure[$N$ ranges from 10 to 1000.]{\includegraphics[width=0.5\textwidth]{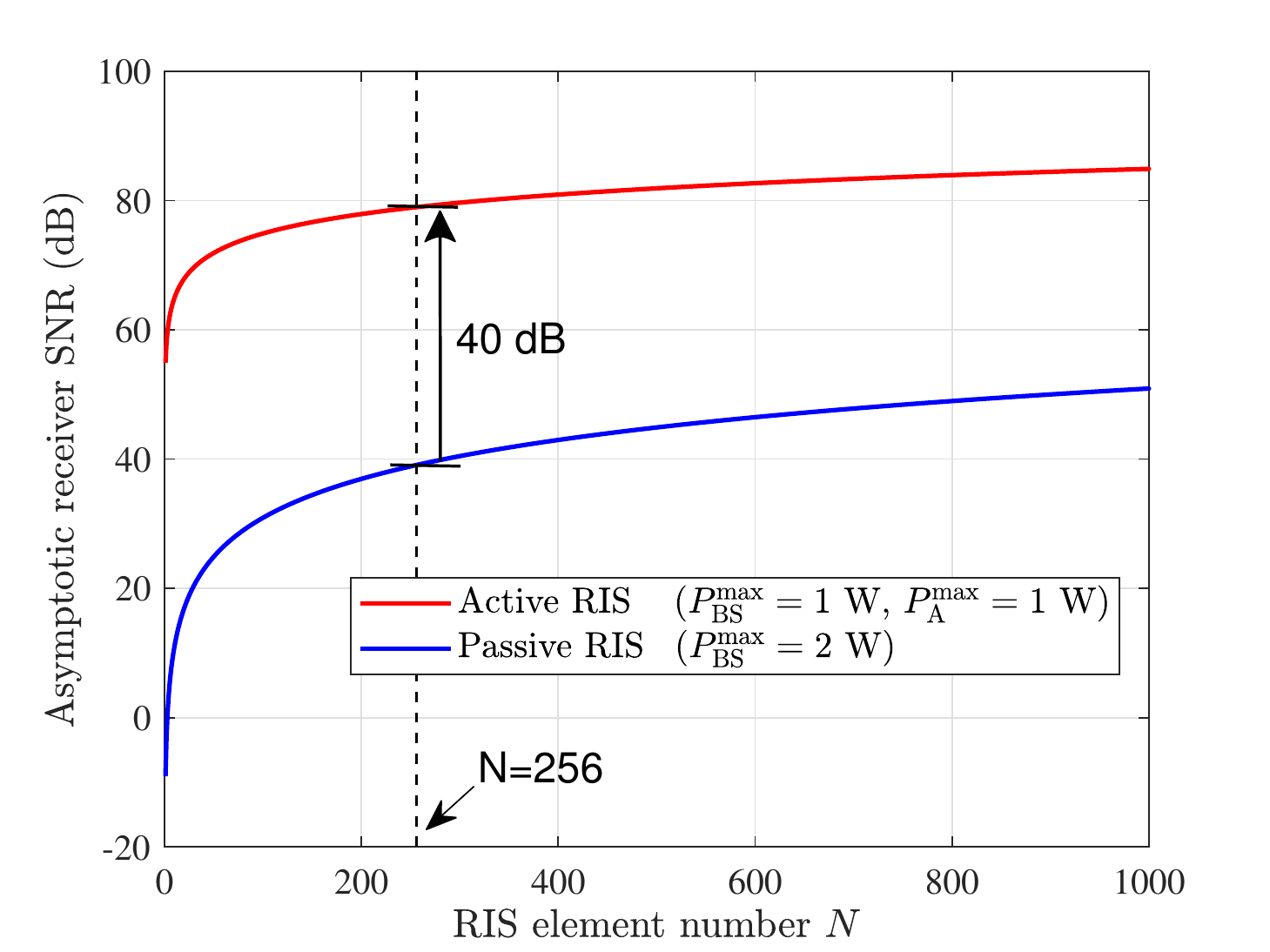}}
	\hspace{-6mm}
	\subfigure[$N$ ranges from $10^4$ to $3\times10^6$.]{\includegraphics[width=0.5\textwidth]{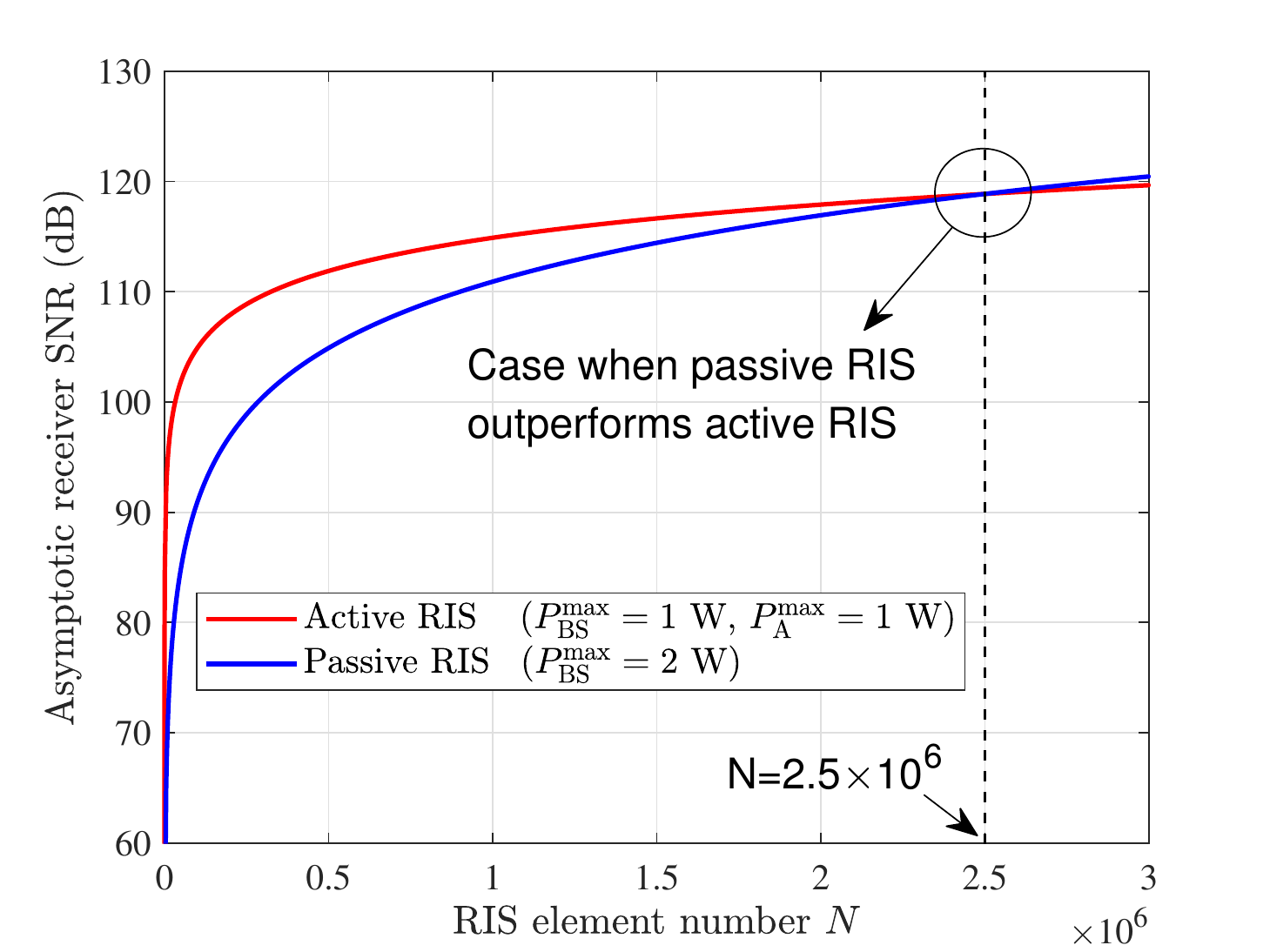}}
	%\vspace*{-0.5em}
	\caption{Asymptotic \ac{snr} as a function of the number of RIS elements $N$ for different ranges of $N$.}
	\label{img:pa}
	%\vspace{-1em}
\end{figure*}

We can observe from {\it Lemma 1} and {\it Lemma 2} that, compared to the asymptotic \ac{snr} for passive RISs $\gamma_{\text{passive}}$ in (\ref{eqn:ag_passive}) which is proportional to $N^2$, the asymptotic \ac{snr} for active RISs $\gamma_{\text{active}}$ in (\ref{eqn:ag_active}) is proportional to $N$ due to the noises additionally introduced by the use of active components. At first glance, it seems that the \ac{snr} proportional to $N^2$ achieved by passive RISs $\gamma_{\text{passive}}$ always exceeds the \ac{snr} proportional to $N$ achieved by active RISs $\gamma_{\text{active}}$. However, this is actually not the case in many scenarios.

The reason behind this counterintuitive behavior is that, different from the denominator of (\ref{eqn:ag_passive}) which depends on the noise power $\sigma^2$, the denominator of (\ref{eqn:ag_active}) is determined by the much smaller multiplicative terms composed of path losses and noise power, i.e., $P_{\text{A}}^{{\max}}\sigma _v^2\varrho _f^2$, $P_{{\text{BS}}}^{{\max}}{\sigma ^2}\varrho _g^2$, and ${\sigma ^2}{\sigma _v^2}$. In this case, the denominator of (\ref{eqn:ag_active}) is usually much smaller than that of (\ref{eqn:ag_passive}). Thus, even if the numerator of (\ref{eqn:ag_active}) is smaller than that of (\ref{eqn:ag_passive}) because of an $N$ gain loss, the SNR gain of active RISs can still be higher than that of passive RISs in many scenarios.

Generally, due to the much smaller denominator of (\ref{eqn:ag_active}), only when $N$ is unaffordably large can passive RISs outperform active RISs. To illustrate this claim, let us consider two different SU-SISO systems, which are aided by an active RIS and a passive RIS, respectively. Then, the following lemma specifies the condition that has to be met for passive RISs to outperform active RISs.

%In essence, the reason that results in the different denominator of (\ref{eqn:ag_passive})
% between active RIS and passive RIS are
%The reason behind this counterintuitive behavior is that, due to the large path loss caused by the ``multiplicative fading'' effect and thanks to the use of the reflection-type amplifiers in active RISs, only when $N$ is unaffordably large can passive RISs outperform active RISs.
\begin{lemma}[Case when passive RISs outperform active RISs]
Assuming the number of RIS elements $N$ is large, the required number of elements $N$ for a passive RIS to outperform an active RIS has to satisfy
\begin{align}\label{eqn:N_>}
	N \ge \frac{{P_{{\text{BS-A}}}^{{\max}}}}{{P_{{\text{BS-P}}}^{{\max}}}}\frac{{P_{\text{A}}^{{\max}}{\sigma ^2}}}{{\left( {P_{\text{A}}^{{\max}}\sigma _v^2\varrho _f^2 + P_{{\text{BS-A}}}^{{\max}}{\sigma ^2}\varrho _g^2 + {\sigma ^2}\sigma _v^2} \right)}},
\end{align}
where ${P^{{\max}}_{\text{BS-A}}}$ denotes the maximum BS transmit power for the active RIS aided system and ${P^{{\max}}_{\text{BS-P}}}$ denotes that for the passive RIS aided system.
\end{lemma}
\begin{IEEEproof}
Please see Appendix B.
\end{IEEEproof}

Next, we consider a specific setup to compare the user's achievable \acp{snr} in the above two systems. For a fair comparison, we constrain the total power consumption $P^{\max}$ of the two systems to $2$ W by setting $P_{\text{BS-P}}^{\max}=2$ W for the passive RIS aided system and $P_{\text{BS-A}}^{\max}=P_{\text A}^{\max}=1$ W for the active RIS aided system, respectively. Therefore, when $\sigma^2=\sigma _{v}^2=-100$ dBm and ${\varrho _f^2}={\varrho _g^2}=-70$ dB, the required number of elements $N$ for the passive RIS to outperform the active RIS is $2.5\times 10^6$ according to (\ref{eqn:N_>}), which is impractical to realize with current technology. Besides, the high overhead for channel estimation \cite{Huchen} and the high complexity for real-time beamforming \cite{Pan'19} also make the application of such a large number of RIS elements impractical \cite{Najafi'20}. Conversely, for a more practical number of elements of $N=256$, according to (\ref{eqn:ag_active}) and (\ref{eqn:ag_passive}), the \ac{snr} achieved by the passive RIS is $\gamma_{\text{passive}}\approx39.0$ dB, while the \ac{snr} achieved by the active RIS is $\gamma_{\text{active}}\approx79.0$ dB, which is about $10,000$ times higher than $\gamma_{\text{passive}}$.

Based on the above parameters, we show the asymptotic \ac{snr} versus the number of RIS elements $N$ for both passive RISs and active RISs in Fig. \ref{img:pa}, where $N$ ranges from $10$ to $1000$ in Fig. \ref{img:pa} (a) and from $10^4$ to $3\times10^6$ in Fig. \ref{img:pa} (b). From this figure we can observe that, when $N$ ranges from $10$ to $1000$, the user's achievable \ac{snr} is about 40 dB higher in an active RIS aided system compared to a passive RIS aided system. Only when $N=2.5\times 10^6$ becomes the performance gain achieved by the passive RIS comparable to that achieved by the active RIS, which agrees well with our above analysis.
\par
\begin{remark}
From the above comparisons we find that, although additional thermal noise is introduced by the active components, active RISs can still achieve a higher \ac{snr} gain than passive RISs. This is due to the fact that, the desired signals reflected by different active RIS elements are coherently added with the same phase at the user, while the introduced noises are not. Besides, when these introduced noises are received by the user, they have become much small due to the RIS-user path loss. In addition, different from the passive RIS aided system that all radiation power suffers from the multiplicative path loss of reflected links, the power radiated by active RISs only experiences the large fading of RIS-user link, thus the power attenuation is much less and the “multiplicative fading” effect can be well overcome. %It is also worth noting that, for the BS-RIS-user channel estimations, since active RISs have no ability to process baseband signals, only the cascaded channels can be estimated \cite{Huchen}. Thus, the estimation overhead for active RISs is almost same as that for passive RISs. However, since the receiver in active RIS aides systems is expected to receive pilots with higher SNR, compared to passive RISs, the system assisted by active RISs can achieve higher channel estimation accuracy. The trade-off between overhead and accuracy can be a potential research direction in future works
\end{remark}

\subsection{Impact of Distances on RIS Performances}\label{subsec:PA3}
According to (\ref{eqn:ag_passive}) and (\ref{eqn:ag_active}), the path losses of the wireless links are the key parameters influencing the RIS performances. Since the path losses highly depend on the inter-device distances, in this section, we analyze the impact of distances on the SNR gain of active RISs and passive RISs. 

To characterize the relationship between distances and path losses, without loss of generality, we assume that the large-scaling fading of BS-RIS channel $\bf g$ and RIS-user channel $\bf f$ follow the far-field spherical-wave propagation model, which is widely used in existing works such as \cite{Wu'19,Pan'19}. Thus, the BS-RIS path loss ${\varrho _g^2}$ and the RIS-user path loss ${\varrho _f^2}$ can be rewritten as:
\begin{align}\label{eqn:g_and_f}
	\varrho _g^2 = {L_0}{d_{\rm t}}^{ - \alpha } ~~{\text{and}}~~~
	\varrho _f^2 = {L_0}{d_{\rm r}}^{ - \beta },
\end{align}
where $L_0$ is the path loss at the reference distance of 1 m, which is usually set to $L_0=-30$ dB \cite{Wu'19}; $d_{\rm t}$ and $d_{\rm r}$ denotes the BS-RIS distance and the RIS-user distance, respectively; $\alpha$ and $\beta$ denote the path loss exponents of BS-RIS channel and RIS-user channel, respectively, whose values usually range from 2 to 4. To find more insights, here we assume that ${\sigma^2}={\sigma_v^2}$, ${P_{{\text{BS-P}}}^{\max}}={P^{\max}}$, and ${P_{{\text{BS-A}}}^{\max}}={P_{{\text{A}}}^{\max}}={P^{\max}}/{2}$, wherein $P^{\max}$ denotes the total radiation power. Then, we obtain the following lemma.

\begin{lemma}[Scenario where active RISs outperform passive RISs]
	Given a large number of RIS elements $N$, the scenario where an active RIS can outperform a passive RIS should satisfy
	\begin{align}\label{eqn:scenario_outperform}
\frac{1}{{{d_{\rm t}}^{ - \alpha } + {d_{\rm r}}^{ - \beta }}} \ge \frac{{2N{P^{\max }}{L_0}}}{{{P^{\max }} - 4N{\sigma ^2}}}.
	\end{align}
\end{lemma}
\begin{IEEEproof}
	Substitute (\ref{eqn:g_and_f}) into (\ref{eqn:double_snr}) in Appendix B and then solve $\gamma_{\text{passive}}\le\gamma_{\text{active}}$ by exploiting $\frac{1}{N}\sum\nolimits_{n = 1}^N {{{\left| {{f_n}} \right|}^2}} \approx \varrho _f^2$ and $\frac{1}{N}\sum\nolimits_{n = 1}^N {{\left| {{g_n}} \right|}^2} \approx \varrho _g^2$. This completes the proof.
\end{IEEEproof}

From (\ref{eqn:scenario_outperform}) one can notice that, active RISs can outperform passive RISs in many scenarios. The reason is that, distances $d_{\rm t}$ and $d_{\rm r}$ are usually large, which makes the left part of (\ref{eqn:scenario_outperform}) very large. By contrast, due to the large path loss $L_0=-30$ dB, the right part of (\ref{eqn:scenario_outperform}) is usually small, which results in the fact that the inequality (\ref{eqn:scenario_outperform}) follows in many practical scenarios. To see the above fact, here we fix the BS-RIS distance as $d_{\rm t}=20$ m and consider the following parameters:  $L_0=-30$ dB, $\alpha=\beta=2$, $P^{\max}=2$ W, $\sigma^2=-100$ dBm, and $N=1024$. Then, we can calculate from (\ref{eqn:scenario_outperform}) that, active RISs can outperform passive RISs as long as the RIS-user distance $d_{\rm r}$ satisfy ${d_{\rm r}} \ge {\left( {\frac{{{P^{\max }} - 4N{\sigma ^2}}}{{2N{P^{\max }}{L_0}}} - {d_{\rm t}}^{ - \alpha }} \right)^{ - \frac{1}{\beta} }} = 1.43$ m, which nearly covers the whole wireless communication region. In other words, to achieve the same performance, active RISs can be located much far away from terminals compared to passive RISs, which is one more advantage of using active RISs.

\section{Joint Transmit Beamforming and Reflect Precoding Design}\label{sec:Precoding}
To investigate the capacity gain enabled by active RISs in typical communication scenarios, in this section, we consider a more general active RIS aided MU-MISO system. Specifically, in Subsection \ref{subsec:IV-A}, we formulate the problem of sum-rate maximization. Then, in Subsection \ref{subsec:IV-B}, a joint transmit beamforming and reflect precoding scheme is proposed to solve the problem.
	
\subsection{Sum-Rate Maximization Problem Formulation}\label{subsec:IV-A}	
According to the MU-MISO transmission model in (\ref{eqn:signal}), the \ac{sinr} at user $k$ can be obtained as
	\begin{equation}
		{\gamma _k} = \frac{{{{\left| {{{\bf{\bar h}}^{\rm H}_k}{{\bf{w}}_k}} \right|}^2}}}{{\sum\nolimits_{j = 1,j \ne k}^K {{{\left| {{{\bf{\bar h}}^{\rm H}_k}{{\bf{w}}_j}} \right|}^2}}  + {{\left\| {{{\bf{f}}_k^{\rm H}}{{\bm\Psi}}} \right\|}^2}\sigma _{v}^2 + {\sigma ^2}}},
	\end{equation}
	wherein ${{\bf{\bar h}}^{\rm H}_k} = {{\bf{h}}^{\rm H}_k} + {{\bf{f}}^{\rm H}_k}{{\bm\Psi}}{\bf{G}}\in{\mathbb{C}^{1\times M}}$
	is the equivalent channel from the \ac{bs} to user $k$, which includes both the direct link and the reflected link. By solving the expectation of the squared Euclidean norm of the radiated signals, the \ac{bs} transmit power, $P_{\text{BS}}$, and the reflect power of the active RIS, $P_{\text{A}}$, can be respectively derived as
\begin{subequations}\label{eqn:power_consumption}
	\begin{align}
		P_{\text{BS}}&= {\mathbb E}\left\{ {{{\left\| {\sum\limits_{k = 1}^K {{{\bf{w}}_k}{s_k}} } \right\|}^2}} \right\}=\underbrace{\sum\limits_{k = 1}^K {{{\left\| {{{\bf{w}}_k}} \right\|}^2}}}_{\text{Desired signal power}}, \\
		P_{\text{A}}&= {\mathbb E}\left\{ {{{\left\| {{\bf{\Psi G}}\sum\limits_{k = 1}^K {{{\bf{w}}_k}{s_k} + } {\bf{\Psi v}}} \right\|}^2}} \right\} \notag \\&=
		 \underbrace{\sum\limits_{k = 1}^K {{{\left\| {{{\bm\Psi}}{\bf{G}}{{\bf{w}}_k}} \right\|}^2}}}_{\text{Desired signal power}}  +  \underbrace{{\left\| {{\bm\Psi}} \right\|^2_{\rm F}}\sigma_{v}^2}_{\text{Amplified noise power}}.
	\end{align}
\end{subequations}
Note that, different from the BS transmit power $P_{\text{BS}}$ which only includes the desired signal power, since the active RIS amplifies the noises as well, the additional power consumption due to the noise amplification should be taken into account in the reflect power of active RIS $P_{\text{A}}$. %Similar definition of $P_{\text{A}}$ in (\ref{eqn:power_consumption}) can be found in many works on active scatters, such as multi-antenna AF relays \cite[Eqn. (9b)]{Yang-wen'11}.

Therefore, the original problem of sum-rate maximization, subject to the power constraints at the \ac{bs} and the active RIS, can be formulated as follows:
\begin{subequations}\label{eqn:problem}
\begin{align}
	\!\!\!\!{\cal P}_o: \mathop {\max }\limits_{{\bf{w}},{{\bm\Psi}}} ~~~ &R_{\rm{sum}}({\bf{w}},{{\bm\Psi}})= \sum\limits_{k = 1}^K {{{\log }_2}\left( {1 + {\gamma _k}} \right)}, \label{eqn:sum-rate} \\
	{\rm s.t.}~~~~ &{\rm C_1}\!:\sum\limits_{k = 1}^K {{{\left\| {{{\bf{w}}_k}} \right\|}^2}} \le {P^{\max}_{\text{BS}}},  \\
	&{\rm C_2}\!: \sum\limits_{k = 1}^K\! {{{\left\| {{{\bm\Psi}}{\bf{G}}{{\bf{w}}_k}} \right\|}^2}  + \left\| {{\bm\Psi}} \right\|^2_{\rm F}\sigma_{v}^2}\le {P^{\max}_{\text{A}}}, 	\label{eqn:problem_C2} 
\end{align}
\end{subequations}
	where ${\bf{w}} := {\left[ {{\bf{w}}_1^{\rm T}, \cdots ,{\bf{w}}_K^{\rm T}} \right]^{\rm T}}$ is the overall transmit beamforming vector for the $K$ users; ${\rm C_1}$ and ${\rm C_2}$ are the power constraints at the BS and active RIS, respectively. 
	
	Due to the non-convexity and highly coupled variables in problem ${\cal P}_o$ in (\ref{eqn:problem}), the joint design of ${\bf{w}}$ and ${\bm \Psi}$ is challenging. Specifically, the introduction of the active RIS brings many difficulties to the beamforming design, such as the additional power constraint, the power allocation among active elements, the cancellation of multi-user interference, and the amplified noise power. Therefore, to efficiently solve this problem, we develop a joint beamforming and precoding scheme based on alternating optimization and fractional programming (FP), as provided
	in the next subsection.
\begin{algorithm}[!t] 
	\caption{Proposed joint transmit beamforming and reflect precoding scheme} 
	\begin{algorithmic}[1] %这个1 表示每一行都显示数字
		\REQUIRE ~~ %算法的输入参数：Input
		Channels ${\bf{G}}$, ${\bf{h}}_k$, and ${\bf{f}}_k$, $\forall k\in\{1,\cdots,K\}$.
		\ENSURE ~~ %算法的输出：Output
		Optimized \ac{bs} beamforming vector $\bf{w}$, optimized RIS precoding matrix of active RIS $\bm{\Psi}$, and optimized sum-rate $R_{\rm{sum}}$.	
		\STATE Randomly initialize $\bf{w}$ and $\bm{\Psi}$;
		\WHILE {no convergence of $R_{\rm{sum}}$}
		\STATE Update ${\bm \rho}$ by (\ref{eqn:rho_update});
		\STATE Update $\bm{\varpi}$ by (\ref{eqn:update_varpi});
		\STATE Update ${\bf w}$ by solving (\ref{eqn:problem_9});		
		\STATE Update ${\bm \Psi}$ by solving (\ref{eqn:problem_10});
		\STATE Update $R_{\rm{sum}}$ by (\ref{eqn:sum-rate});
		\ENDWHILE	
		\RETURN Optimized $\bf{w}$, $\bm{\Psi}$, and $R_{\rm{sum}}$. %算法的返回值
	\end{algorithmic}
\end{algorithm}
	\subsection{Proposed Joint Beamforming and Precoding Scheme}\label{subsec:IV-B}
	To solve the problem efficiently, we reformulate the problem first. For simplicity, here we refer to ${\bf w}$ and $\bm \Psi$ as the BS beamforming vector and the RIS precoding matrix, respectively. In order to deal with the non-convex sum-of-logarithms and fractions in (\ref{eqn:problem}), we exploit FP methods proposed in \cite{Shen'18'1} to decouple the variables in problem ${\cal P}_o$ in (\ref{eqn:problem}), so that multiple variables can be optimized separately. This leads to the following lemma.
	\begin{lemma}[Equivalent problem for sum-rate maximization]
		 By introducing auxiliary variables ${\bm\rho}:=\left[\rho_1,\cdots,\rho_K\right]\in{\mathbb R}^{K}_+$ and  $\bm \varpi:=\left[\varpi_1,\cdots, \varpi_K\right]\in{\mathbb C}^{K}$, the original problem ${\cal P}_o$ in (\ref{eqn:problem}) can be equivalently reformulated as follows
	\begin{equation}\label{eqn:problem_6}
		\begin{aligned}
		{\cal P}_1:	\mathop {\max }\limits_{{\bf{w}},{\bm{\Psi}}, {\bm{\rho}},{\bm\varpi}}  &R_{\rm sum}'({\bf{w}},{\bm{\Psi}},{\bm \rho},{\bm \varpi}) = \sum\limits_{k = 1}^K {{{\ln }}\left( {1 + {\rho _k}} \right)}  -\\&~~~~~~~~~~~~ \sum\limits_{k = 1}^K {\rho _k} +\sum\limits_{k = 1}^K {g({\bf{w}},{\bm{\Psi}},{\rho_k},{\varpi_k})},  \\
			~~{\rm s.t.}~~ &{\rm C_1}\!: {{{\left\| {{{\bf{w}}}} \right\|}^2}} \le {P^{\max}_{\text{BS}}},\\
			\quad \,\,\,\,\,\quad &{\rm C_2}\!: \sum\limits_{k = 1}^K {{{\left\| {{{\bm{\Psi}}}{\bf{G}}{{\bf{w}}_k}} \right\|}^2}  + \left\| {{{\bm{\Psi}}}} \right\|^2_{\rm F}\sigma _{v}^2}\le {P^{\max}_{\text{A}}},
		\end{aligned}
	\end{equation}
	where function $g({\bf{w}},{\bm{\Psi}},{\rho_k},{\varpi_k})$ is defined as
	\begin{equation}
		\begin{aligned}
			g({\bf{w}},{\bm{\Psi}},{\rho_k},{\varpi_k})= 2\sqrt {\left( {1 + {\rho _k}} \right)} {\mathop{{\mathfrak R}}\nolimits} \left\{ {{\varpi_k ^*}{\bf{\bar h}}_k^{\rm H}{{\bf{w}}_k}} \right\} -\\ {\left| \varpi_k \right|^2}\left( {\sum\limits_{j = 1}^K {{{\left| {{\bf{\bar h}}_k^{\rm H}{{\bf{w}}_j}} \right|}^2}}  + {{\left\| {{{\bf{f}}_k^{\rm H}}{{\bf{\Psi }}}} \right\|}^2}\sigma _{v}^2 + {\sigma ^2}} \right).
		\end{aligned}
	\end{equation}
	\end{lemma}
	\begin{IEEEproof}
		Constructive proof can be found in \cite[Subsection III-C]{Shen'18'1}.
	\end{IEEEproof}
	\par
	Strong convergence of the FP methods was proved in \cite{Shen'18'1}. Thus, if the updates in each iteration step of the \ac{bs} beamforming vector ${\bf{w}}$, \ac{ris} precoding matrix ${\bm{\Psi}}$, auxiliary variables ${\bm{\rho}}$ and ${\bm\varpi}$ in (\ref{eqn:problem_6}) are all optimal, a locally optimal solution to (\ref{eqn:problem_6}) can be obtained by alternately optimizing these variables until $R_{\rm{sum}}$ converges.  For clarity, we summarize the proposed joint beamforming and precoding scheme in {\bf Algorithm 1}, and the specific optimal solutions for variables ${\bf{w}}$, ${\bm{\Psi}}$, ${\bm{\rho}}$, and ${\bm\varpi}$ are given in the following four steps, respectively.
	
	\subsubsection{Fix $\left({\bf{w}},{\bf{\Psi}},{\bm \varpi}\right)$ and then optimize ${\bm \rho}$} 
	After fixing \ac{bs} beamforming vector ${\bf{w}}$, \ac{ris} precoding matrix ${\bf{\Psi}}$, and auxiliary variable ${\bm \varpi}$, the optimal $\bm \rho$ can be obtained by solving $\frac{{\partial {R_{{\rm{sum}}}'}}}{{\partial {\rho _k}}} = 0$ as 
	\begin{equation}\label{eqn:rho_update}
		\begin{aligned}
			\rho^{\rm opt}_k  = \frac{{{\xi_k ^2} + \xi_k \sqrt {{\xi_k ^2} + 4} }}{2},\quad\forall k\in \{1,\cdots,K\},
		\end{aligned}
	\end{equation}
	where $\xi_k  = \Re \left\{ {\varpi _k^*{\bf{\bar h}}_k^{\rm H}{{\bf{w}}_k}} \right\}$.
	\subsubsection{Fix $\left({\bf w},{\bf{\Psi}},{\bm \rho}\right)$ and then optimize $\bm \varpi$}
	After fixing the \ac{bs} beamforming vector ${\bf{w}}$, \ac{ris} precoding matrix ${\bf{\Psi}}$, and auxiliary variable ${\bm \rho}$, the optimal $\bm \varpi$ can be derived by solving $\frac{\partial {R_{{\rm{sum}}}'}}{{\partial \varpi_k }} = 0$ as
	\begin{equation}\label{eqn:update_varpi}
		\begin{aligned}
			{\varpi_k^{{\rm{opt}}}} =& \frac{{\sqrt {\left( {1 + {\rho _k}} \right)} {\bf{\bar h}}_k^{\rm H}{{\bf{w}}_k}}}{{\sum\nolimits_{j = 1}^K {{{\left| {{\bf{\bar h}}_k^{\rm H}{{\bf{w}}_j}} \right|}^2}}  + {{\left\| {{\bf{f}}_k^{\rm H}{{\bf{\Psi }}}} \right\|}^2}\sigma _{v}^2 + {\sigma ^2}}}, \\ & \forall k\in \{1,\cdots,K\}.
		\end{aligned}
	\end{equation}
	\subsubsection{Fix $\left({\bf{\Psi}},{\bm \rho},{\bm \varpi}\right)$ and then optimize ${\bf w}$}
	To simplify the notations, we first introduce the following definitions:
\begin{subequations}\label{eqn:varibles}
	\begin{align}
		&{\bf{b}}_k^{\rm H} = \sqrt {\left( {1 + {\rho _k}} \right)} \varpi _k^*{\bf{\bar h}}_k^{\rm H},~~
		{\bf b} = \left[{{\bf b}_{1}^{\rm T}},{{\bf b}_{2}^{\rm T}}, \cdots, {{\bf b}_{N}^{\rm T}}\right]^{\rm T},\\
		&{\bf{A}} \!=\! {{\bf{I}}_K} \!\otimes\! \sum\limits_{k = 1}^K {{\left| {{\varpi _k}} \right|^2}{{\bf{\bar h}}_k}{\bf{\bar h}}_k^{\rm H}},
		~{\bf{\Xi }} \!=\! {{\bf{I}}_K} \otimes \left( {{{\bf{G}}^{\rm H}}{{\bf{\Psi }}^{\rm H}}{\bf{\Psi G}}} \right), \\&P_{\text m}^{\max } = P_{{\text{A}}}^{{\max}} - {\left\| {{{\bf{\Psi }}}} \right\|^2_{\rm F}}\sigma _{v}^2.
	\end{align}
\end{subequations}
Then, for fixed RIS precoding matrix ${\bf{\Psi}}$ and auxiliary variables ${\bm \rho}$ and ${\bm \varpi}$, problem ${\cal P}_1$ in (\ref{eqn:problem_6}) can be reformulated as follows
	\begin{equation}\label{eqn:problem_9}
		\begin{aligned}
		{\cal P}_2:~~	\mathop {\max }\limits_{{\bf{w}}} ~~ & {\mathop{\mathfrak R}\nolimits} \left\{2 {{{\bf{b}}^{\rm H}}{\bf{w}}} \right\} - {{\bf{w}}^{\rm H}}{\bf{Aw}},  \\
			{\rm s.t.}~~~ &{\rm C_1}: {{{\left\| {{{\bf{w}}}} \right\|}^2}} \le {P^{\max}_{\text{BS}}},\\
			\quad ~~~~ &{\rm C_2}: {{\bf{w}}^{\rm H}}{\bf{\Xi w}} \le P_{\text m}^{\max}.
		\end{aligned}
	\end{equation}
Since ${\cal P}_2$ in (\ref{eqn:problem_9}) is a standard quadratic constraint quadratic programming (QCQP) problem, by adopting the Lagrange multiplier method \cite{admm}, the optimal solution ${\bf w}^{\rm opt}$ to ${\cal P}_2$ in (\ref{eqn:problem_9}) can be obtained as follows
	\begin{equation}\label{eqn:opt_w}
		\begin{aligned}
			{{\bf{w}}^{\rm opt}} = {\left( {{\bf{A}} + {\lambda _1}{{\bf{I}}_{MK}} + {\lambda _2}{\bf{\Xi }}} \right)^{ - 1}}{\bf{b}},
		\end{aligned}
	\end{equation}
	where $\lambda _1$ and $\lambda _2$ are the Lagrange
	multipliers, which should be chosen such that the complementary slackness conditions of power constrains ${\rm C_1}$ and ${\rm C_2}$ are satisfied. The optimal Lagrange
	multipliers $\lambda _1^{\rm opt}$ and $\lambda_2^{\rm opt}$ can be obtained via a two-dimensional grid search \cite{admm}.
	
	\subsubsection{Fix $\left({\bf w},{\bm \rho},{\bm \varpi}\right)$ and then optimize ${\bf{\Psi}}$}
	Define ${\bm{\psi}}=\left[p_1e^{j\theta_1},\cdots,p_Ne^{j\theta_N}\right]^{\rm H}$ as the vectorized RIS precoding matrix ${\bm{\Psi}}$, i.e., ${\rm diag}\left({\bm{\psi}}^{\rm H}\right):= {\bf{\Psi}}$. Thus, the equivalent channel ${{\bf{\bar h}}^{\rm H}_k}$ can be rewritten as follows:
\begin{equation}\label{eqn:equivalent_channel}
	\begin{aligned}
{\bf{\bar h}}_k^{\rm{H}} = {\bf{h}}_k^{\rm{H}} + {{\bf{f}}^{\rm{H}}_k}{\bf{\Psi G}} = {\bf{h}}_k^{\rm{H}} + {{\bm{\psi }}^{\rm{H}}}{\rm{diag}}\left( {{{\bf{f}}^{\rm{H}}_k}} \right){\bf{G}}.
	\end{aligned}
\end{equation}	
Utilizing (\ref{eqn:equivalent_channel}), while fixing \ac{bs} beamforming vector ${\bf w}$ and auxiliary variables ${\bm \rho}$ and ${\bm \varpi}$, problem ${\cal P}_1$ in (\ref{eqn:problem_6}) can be reformulated as follows:
	\begin{equation}\label{eqn:problem_10}
		\begin{aligned}
		{\cal P}_3:~~\mathop {\max }\limits_{{\bm{\psi}}} \,\,\,\, &{\mathop{\mathfrak R}\nolimits} \left\{2 {{{\bm{\psi }}^{\rm H}}{\bm{\upsilon }}} \right\} - {{\bm{\psi }}^{\rm H}}{\bm{\Omega \psi }},  \\
			{\rm s.t.}~~ &{\rm C_2}: {{\bm{\psi }}^{\rm H}}{\bm{\Pi \psi }} \le P_{{\text{A}}}^{{\max}},
		\end{aligned}
	\end{equation} 
	wherein
\begin{subequations}
	\begin{align}
	{{\bm{\upsilon }}} &= \sum\limits_{k = 1}^K\sqrt {\left( {1 + {\rho _k}} \right)} {\rm{diag}}\left( {\varpi _k^*{\bf{f}}_k^{\rm H}} \right){\bf{G}}{{\bf{w}}_k} -\notag\\&~~~~~~~~~~~ \sum\limits_{k = 1}^K{\left| {{\varpi _k}} \right|^2}{\rm{diag}}\left( {{\bf{f}}_k^{\rm H}} \right){\bf{G}}\sum\limits_{j = 1}^K {{{\bf{w}}_j}{\bf{w}}_j^{\rm H}} {{\bf{h}}_k},\\
	{{\bm{\Omega }}} &= \sum\limits_{k = 1}^K{\left| {{\varpi _k}} \right|^2}{\rm{diag}}\left( {{\bf{f}}_k^{\rm H}} \right){\rm{diag}}\left( {{{\bf{f}}_k}} \right)\sigma _{v}^2 +\notag\\&~~~~ \sum\limits_{k = 1}^K{\left| {{\varpi _k}} \right|^2}\sum\limits_{j = 1}^K {{\rm{diag}}\left( {{\bf{f}}_k^{\rm H}} \right){\bf{G}}{{\bf{w}}_j}{\bf{w}}_j^{\rm H}{{\bf{G}}^{\rm H}}{\rm{diag}}\left( {{{\bf{f}}_k}} \right)},\\
	{\bf{\Pi }} &= \sum\limits_{k = 1}^K {{\rm{diag}}\left( {{\bf{G}}{{\bf{w}}_k}} \right){{\left( {{\rm{diag}}\left( {{\bf{G}}{{\bf{w}}_k}} \right)} \right)}^{\rm H}}}  + \sigma _{v}^2{{\bf{I}}_N}.
	\end{align}
\end{subequations}

Note that problem ${\cal P}_3$ in (\ref{eqn:problem_10}) is also a standard QCQP problem. Thus, the optimal solution ${\bm{\psi}}^{\rm opt}$ can be obtained by adopting the Lagrange multiplier method and is given by
\begin{align}\label{eqn:psi_clos}
{\bm{\psi}^{\rm opt}}={\left( {{\bm{\Omega}  + \mu \bm{\Pi }}} \right)^{ - 1}}{\bm{\upsilon }},
\end{align}	
where $\mu$ is the Lagrange
multiplier, which should be chosen such that the complementary slackness condition of power constrain ${\rm C_2}$ is satisfied.	Similarly, the optimal Lagrange multiplier $\mu^{\rm opt}$ can be obtained via a binary search \cite{admm}.

% In summary, {\bf Algorithm 1} focuses on tacking algorithmic challenges of the sum-rate maximization problem ${\cal P}_o$ in (\ref{eqn:problem}), which aims to address the most basic issue, i.e., the joint design the BS beamforming and the active RIS precoding. Since the considered channels $\bf G$, ${\bf h}_k$, and ${\bf f}_k$ are arbitrary, regardless of the specific channel state information (CSI), the proposed algorithm works as a feasible solution to improve the performance gain of active RISs in communication systems. Besides, since the proposed algorithm decouples different variables to be optimized, this algorithm has a good expansibility, which can serve as a framework for the possible algorithmic improvements in future works, such as addressing non-ideal CSI and reducing computational complexity.

\section{Self-Interference Suppression for Active RISs}\label{sec:IV-SI}
Since active RISs work in full-duplex (FD) mode, the self-interference of active RISs occurs in practical systems. In this section, we extend the studied joint beamforming and precoding design to the practical system with the self-interference of active RISs. Specifically, in Subsection \ref{subsec:iV-1}, we first model the self-interference of active RISs, which allows us to account for the self-interference suppression in the beamforming design. In Subsection \ref{subsec:iV-2}, we formulate a mean-squared error minimization problem to suppress the self-interference of active RISs. In Subsection \ref{subsec:iV-3}, by utilizing ADMM \cite{admm} and SUMT \cite{fiacco1990nonlinear}, an alternating optimization scheme is proposed to solve the formulated problem.
\subsection{Self-Interference Modeling}\label{subsec:iV-1}
The self-interference of FD relays and that of active RISs are quite different. Specifically, due to the long processing delay at relays, the self-interference of FD relay originates from the different symbols that transmitted in the adjacent timeslot \cite{suraweera2014low,lioliou2010self,xing2016self}. In this case, the self-interference at relays is usually viewed as colored Gaussian noise, which can be canceled by a zero-forcing suppression method \cite{lioliou2010self}. Differently, since active RISs have nanosecond processing delay, the incident and reflected signals carry the same symbol in a timeslot. Due to the non-ideal inter-element isolation of practical arrays, part of the reflected signals may be received again by the active RIS. In this case, the feedback-type self-interference occurs, which cannot be viewed as Gaussian noise anymore. 

To distinguish the RIS precoding matrix in the ideal case $\bf \Psi$, we denote the RIS precoding matrix in the non-ideal case with self-interference as ${\bm \Phi }:={\rm diag}\left(p_1e^{j\theta_1},\cdots,p_Ne^{j\theta_N}\right)$. Recalling (\ref{eqn:active_model}) and ignoring the negligible static noise for simplicity, the reflected signal of active RISs in the presence of self-interference can be modeled as follows:
\begin{equation}\label{eqn:active_model_SI}
	\begin{aligned}
{\bf{y}} = \underbrace {{\bf{\Phi x}}}_{{\text{Desired signal}}} + \underbrace {{\bf{\Phi Hy}}}_{{\text{Self-interference}}} + \underbrace {{\bf{\Phi v}}}_{{\text{Dynamic noise}}},
	\end{aligned}
\end{equation}
where ${\bf H}\in{\mathbb C}^{N\times N}$ denotes the self-interference matrix \cite{suraweera2014low}. In the general case without self-excitation (determinant of $({{\bf{I}}_N} - {\bf{\Phi H}})$ is not zero), model (\ref{eqn:active_model_SI}) is a standard self-feedback loop circuit, of which the output $\bf y$ naturally converges to the following steady state:
\begin{equation}\label{eqn:active_model_SI_2}
	\begin{aligned}
		{\bf{y}} = \underbrace{{\left( {{{\bf{I}}_N} - {\bf{\Phi H}}} \right)^{ - 1}}{\bf{\Phi }}}_{\text{Equivalent RIS precoding matrix}}\!\!\!\!\!\!\!\!\left( {{\bf{x}} + {\bf{v}}} \right).
	\end{aligned}
\end{equation}
Comparing (\ref{eqn:active_model_SI_2}) and (\ref{eqn:active_model}), one can observe that the difference is that the RIS precoding matrix $\bm \Psi$ in (\ref{eqn:active_model}) is replaced by ${\left( {{{\bf{I}}_N} - {\bf{\Phi H}}} \right)^{ - 1}}{\bf{\Phi }}$. In particular, when all elements in $\bf H$ are zero, the equivalent RIS precoding matrix ${\left( {{{\bf{I}}_N} - {\bf{\Phi H}}} \right)^{ - 1}}{\bf{\Phi }}$ is equal to diagonal matrix ${\bf \Phi}$.

\subsection{Problem Formulation}\label{subsec:iV-2}
To account for the self-interference of active RISs in the beamforming design, according to the new signal model (\ref{eqn:active_model_SI_2}), an intuitive way is to replace the RIS precoding matrix $\bm \Psi$ in problem ${\cal P}_1$ in (\ref{eqn:problem_6}) with the equivalent RIS precoding matrix ${\left( {{{\bf{I}}_N} - {\bf{\Phi H}}} \right)^{ - 1}}{\bf{\Phi }}$ and then solve ${\cal P}_1$. Since this operation does not influence the optimizations of ${\bf{w}}$, ${\bm{\rho}}$, and ${\bm\varpi}$, here we focus on the optimization of ${\bm{\Phi}}$.

Consider replacing $\bm \Psi$ in (\ref{eqn:equivalent_channel}) with ${\left( {{{\bf{I}}_N} - {\bf{\Phi H}}} \right)^{ - 1}}{\bf{\Phi }}$, thus the equivalent channel ${\bf{\bar h}}_k^{\rm{H}}$ with self-interference can be written as:
\begin{equation}\label{eqn:equivalent_channel_SI}
	\begin{aligned}
		{\bf{\bar h}}_k^{\rm{H}} = {{\bf h}^{\rm{H}}_k} + {{\bf f}_k^{\rm{H}}}{\left( {{{\bf{I}}_N} - {\bf{\Phi H}}} \right)^{ - 1}}{\bf{\Phi }}{\bf{G}}.
	\end{aligned}
\end{equation}	
However, due to the existence of self-interference matrix $\bf H$, $\bm \Phi$ to be optimized exists in an inversion, thus the equivalent channel ${\bf{\bar h}}_k^{\rm{H}}$ cannot be processed like  (\ref{eqn:equivalent_channel}), which makes $\bm \Phi$ hard to be optimized. To address this challenge, we introduce the first-order Taylor expansion\footnote{ This approximation requires that the self-interference is not too strong (i.e., the values in $\bf H$ are small), so that the high-order expansion items can be reasonably ignored.} to approximate ${\left( {{{\bf{I}}_N} - {\bf{\Phi H}}} \right)^{ - 1}} \approx {{{\bf{I}}_N} + {\bf{\Phi H}}}$, thus (\ref{eqn:equivalent_channel_SI}) can be rewritten as follows:
\begin{equation}\label{eqn:Equivalent_channel_SI_Approx}
	\begin{aligned}
{\bf{\bar h}}_k^{\rm{H}} \approx & {\bf{h}}_k^{\rm{H}} + {{\bf f}_k^{\rm{H}}}\left( {{{\bf{I}}_N} + {\bf{\Phi H}}} \right){\bf{\Phi G}} \\ \stackrel{(a)}{=}& {\bf{h}}_k^{\rm{H}} + \left( {{{\bf f}_k^{\rm{H}}} + {{\bm \phi} ^{\rm{H}}}{\rm{diag}}\left( {{\bf f}_k^{\rm{H}}} \right){\bf{H}}} \right){\bf{\Phi G}} \\ \stackrel{(b)}{=} & {\bf{h}}_k^{\rm{H}} + \underbrace{\left( {{{\bm \phi} ^{\rm{H}}} + {{\bm \phi} ^{\rm{H}}}{{\bf{H}}}_k{\rm{diag}}\left( {{{\bm \phi} ^{\rm{H}}}} \right)} \right)}_{\text{Equivalent precoding vector for user $k$}}{\rm{diag}}\left( {{\bf f}_k^{\rm{H}}} \right){\bf{G}},
	\end{aligned}
\end{equation}	
wherein RIS precoding vector ${\bm \phi}$ satisfies ${\bm \Phi} = {\rm diag}({\bm \phi}^{\rm H})$; $(a)$ holds since ${{\bm \phi} ^{\rm{H}}}{\rm{diag}}({{\bf f}_k^{\rm{H}}} ) = {{\bf f}_k^{\rm{H}}}{\rm{diag}}\left( {{{\bm \phi} ^{\rm{H}}}} \right)$; $(b)$ holds by defining ${{\bf{H}}}_k={\rm diag}({{\bf f}_k ^{\rm{H}}}){\bf H}({\rm diag}({{\bf f}_k ^{\rm{H}}}))^{-1}$. 
\par
Comparing (\ref{eqn:Equivalent_channel_SI_Approx}) and (\ref{eqn:equivalent_channel}), the difference is that the RIS precoding vector $\bm \psi$ in (\ref{eqn:equivalent_channel}) is replaced by the equivalent precoding vector ${{\bm{\phi }}} + {\rm{diag}}\left( {{{\bm{\phi }}}} \right){\bf{H}}_k^{\rm H}{{\bm{\phi }}}$ for user $k$. Therefore, an efficient way to eliminate the impact of self-interference is to design a ${\bm \phi}$ to make all ${{\bm{\phi }}} + {\rm{diag}}\left( {{{\bm{\phi }}}} \right){\bf{H}}_k^{\rm H}{{\bm{\phi }}}$ approach the ideally optimized RIS precoding vector ${\bm \psi}^{\rm opt}$ as close as possible. To achieve this, we temporarily omit the power constraint of active RISs in (\ref{eqn:problem_C2}) and formulate the following mean-squared error minimization problem:
	\begin{equation}\label{eqn:problem_4}
	\begin{aligned}
		{\cal P}_4:\mathop {\min }\limits_{{\bm{\phi}}} \, f\left({\bm{\phi}} \right)=\frac{1}{K}\sum\limits_{k = 1}^K {{{\left\| {\left({{\bm{\phi }}} + {\rm{diag}}\left( {{{\bm{\phi }}}} \right){\bf{H}}_k^{\rm H}{{\bm{\phi }}}\right) -\ {{\bm{\psi }}^{\rm opt}}} \right\|^2}}},
	\end{aligned}
\end{equation} 
where objective $f\left({\bm{\phi}} \right)$ is the cost function, defined as the mean of the squared approximation errors.

\subsection{Proposed Self-Interference Suppression Scheme}\label{subsec:iV-3}
To ensure the communication performance of active RIS aided systems, in this subsection, we propose a self-interference suppression scheme to solve problem ${\cal P}_4$ in (\ref{eqn:problem_4}).

Obviously, in the ideal case without self-interference (i.e., self-interference matrix $\bf H$ is zero matrix), the optimal solution to problem ${\cal P}_4$ in (\ref{eqn:problem_4}) is ${\bm \phi}={\bm \psi}^{\rm opt}$ and satisfies $f\left({\bm{\phi}} \right)=0$. Here we focus on the non-ideal case with a non-zero $\bf H$. In this case, problem ${\cal P}_4$ is challenging to solve due to the three reasons. Firstly, the objective $ f\left({\bm{\phi}} \right)$ is usually non-convex since ${\bf{H}}_k$ is asymmetric and indefinite. Secondly, $f\left({\bm{\phi}} \right)$ is in quartic form with respect to $\bm \phi$ thus ${\cal P}_4$ has generally no closed-form solution. Finally, the coupled term ${\rm{diag}}\left( {{{\bm{\phi }}}} \right){\bf{H}}_k^{\rm H}{{\bm{\phi }}}$ is a non-standard quadratic thus is hard to be preprocessed and optimized like (\ref{eqn:problem_10}). 

To tackle this issue, inspired by ADMM \cite{admm} and SUMT \cite{fiacco1990nonlinear}, we turn to find a feasible solution to problem ${\cal P}_4$ by alternating optimization, as shown in {\bf Algorithm 2}.
\begin{algorithm}[!t] 
	\caption{Proposed self-interference suppression scheme} 
	\begin{algorithmic}[1] %这个1 表示每一行都显示数字
		\REQUIRE ~~ %算法的输入参数：Input
		Ideally optimized active RIS precoding vector ${\bm \psi}^{\rm opt}$, self-interference matrix ${\bf{H}}$, and channel ${\bf{f}}_k$, $\forall k\in\{1,\cdots,K\}$.
		\ENSURE ~~ %算法的输出：Output
		Active RIS precoding matrix ${\bm{\Phi}}$ in the non-ideal case with self-interference.	
		\STATE Initialization: $\bm{\phi} \leftarrow {\bm \psi}^{\rm opt}$, $\bm{\phi}' \leftarrow {\bm \psi}^{\rm opt}$, and $\zeta\leftarrow10^{-3}$;
		\WHILE {no convergence of ${q}\left({\bm{\phi}},{\bm{\phi}}' \right)$}
		\STATE Update ${\bm \phi}$ by (\ref{eqn:f_phi1});
		\STATE Update $\bm{\phi}'$ by (\ref{eqn:f_phi2});
		\STATE Update $\zeta$ by $\zeta \leftarrow 2\times\zeta$;		
		\ENDWHILE
		\STATE 	${\bm{\Phi}} \leftarrow {\rm diag}({\bm{\phi}^{\rm H}})$;
		\RETURN Optimized active RIS precoding matrix ${\bm{\Phi}}$. %算法的返回值
	\end{algorithmic}
\end{algorithm}
The key idea of this algorithm includes two aspects: i) ADMM: Fix some variables and then optimize the others, so that $f({\bm \phi})$ becomes temporarily convex thus can be minimized by alternating optimizations. ii) SUMT: Introduce an initially small but gradually increasing penalty term into the objective, so that the variables to be optimized can converge as an achievable solution to the original problem. 

Following this idea, ${\cal P}_4$ in (\ref{eqn:problem_4}) can be reformulated as
\begin{equation}\label{eqn:problem_5}
	\begin{aligned}
		{\cal P}_5:~\mathop {\min }\limits_{{\bm{\phi}},{\bm{\phi}}'} ~~{q}\left({\bm{\phi}},{\bm{\phi}}' \right) =  f\left({\bm{\phi}},{\bm{\phi}}' \right) + \underbrace{\zeta{\left\| {\bm{\phi}}' - {\bm{\phi}} \right\|^2}}_{\text{Penalty term}},
	\end{aligned}
\end{equation} 
wherein $f\left({\bm{\phi}},{\bm{\phi}}' \right)$ is defined as 
\begin{equation}\label{eqn:f_phi1_phi2}
	\begin{aligned}
f\left({\bm{\phi}},{\bm{\phi}}' \right)=\frac{1}{K}\sum\limits_{k = 1}^K {{{\left\| {\left({{\bm{\phi }}} + {\rm{diag}}\left( {\bm{\phi }}' \right){\bf{H}}_k^{\rm H}{{\bm{\phi }}}\right) - {{\bm{\psi }}^{\rm opt}}} \right\|^2}}}
	\end{aligned}
\end{equation} 
and $\zeta>0$ is the penalty coefficient that increases in each iteration. For simplicity, here we assume $\zeta$ doubles in each update. In particular, when $\zeta \to \infty$, problem ${\cal P}_5$ in (\ref{eqn:problem_5}) is equivalent to ${\cal P}_4$ in (\ref{eqn:problem_4}).

Observing (\ref{eqn:f_phi1_phi2}), we note that $q\left({\bm{\phi}},{\bm{\phi}}' \right)=f\left({\bm{\phi}},{\bm{\phi}}' \right)=f\left({\bm{\phi}} \right)$ when ${\bm{\phi}}'={\bm{\phi}}$. Particularly, when ${\bm{\phi}}$ (or ${\bm{\phi}}'$) is fixed, objective $q\left({\bm{\phi}},{\bm{\phi}}' \right)$ becomes a convex quadratic as a function of ${\bm{\phi}}'$ (or ${\bm{\phi}}$). Therefore, for a given $\zeta$, $q\left({\bm{\phi}},{\bm{\phi}}' \right)$ in (\ref{eqn:problem_5}) can be minimized by optimizing ${\bm{\phi}}$ and ${\bm{\phi}}'$ alternatively. By solving ${{\partial q\left( {{\bm \phi} ,{{\bm \phi}' }} \right)}}/{{\partial {\bm \phi} }}=0$ and  ${{\partial q\left( {{\bm \phi} ,{{\bm \phi}' }} \right)}}/{{\partial {\bm \phi}' }}=0$, we obtain the updating formulas of ${{\bm{\phi }}}$ and ${{\bm{\phi }}}'$ respectively, given by
\begin{equation}\label{eqn:f_phi1}
{{\bm{\phi }}} = {\left( \frac{1}{K}{\sum\limits_{k = 1}^K {{\bf{B}}_k^{\rm H}{{\bf{B}}_k} + \zeta {\bf{I}}_N} } \right)^{ - 1}}\left( {\zeta {{\bm{\phi }}'} + \frac{1}{K}\sum\limits_{k = 1}^K {{\bf{B}}_k^{\rm H}{{\bm{\psi}}^{\rm opt}}} } \right),
\end{equation} 
\begin{equation}\label{eqn:f_phi2}
	{{\bm{\phi }}}' \!=\! {\left( \frac{1}{K}{\sum\limits_{k = 1}^K \!{{\bf{D}}_k^{\rm H}{{\bf{D}}_k} \!+\! \zeta {\bf{I}}}_N } \right)^{\!\! - 1}}\!\!\left( {\zeta {{\bm{\phi }}} \!+\! \frac{1}{K}\sum\limits_{k = 1}^K\! {{\bf{D}}_k^{\rm H}{\left({\bm{\psi}}^{\rm opt}\!-\!{\bm{\phi }}\right)}} }\! \right),
\end{equation} 
where ${\bf{B}}_k = {{\bf{I}}_N + {\rm{diag}}\left( {{{\bm{\phi }}'}} \right){\bf{H}}_k^{\rm H}}$ and ${\bf{D}}_k = {{\rm{diag}}\left( {{\bf{H}}_k^{\rm H}{{\bm{\phi }}}} \right)}$. Besides, due to the existence of penalty term $\zeta{\left\| {\bm{\phi}}' - {\bm{\phi}} \right\|^2}$, as $\zeta$ increases, the converged solution to ${\cal P}_5$ in (\ref{eqn:problem_5}) tends to satisfy ${\bm{\phi}}'={\bm{\phi}}$. After several alternating updates, ${\bm{\phi}}$ and ${\bm{\phi}}'$ will converge to the same value (${\bm{\phi}}={\bm{\phi}}'$), thus we obtain the desired RIS precoding matrix ${\bm \Phi} = {\rm diag}({\bm \phi}^{\rm H})$, which is exactly the output of {\bf Algorithm 2}. 

Recall that we temporarily omitted the power constraint in (\ref{eqn:problem_C2}) while optimizing ${\bm{\phi}}$. Here we introduce a scaling factor $\tau>0$ for ${\bm{\Phi}}$ to satisfy (\ref{eqn:problem_C2}), leading to the final solution ${\bm{\Phi}}^{\rm opt}$, i.e.,
\begin{equation}
{\bm{\Phi}}^{\rm opt} = \tau{\bf{\Phi}}.
\end{equation} 
According to (\ref{eqn:active_model_SI_2}) and (\ref{eqn:power_consumption}), $\tau$ can be obtained by replacing $\bm \Psi$ in $P_{\text A}$ in (\ref{eqn:power_consumption}) with ${\left( {{{\bf{I}}_N} - {\bm{\Phi}}^{\rm opt}{\bf{H}}} \right)^{ - 1}}{{\bm{\Phi}}^{\rm opt}}$ and then doing a binary search to find a proper $\tau$ that satisfies $P_{{\text{A}}} = P_{{\text{A}}}^{{\max}}$. This completes the proposed self-interference suppression scheme.

\section{Convergence and Complexity}\label{subsec:IV-C}
%In this subsection, we analyze the convergence and the computational complexity of the proposed {\bf Algorithm 1} and {\bf Algorithm 2}, respectively.
\subsection{Convergence Analysis}
{\bf Algorithm 1} converges to a local optimal point after several iterations, since the updates in each iteration step of the algorithm are all optimal solutions to the respective subproblems. To prove this, here we introduce superscript $t$ as the iteration index, e.g., ${\bf w}^t$ refers to the transmit beamforming vector at the end of the $t$-th iteration. Then, {\bf Algorithm 1} converges as 
	\begin{equation}
		\begin{aligned}
	&{R_{{\rm{sum}}}'}({{\bf{w}}^{t + 1}},{{\bm \Psi} ^{t + 1}},{{\bm \rho} ^{t + 1}},{{\bm\varpi} ^{t + 1}})
	\stackrel{(a)}{\geq} \\& {R_{{\rm{sum}}}'}({{\bf{w}}^{t + 1}},{{\bm \Psi} ^t},{{\bm \rho} ^{t + 1}},{{\bm\varpi} ^{t + 1}})
		\stackrel{(b)}{\geq} R_{{\rm{sum}}}'({{\bf{w}}^t},{{\bm \Psi} ^t},{{\bm \rho} ^{t + 1}},{{\bm\varpi} ^{t + 1}}) \\&
		\stackrel{(c)}{\geq}  {R_{{\rm{sum}}}'}({{\bf{w}}^t},{{\bm \Psi} ^t},{{\bm \rho} ^{t + 1}},{{\bm\varpi} ^t})
		\stackrel{(d)}{\geq} {R_{{\rm{sum}}}'}({{\bf{w}}^t},{{\bm \Psi} ^t},{{\bm \rho} ^t},{{\bm\varpi} ^t}),
		\end{aligned}
	\end{equation} 
	where $(a)$ and $(b)$ follow since the updates of ${\bm \Psi}$ and ${\bf w}$ are the optimal solutions to subproblems ${\cal P}_3$ in (\ref{eqn:problem_10}) and ${\cal P}_2$ in (\ref{eqn:problem_9}), respectively; $(c)$ and $(d)$ follow because the updates of ${\bm \varpi}$ and ${\bm \rho}$ maximize $R_{\rm{sum}}'$ when the other variables are fixed, respectively. Therefore, the objective $R_{{\rm{sum}}}'$ is monotonically non-decreasing in each iteration. Since the value of $R_{{\rm{sum}}}'$ is upper-bounded due to power constrains ${\rm C_1}$ and ${\rm C_2}$, {\bf Algorithm 1} will converge to a local optimum.
	
	As an exterior point method, {\bf Algorithm 2} meets two standard convergence conditions \cite{fiacco1990nonlinear}, which determines that it converges to a local optimal point where ${\bm \phi} ={\bm \phi}'$ and ${q}\left({\bm{\phi}},{\bm{\phi}}' \right)={f}\left({\bm{\phi}},{\bm{\phi}}' \right)={f}\left({\bm{\phi}} \right)$. Firstly, for a given penalty coefficient $\zeta$ in each iteration, the value of ${q}\left({\bm{\phi}},{\bm{\phi}}' \right)$ in (\ref{eqn:problem_5}) is lower-bounded by zero and experiences the following monotonically non-increasing update:
	\begin{equation}
		\begin{aligned}
		{q}\left({\bm{\phi}}^{t+1},({\bm{\phi}}')^{t+1} \right) 	\stackrel{(a)}{\leq} {q}\left({\bm{\phi}}^{t+1},({\bm{\phi}}')^{t} \right) 	\stackrel{(b)}{\leq} q\left({\bm{\phi}}^{t},({\bm{\phi}}')^{t} \right),
		\end{aligned}
	\end{equation} 	
	where $(a)$ follows because the update of ${\bm \phi}'$ minimizes ${q}\left({\bm{\phi}},{\bm{\phi}}' \right)$ in ${\cal P}_5$ in (\ref{eqn:problem_5}) when ${\bm \phi}$ is fixed and $(b)$ follows since the update of ${\bm \phi}$ minimizes ${q}\left({\bm{\phi}},{\bm{\phi}}' \right)$ when ${\bm \phi}'$ is fixed. Secondly, as penalty coefficient $\zeta$ increases to be sufficiently large ($\zeta \to \infty$), ${q}\left({\bm{\phi}},{\bm{\phi}}' \right)$ in (\ref{eqn:problem_5}) is dominated by the penalty term $\zeta{\left\| {\bm{\phi}}' - {\bm{\phi}} \right\|^2}$. The updating formulas (\ref{eqn:f_phi1}) becomes ${\bm \phi}={\bm \phi}'$ and (\ref{eqn:f_phi2}) becomes ${\bm \phi}'={\bm \phi}$. It indicates that,  ${\bm \phi}$ and ${\bm \phi}'$ do not update anymore and ${\bm \phi}={\bm \phi}'$ always holds. As a result, penalty term $\zeta{\left\| {\bm{\phi}}' - {\bm{\phi}} \right\|^2}$ is equal to zero and the converged objective ${q}\left({\bm{\phi}},{\bm{\phi}}' \right)$ finally satisfies ${q}\left({\bm{\phi}},{\bm{\phi}}' \right)={f}\left({\bm{\phi}},{\bm{\phi}}' \right)={f}\left({\bm{\phi}} \right)$.

	\subsection{Computational Complexity Analysis}
	The computational complexity of {\bf Algorithm 1} is mainly determined by the updates of the four variables $\bm \rho$, $\bm \varpi$, $\bf w$, and $\bm \Psi$ via (\ref{eqn:rho_update}), (\ref{eqn:update_varpi}), (\ref{eqn:problem_9}), and (\ref{eqn:problem_10}), respectively. Specifically, the computational complexity of updating $\bm \rho$ is ${\cal O}\left(KM\right)$. The complexity of updating $\bm \varpi$ is ${\cal O}\left(K^2M+KN\right)$. Considering the complexity of solving standard convex QCQP problem, for a given accuracy tolerance $\varepsilon$, the computational complexity of updating $\bf w$ is ${\cal O}\left( {{{\log }_2}\left( {{1 \mathord{\left/{\vphantom {1 \varepsilon }} \right.\kern-\nulldelimiterspace} \varepsilon }} \right)\sqrt {MK + 2} \left( {1 + MK} \right){M^3}{K^3}} \right)$. Similarly, the computational complexity of updating $\bm \Psi$ is ${\cal O}\left( {{{\log }_2}\left( {{1 \mathord{\left/{\vphantom {1 \varepsilon }} \right.\kern-\nulldelimiterspace} \varepsilon }} \right)\sqrt {N + 1} \left( {1 + 2N} \right){N^3}} \right)$. Thus, the overall computational complexity of {\bf Algorithm 1} is given by ${\cal O}\left( {{{\log }_2}\left( {{1 \mathord{\left/{\vphantom {1 \varepsilon }} \right.\kern-\nulldelimiterspace} \varepsilon }} \right){{I_o}}\left( {{M^{4.5}}{K^{4.5}} + {N^{4.5}}} \right)} \right)$, wherein $I_{o}$ denotes the number of iterations required by {\bf Algorithm 1} for convergence.
	
	Similarly, the computational complexity of {\bf Algorithm 2} is mainly determined by updating ${\bm \phi}$ and ${\bm \phi}'$ via (\ref{eqn:f_phi1}) and (\ref{eqn:f_phi2}), respectively. As closed-form updating formulas, their computational complexity are both ${\cal O}\left(\left(K+1\right)N^3+\left(K+1\right)N^2\right)$, which are mainly caused by matrix inversions. Thus, the overall computational complexity of {\bf Algorithm 2} is ${\cal O}\left(I_{s}KN^3\right)$, wherein $I_{s}$ is the number of iterations required by {\bf Algorithm 2} for convergence. 
	
	\section{Validation Results}\label{sec:sim}
	In this section, we present validation results. To validate the signal model (\ref{eqn:active_model}), in Subsection \ref{sub:vr:sm}, we present experimental results based on a fabricated active RIS element. Then, in Subsection \ref{sub:vr:2}, simulation results are provided to evaluate the sum-rate of active RIS aided MU-MISO systems. Finally, in Subsection \ref{sub:vr:3}, the impact of active RIS self-interference on system performance is discussed.
	
	\begin{figure*}[!t]
		\centering
		\includegraphics[width = 1\textwidth]{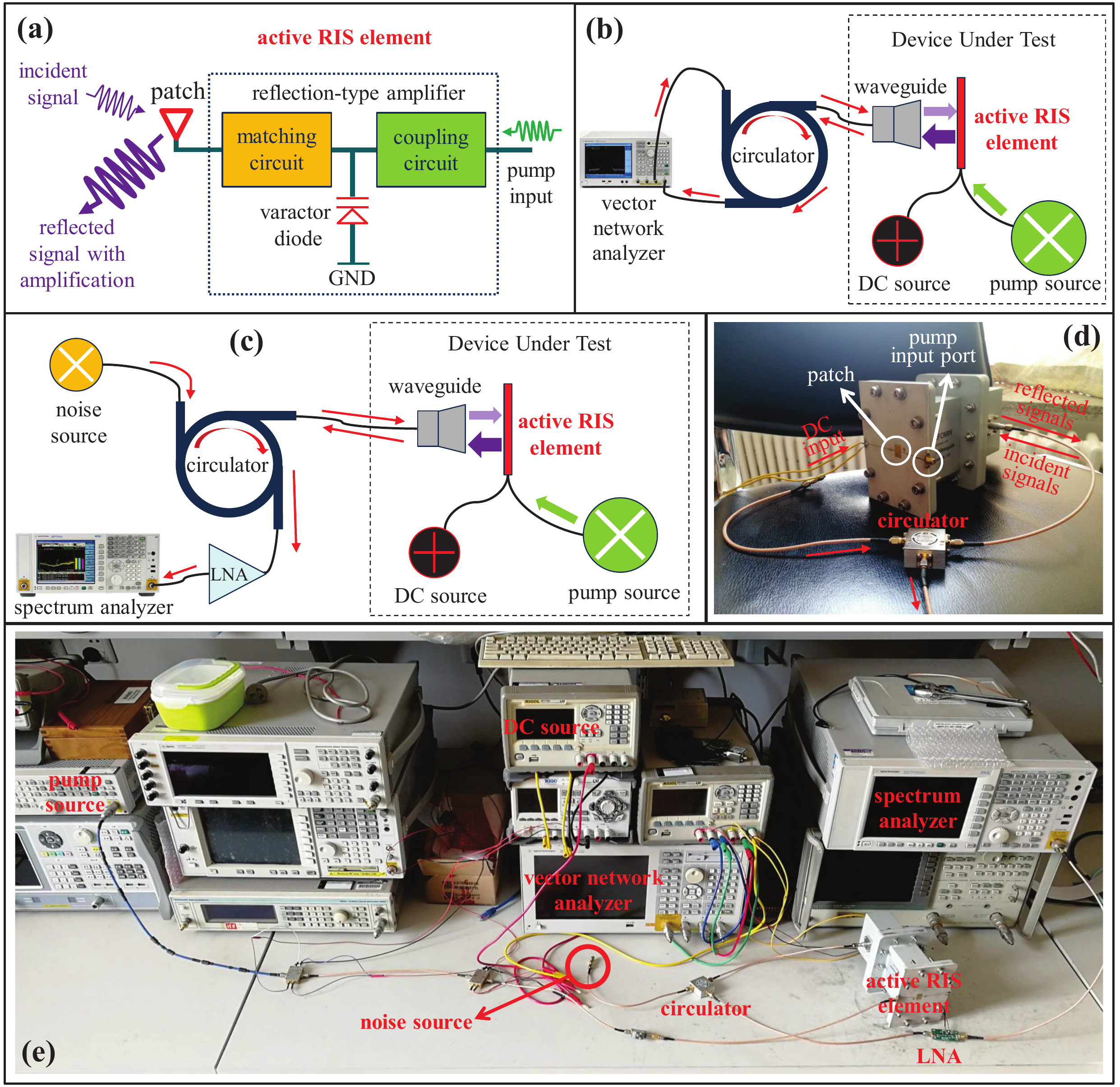}
		%\vspace*{-2em}
		\caption{\color{black}The experimental devices and environment for validating the signal model (\ref{eqn:active_model}) of the active RIS.
		}
		\label{img:test_environment}
		%\vspace*{-1em}
	\end{figure*}

	\subsection{Validation Results for Signal Model}\label{sub:vr:sm}
	\begin{figure}[!t]
		\centering
		\includegraphics[width=0.5\textwidth]{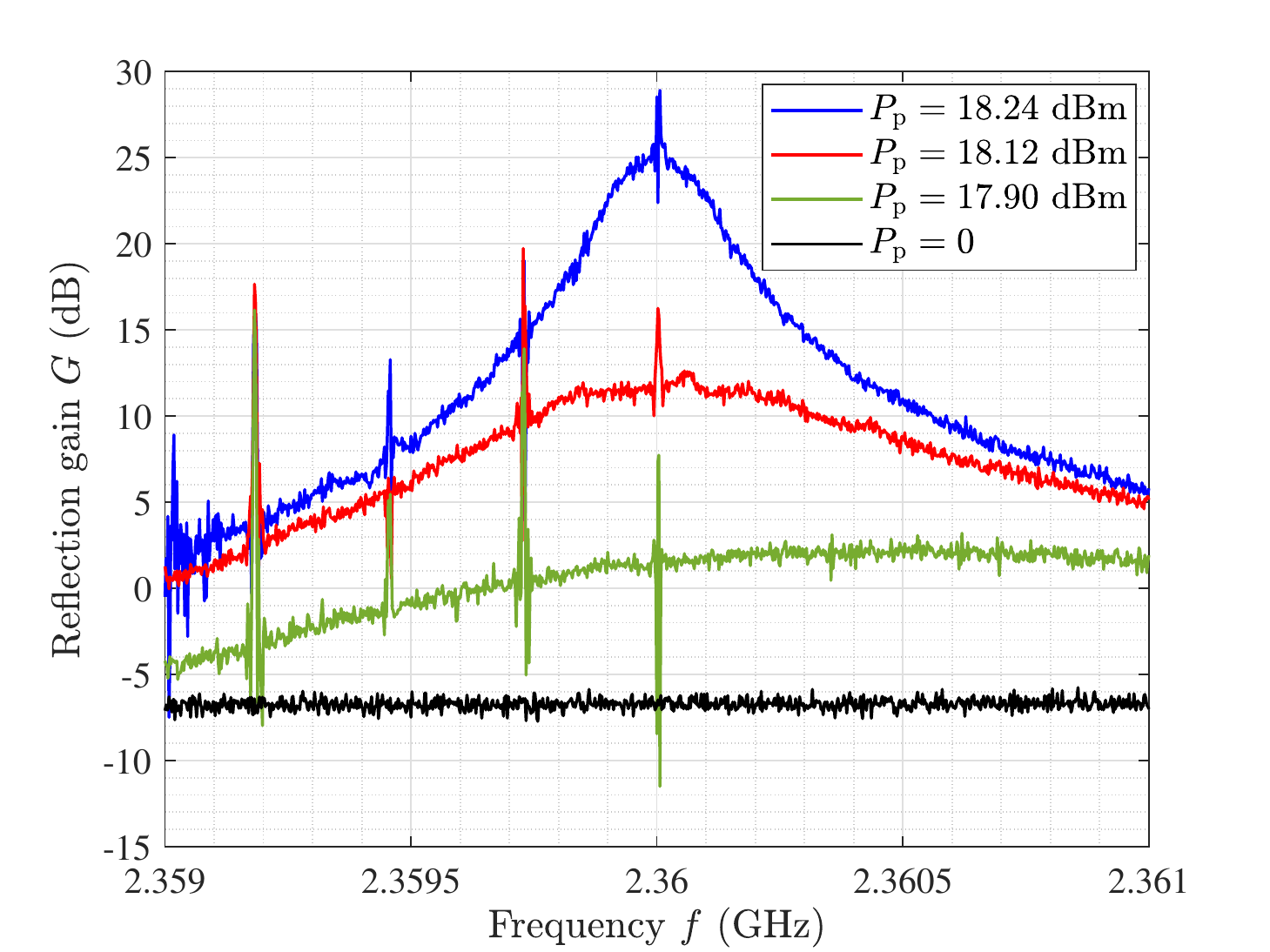}	
		%\vspace*{-1em}
		\caption{Experimental measurement result for reflection gain $G$ versus signal frequency $f$.}		
		\label{img:val:signal1}
		%\vspace*{-1em}
	\end{figure}
	To validate the signal model (\ref{eqn:active_model}), we designed and fabricated an active RIS element with an integrated reflection-type amplifier for experimental measurements in \cite{Xibi'20}. Particularly, since the phase-shifting ability of \acp{ris} has been widely verified \cite{Yang'17}, we focus on studying the reflection gain and the noise introduced by an active RIS element. Thus, the validation of signal model (\ref{eqn:active_model}) is equivalent to validating
	\begin{equation}\label{eqn:active_model_power}
		\begin{aligned}
			{P_y} = \underbrace {G{P_x}}_{{\text{Desired-signal power}}} + \underbrace {{G}\sigma _v^2 + \sigma _s^2}_{{\text{noise power}}},
		\end{aligned}
	\end{equation}
	where $P_y$ is the power of the signals reflected by the active RIS element; $P_x$ is the power of the incident signal; $ G:=p^2$ is the reflection gain of the active RIS element; ${G}\sigma _v^2$ and $\sigma _s^2$ are the powers of the dynamic noise and static noise introduced by the active RIS element, respectively.

	\subsubsection{Hardware platform}
	To validate the model in (\ref{eqn:active_model_power}), we first establish the hardware platform used for our experimental measurements, see Fig. \ref{img:test_environment}. Specifically, we show the following aspects:
	\begin{itemize}
		\item Fig. \ref{img:test_environment} (a) illustrates the structure of the fabricated active RIS element operating at a frequency of 2.36 GHz \cite{Xibi'20}. A pump input at a frequency of 4.72 GHz is used to supply the power required by the active RIS element. The incident signal and the pump input are coupled in a varactor-diode-based reflection-type amplifier to generate the reflected signal with amplification.
		\item Fig. \ref{img:test_environment} (b) illustrates the system used for measuring the reflection gain $G$ of the active RIS element.  A direct-current (DC) source is used to provide a bias voltage of 7.25 V for driving the active RIS element, and a controllable pump source is used to reconfigure the reflection gain $G$. A circulator is used to separate the incident signal and the reflected signal, and the reflection gain is directly measured by a vector network analyzer.
		\item Fig. \ref{img:test_environment} (c) illustrates the system for measuring the noises introduced at the active RIS element, where a spectrum analyzer is used to measure the noise power. The noise source is a 50 $\Omega$ impedance, which aims to simulate a natural input noise of -174 dBm/Hz at each patch. The reflected signal is amplified by a low-noise amplifier (LNA) so that the spectrum analyzer can detect it.
		\item Fig. \ref{img:test_environment} (d) shows a photo of the fabricated active RIS element under test, which is connected by a waveguide for incident/reflected signal exchanges.
		\item Fig. \ref{img:test_environment} (e) shows a photo of the experimental environment with the required equipment for device driving and signal measurement.	
	\end{itemize}	
	   	
	\subsubsection{Reflection gain measurement}
	Using the measurement system for the reflection gain depicted in Fig. \ref{img:test_environment} (b), we first investigate the reflection gain $G$ of the active RIS element. The reflection gain $G$ can be reconfigured by the input power of the pump source $P_{\rm p}$. By setting the input power of the vector network analyzer as $P_x=-50$ dBm, the reflection gain $G$ as a function of the signal frequency can be directly measured via the vector network analyzer. Then, in Fig. \ref{img:val:signal1}, we show the measurement results for reflection gain $G$ as a function of signal frequency $f$ for different input powers of the pump source $P_{\rm p}$. We observe that the active RIS element can achieve a reflection gain $G$ of more than 25 dB, when $P_{\rm p}=18.24$ dBm, which confirms the significant reflection gains enabled by active RISs. On the other hand, when $P_{\rm p}=0$, we observe that $G$ falls to $-6$ dB, which is lower than the expected 0 dB. This loss is mainly caused by the inherent power losses of the circulator and transmission lines used for measurement. %One can also observe the frequency selectivity, i.e., the response of active element has some differences for different frequencies, which depends on the operating band of reflection-type amplifier. This indicates that, in the wideband multi-carrier communication systems (different from the single-carrier narrowband system considered in this paper), the frequency-dependent power attenuation on different subcarriers should be taken into account while designing RIS precoding matrix $\bm \Psi$.

	\subsubsection{Noise power measurement}
	
\begin{figure}[!t]
	\centering
	\includegraphics[width=0.5\textwidth]{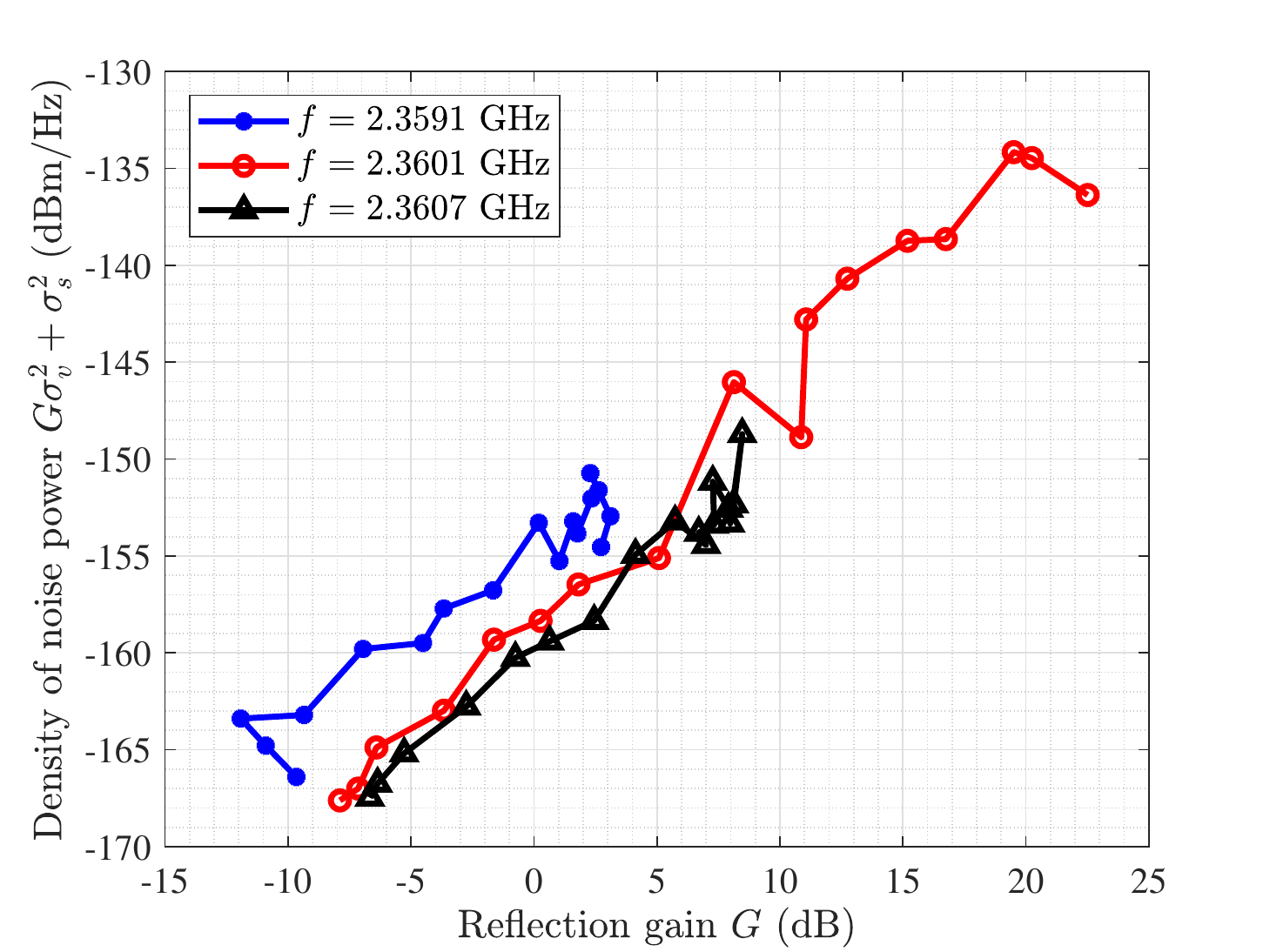}
	%\vspace*{-1em}
	\caption{Experimental measurement result for the density of noise power $G\sigma_v^2+\sigma_s^2$ versus reflection gain $G$.}		
	%\vspace*{-1em}
	\label{img:val:signal2}
\end{figure}

	We further study the noise power introduced and amplified by the active RIS element, i.e., ${G}\sigma _v^2 + \sigma _s^2$ in (\ref{eqn:active_model_power}), where ${G}\sigma _v^2$ and $\sigma _s^2$ are the powers of the dynamic noise and static noise introduced at the active RIS element, respectively. Using the noise measurement system in Fig. \ref{img:test_environment} (c), we show the measurement results for the spectral density of noise power $G\sigma_v^2+\sigma_s^2$ as a function of $G$ for different operating frequencies in Fig. \ref{img:val:signal2}. We can observe that the noise power increases nearly linearly with $G$, which verifies the noise model ${G}\sigma _v^2 + \sigma _s^2$ in (\ref{eqn:active_model_power}). Particularly, for $f=2.3601$ GHz, the spectral density of $\sigma_s^2$ is about $-174$ dBm/Hz, while that of $\sigma_v^2$ is about $-160$ dBm/Hz, which is about $15$ dB higher. The reason for this is that the input noise is amplified by the noise factor \cite{Bousquet'12}, and additional noises are also introduced by the other active components the measurement equipment, such as the leakage noise from the DC source.	 %It is worth mentioning that, after experiencing the large fading of RIS-user channel, the static noise power $\sigma _s^2$ reaching the user will be negligible \cite{Wu'19}. This explains why most existing works on active scatters neglect $\sigma_s^2$ and regard ${G}\sigma _v^2$ as the non-negligible noise power \cite{Bousquet'12}. 

	\subsection{Simulation Results for Joint Beamforming and Precoding Design}\label{sub:vr:2}

	To evaluate the effectiveness of the proposed joint beamforming and precoding design, in this subsection, we present simulation results for passive RIS and active RIS aided MU-MISO systems, respectively.
	\par	
	\subsubsection{Simulation setup}
\begin{figure*}[!t]
	\centering
	%\hspace{-2.5mm}	
	\subfigure[Scenario 1 with a weak direct link. ]{\includegraphics[width=0.41\textwidth]{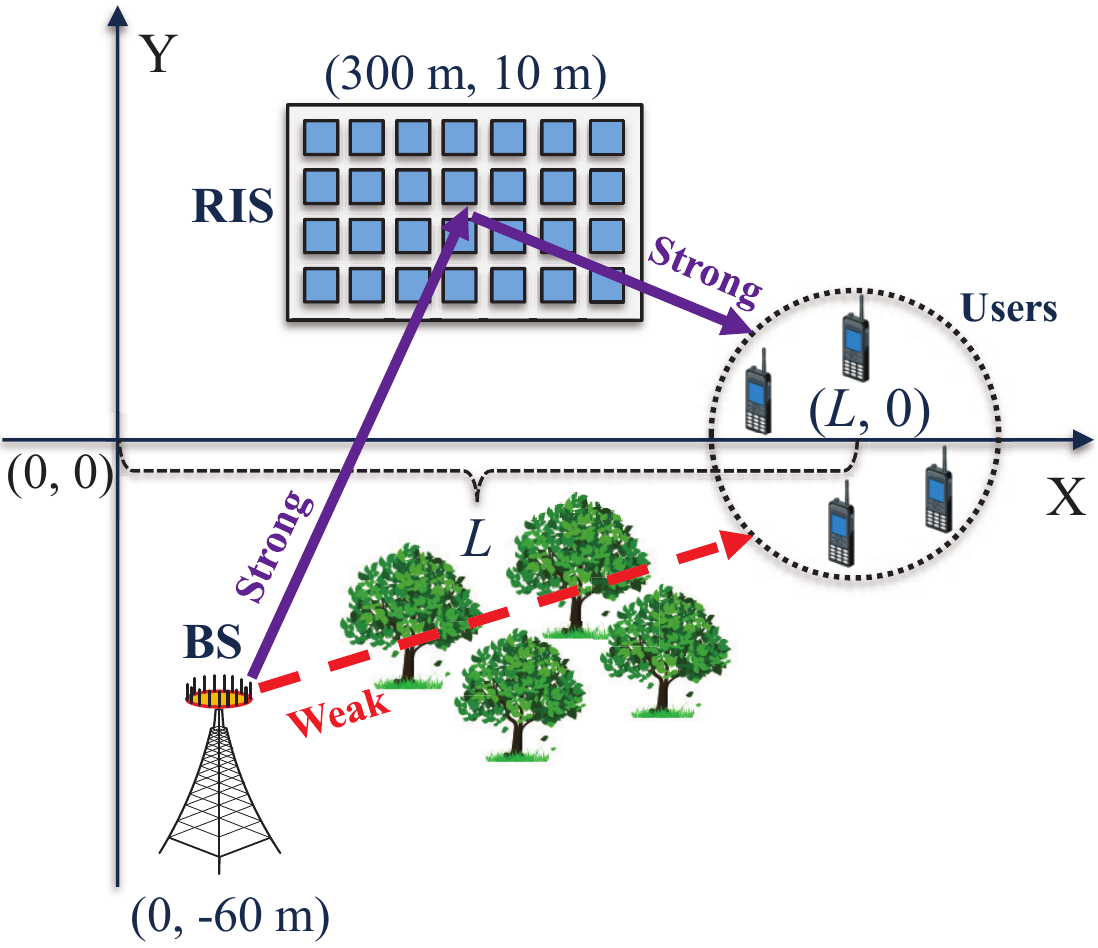}}
	\hspace{12mm}	
	\subfigure[Scenario 2 with a strong direct link.]{\includegraphics[width=0.41\textwidth]{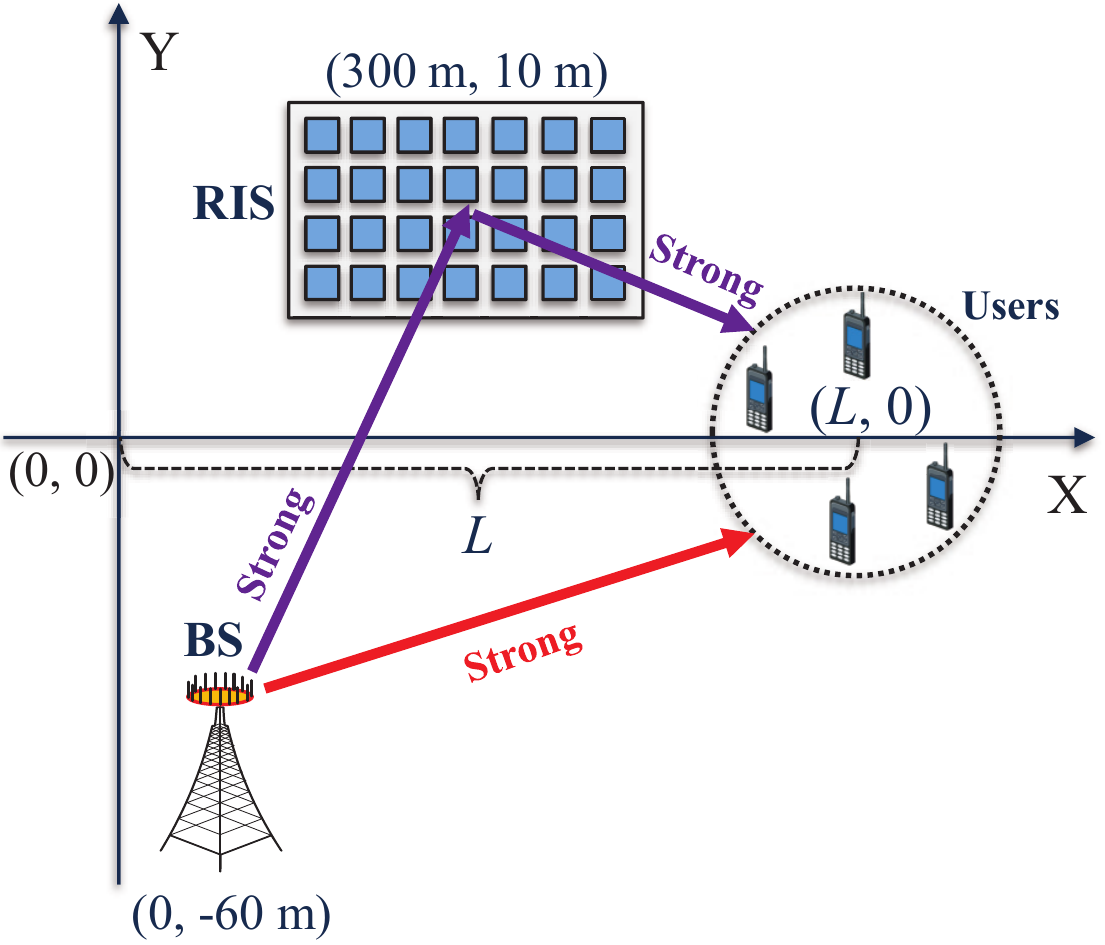}}
	%\vspace*{-0.5em}
	\caption{Two simulation scenarios with different channel conditions, where a BS aided by an active RIS serves four users.}
	\label{img:scenario}
	%\vspace*{-1em}
\end{figure*}	
\begin{figure*}[!t]
	\centering
	%\hspace{-2.5mm}	
	\subfigure[Scenario 1 with a weak direct link. ]{\includegraphics[width=0.5\textwidth]{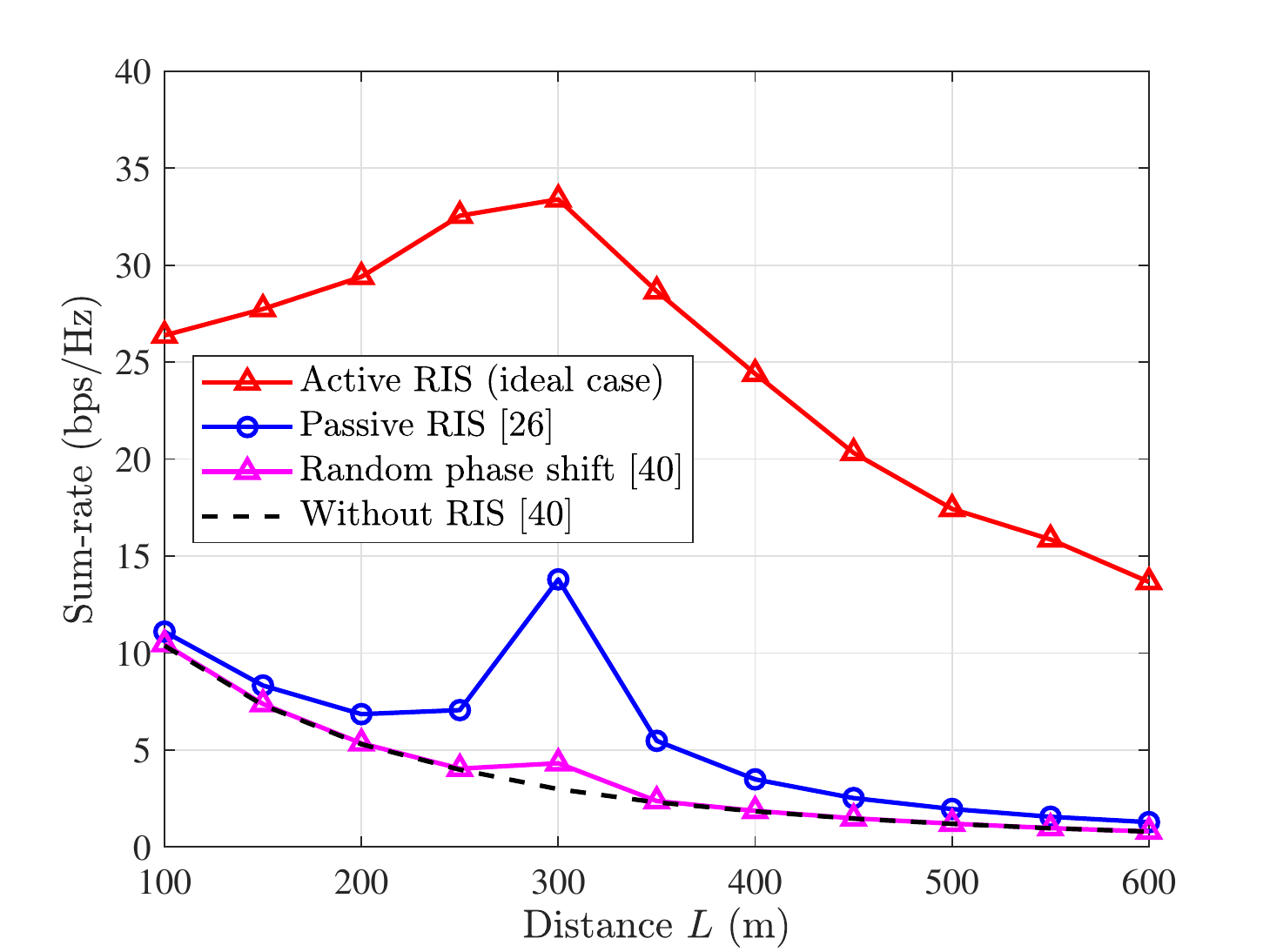}}
	\hspace{-6mm}	
	\subfigure[Scenario 2 with a strong direct link.]{\includegraphics[width=0.5\textwidth]{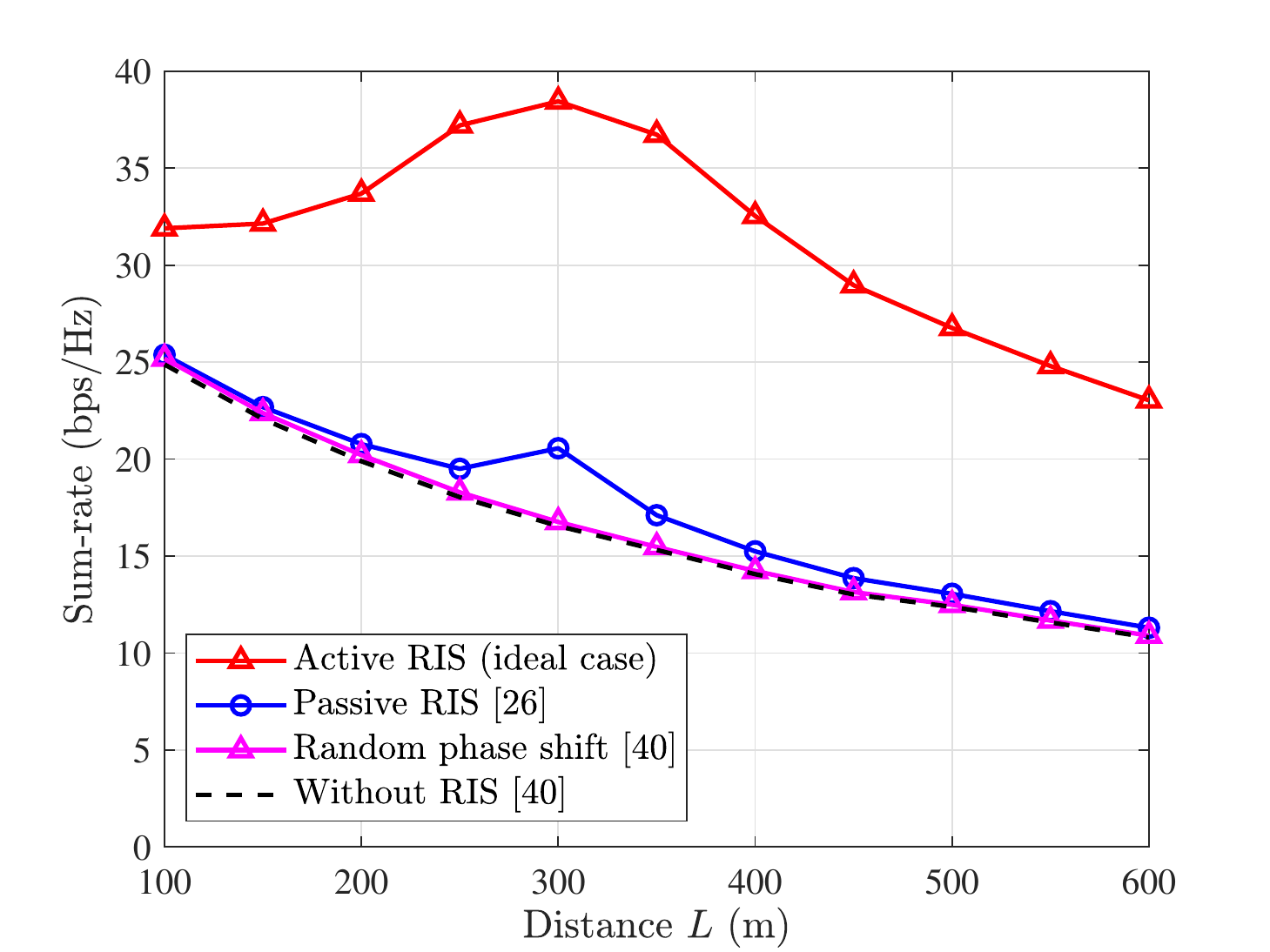}}
	%\vspace*{-0.5em}
	\caption{Simulation results for the sum-rate as a function of distance $L$ in an RIS-aided MU-MISO system.}
	\label{img:val:jpd}
	%\vspace*{-1em}
\end{figure*}		
	For the simulation setup, we consider an active/passive RIS aided MU-MISO system operating at a frequency of 5 GHz as shown in Fig. \ref{img:scenario}. Particularly, we consider two scenarios with different channel conditions. In Fig. \ref{img:scenario} (a), the direct link is weak due to severe obstruction, while the direct link is strong in Fig. \ref{img:scenario} (b). To be specific, two different path loss models from the 3GPP standard \cite[B.1.2.1]{3GPP} are utilized to characterize the large-scale fading of the channels:
	\begin{equation}
	\begin{aligned}
	{\rm{P}}{{\rm{L}}_{s}} &= 37.3 + 22.0\log d,\\
	{\rm{P}}{{\rm{L}}_w} &= 41.2 + 28.7\log d,
	\end{aligned}
	\end{equation}
where $d$ is the distance between two devices. Path loss model ${\rm{P}}{{\rm{L}}_w}$ is used to generate the weak BS-user link in scenario 1, while ${\rm{P}}{{\rm{L}}_s}$ is used to generate the strong BS-user link in scenario 2. For both scenarios in Fig. \ref{img:scenario}, ${\rm{P}}{{\rm{L}}_s}$ is used to generate the BS-RIS and the RIS-user channels. To account for small-scale
fading, following \cite{Guo'19}, we adopt the Ricean fading channel model for all
channels involved. In this way, an arbitrary channel matrix $\bf H$ is generated by
\begin{equation}
	\begin{aligned}
{\bf{H}} = \sqrt{\rm PL} \left(\sqrt {\frac{\kappa}{{\kappa  + 1}}} {{\bf{H}}_{{\rm{LoS}}}} + \sqrt {\frac{1 }{{\kappa  + 1}}} {{\bf{H}}_{{\rm{NLoS}}}}\right),
	\end{aligned}
\end{equation}
where ${\rm PL}$ is the corresponding path loss of $\bf{H}$; $\kappa$ is the Ricean factor; and ${{\bf{H}}_{{\rm{LoS}}}}$ and ${\bf{H}}_{{\rm{NLoS}}}$ represent the deterministic LoS and Rayleigh fading components, respectively. In particular, here we assume $\kappa=1$.

As common settings, the \ac{bs} and the active/passive RIS are located at (0, -60 m) and (300 m, 10 m), respectively. The locations of the four users will be specified later. Unless specified otherwise, the numbers of BS antennas and RIS elements are set as $M=4$ and $N=512$, respectively. The noise power is set as ${\sigma}^2={\sigma}_v^2=-100$ dBm. Let ${P^{\max}_{\text{BS}}}$ denote the maximum transmit power at the \ac{bs} and ${P^{\max}_{\text{A}}}$ denote the maximum reflect power of the active \ac{ris}, which don't include the hardware static power. For fair comparison, we constrain the total power consumption $P^{\max}:=P_{\text{BS}}^{\max}+P_{\text A}^{\max}$ to $10$ dBm by setting $P_{\text{BS}}^{\max}=0.99\times P^{\max}$ and $P_{\text A}^{\max}=0.01\times P^{\max}$ for the active RIS aided system, and $P_{\text{BS}}^{\max}=10$ dBm for the other benchmark systems. To show the effectiveness of beamforming designs, here we consider the following four schemes for simulations:
\begin{itemize}
	\item {\bf Active RIS (ideal case):} In an ideal active RIS-aided MU-MISO system without self-interference, the proposed {\bf Algorithm 1} is employed to jointly optimize the BS beamforming and the precoding at the active RIS.
	
	\item {\bf Passive RIS \cite{Pan'19}:} In a passive RIS-aided MU-MISO system, the algorithm proposed in \cite{Pan'19} is adopted to jointly optimize the BS beamforming and the precoding at the passive RIS.
	
	\item {\bf Random phase shift \cite{Qingjiang'11}:} In a passive RIS-aided MU-MISO system, the phase shifts of all passive RIS elements are randomly set. Then, relying on the equivalent channels from the \ac{bs} to users, the weighted mean-squared error minimization (WMMSE) algorithm from \cite{Qingjiang'11} is used to optimize the BS beamforming.
	
	\item {\bf Without RIS \cite{Qingjiang'11}:} In an MU-MISO system without RIS, the WMMSE algorithm from \cite{Qingjiang'11} is adopted to optimize the BS beamforming.
\end{itemize}

\begin{figure*}[!t]
	\centering
	%	\hspace{-2.5mm}	
	\subfigure[Scenario 1 with a weak direct link. ]{\includegraphics[width=0.5\textwidth]{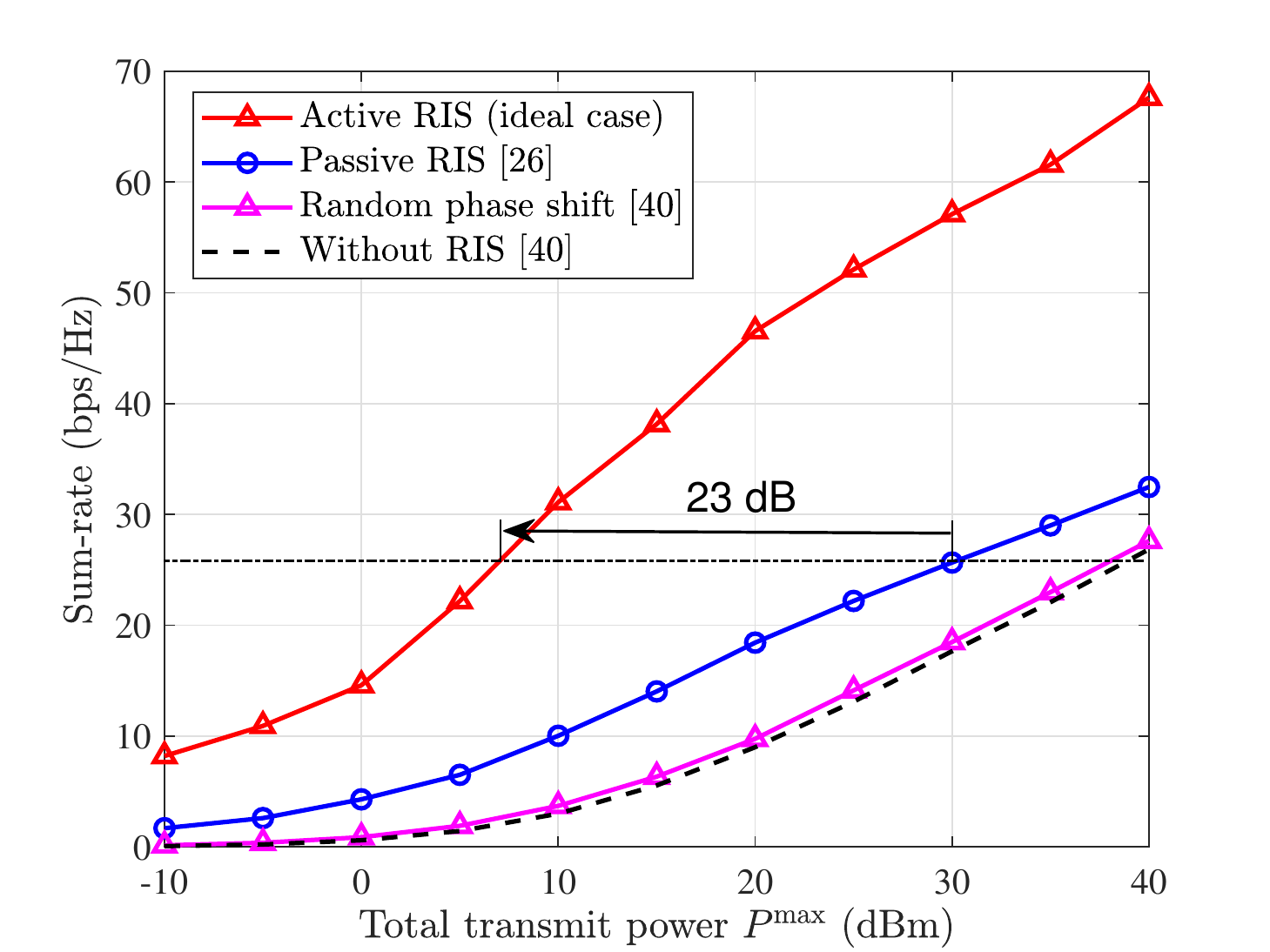}}
	\hspace{-6mm}	
	\subfigure[Scenario 2 with a strong direct link.]{\includegraphics[width=0.5\textwidth]{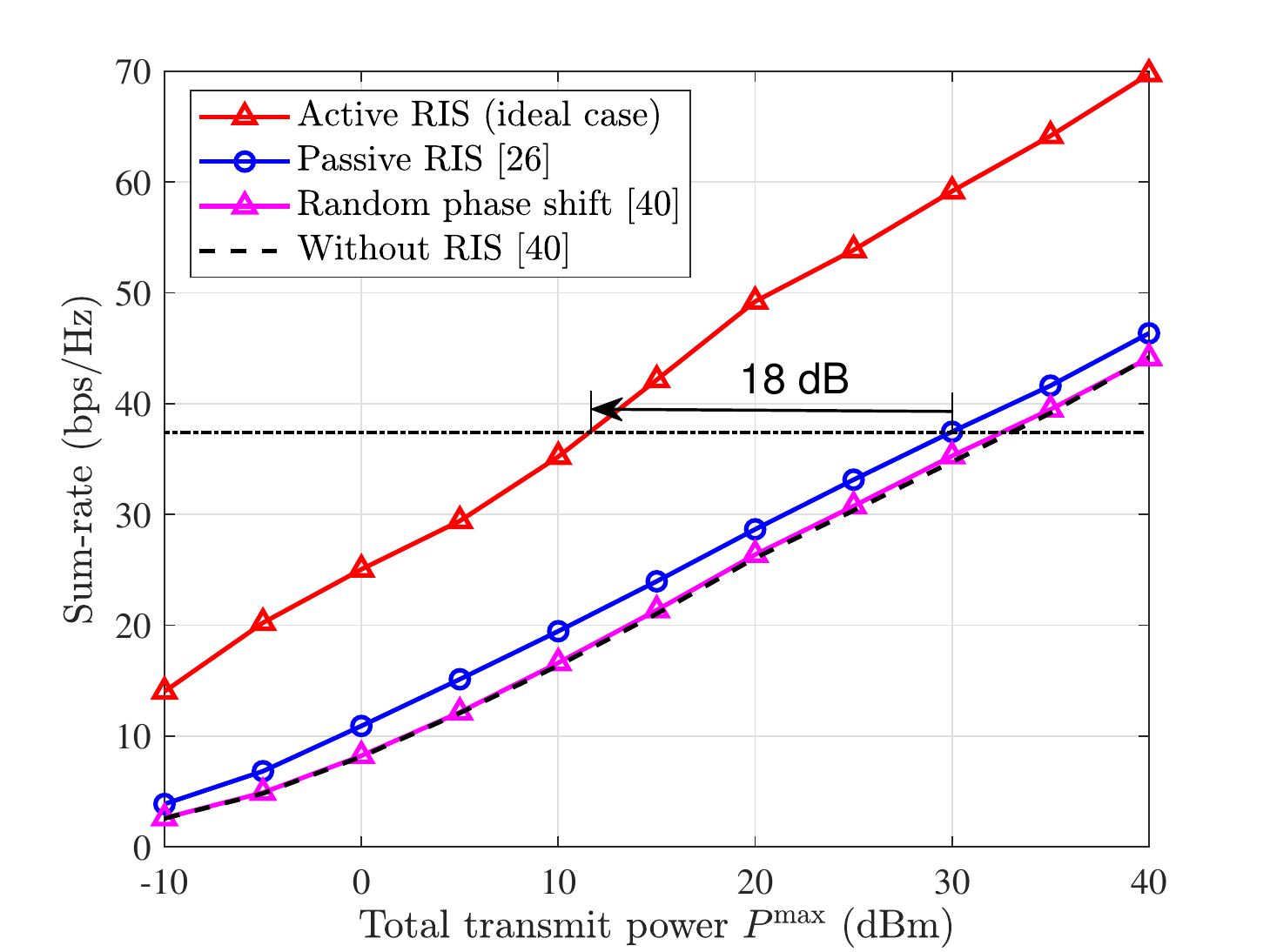}}
	%\vspace*{-0.5em}
	\caption{Simulation results for the sum-rate versus total power consumption $P^{\max}$ in an RIS-aided MU-MISO system.}	
	\label{img:val:sumratevsP}
	%\vspace{-1em}
\end{figure*}	

\subsubsection{Coverage performance of active RISs}
To observe the coverage performance of active RISs, we assume the four users are randomly located in a circle with a radius of 5 m from the center ($L$, 0). In Fig. \ref{img:val:jpd} (a) and (b), we plot the sum-rate versus distance $L$ for the two considered scenarios, where the direct link is weak and strong, respectively. Based on these results, we have two observations. Firstly, in scenario 1 with a weak direct link, the passive RIS can indeed achieve an obvious performance improvement, while the active RIS achieves a much higher sum-rate gain. Secondly, in scenario 2 with a strong direct link, the passive RIS only achieves a limited sum-rate gain, while the active RIS still realizes a noticeable sum-rate gain. For example, when $L=300$ m, the capacities without RIS, with passive RIS, and with active RIS in scenario 1 are 2.98 bps/Hz, 13.80 bps/Hz, and 33.39 bps/Hz respectively, while in scenario 2, these values are 16.75 bps/Hz, 20.56 bps/Hz, and 38.45 bps/Hz, respectively. For this position, the passive RIS provides a 363\% gain in scenario 1 and a 22\% gain in scenario 2. By contrast, the active RIS achieves noticeable sum-rate gains of 1020\% in scenario 1 and 130\% in scenario 2, which are much higher than those achieved by the passive RIS in the corresponding scenarios. These results demonstrate that, compared with the passive RIS, the active RIS can overcome the “multiplicative fading” effect and achieve noticeable sum-rate gains even when direct link is strong. 	
\begin{figure*}[!t]
	\centering
	%	\hspace{-2.5mm}	
	\subfigure[Scenario 1 with a weak direct link. ]{\includegraphics[width=0.5\textwidth]{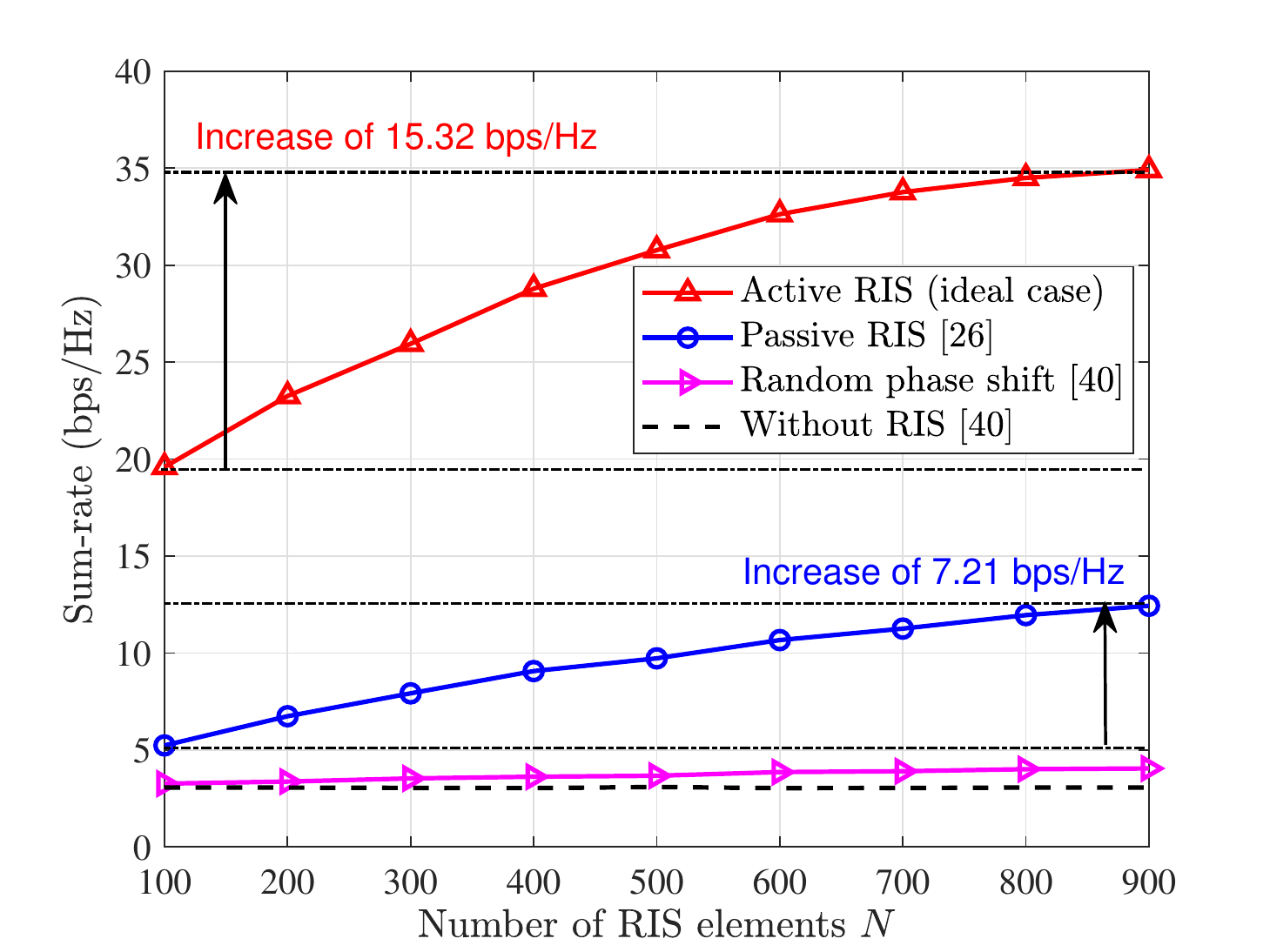}}
	\hspace{-6mm}	
	\subfigure[Scenario 2 with a strong direct link.]{\includegraphics[width=0.5\textwidth]{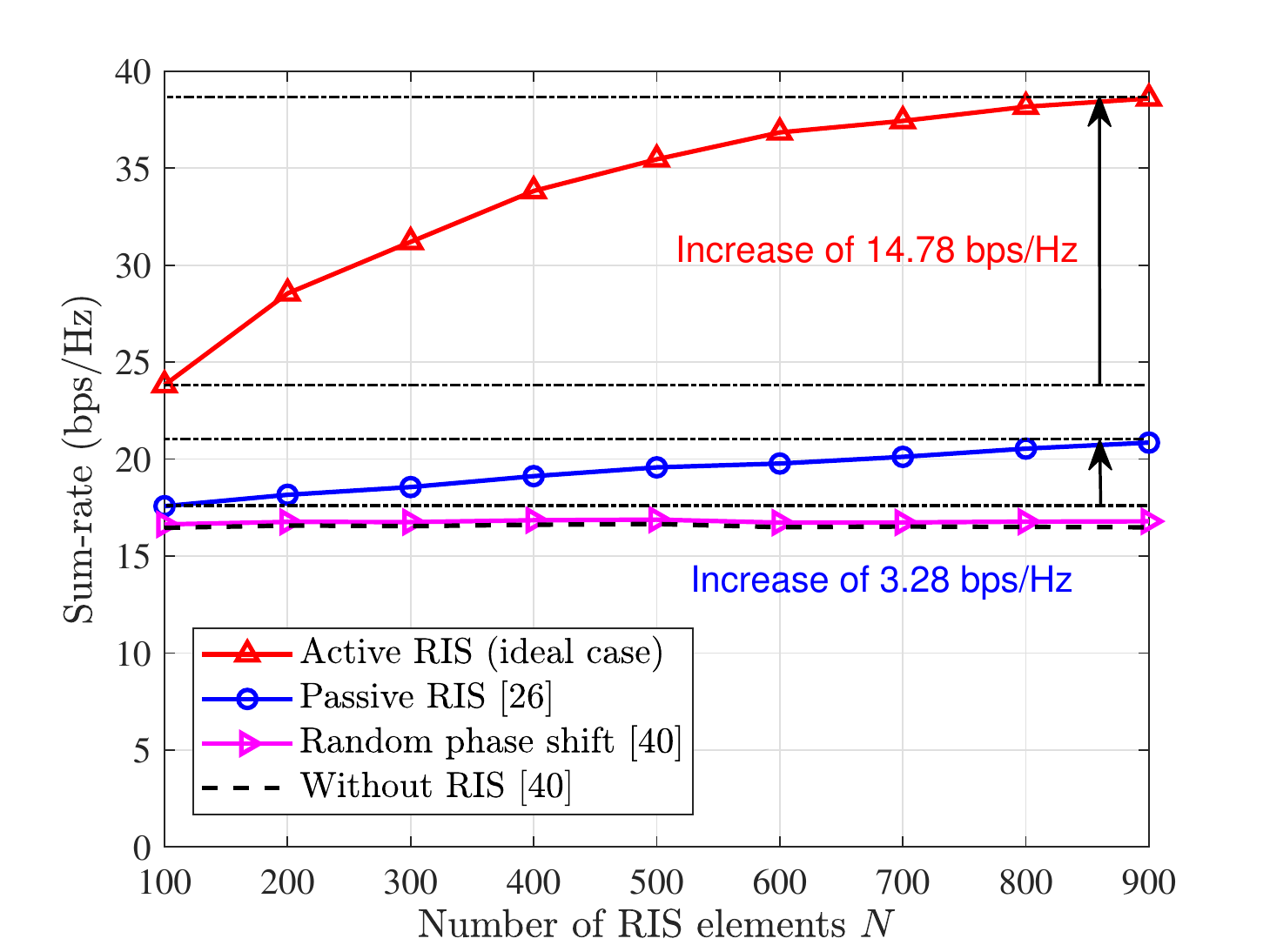}}
	%\vspace*{-0.5em}
	\caption{Simulation results for the sum-rate versus the number of RIS elements $N$ in an RIS-aided MU-MISO system.}	
	\label{img:val:sumratevsN}
	%\vspace{-1em}
\end{figure*}		
\begin{figure*}[!t]
	\centering
	%	\hspace{-2.5mm}	
	\subfigure[Scenario 1 with a weak direct link. ]{\includegraphics[width=0.5\textwidth]{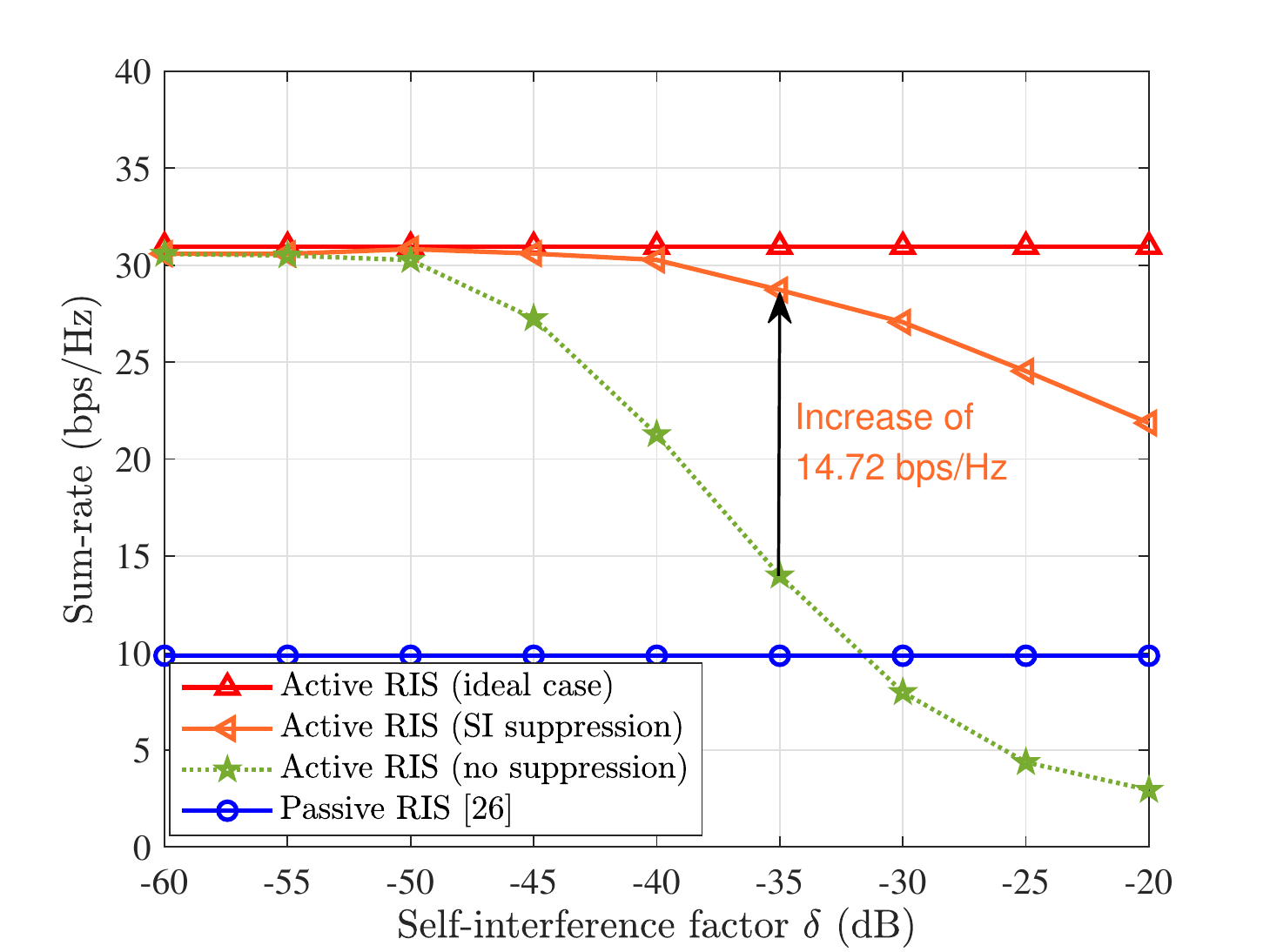}}
	\hspace{-6mm}	
	\subfigure[Scenario 2 with a strong direct link.]{\includegraphics[width=0.5\textwidth]{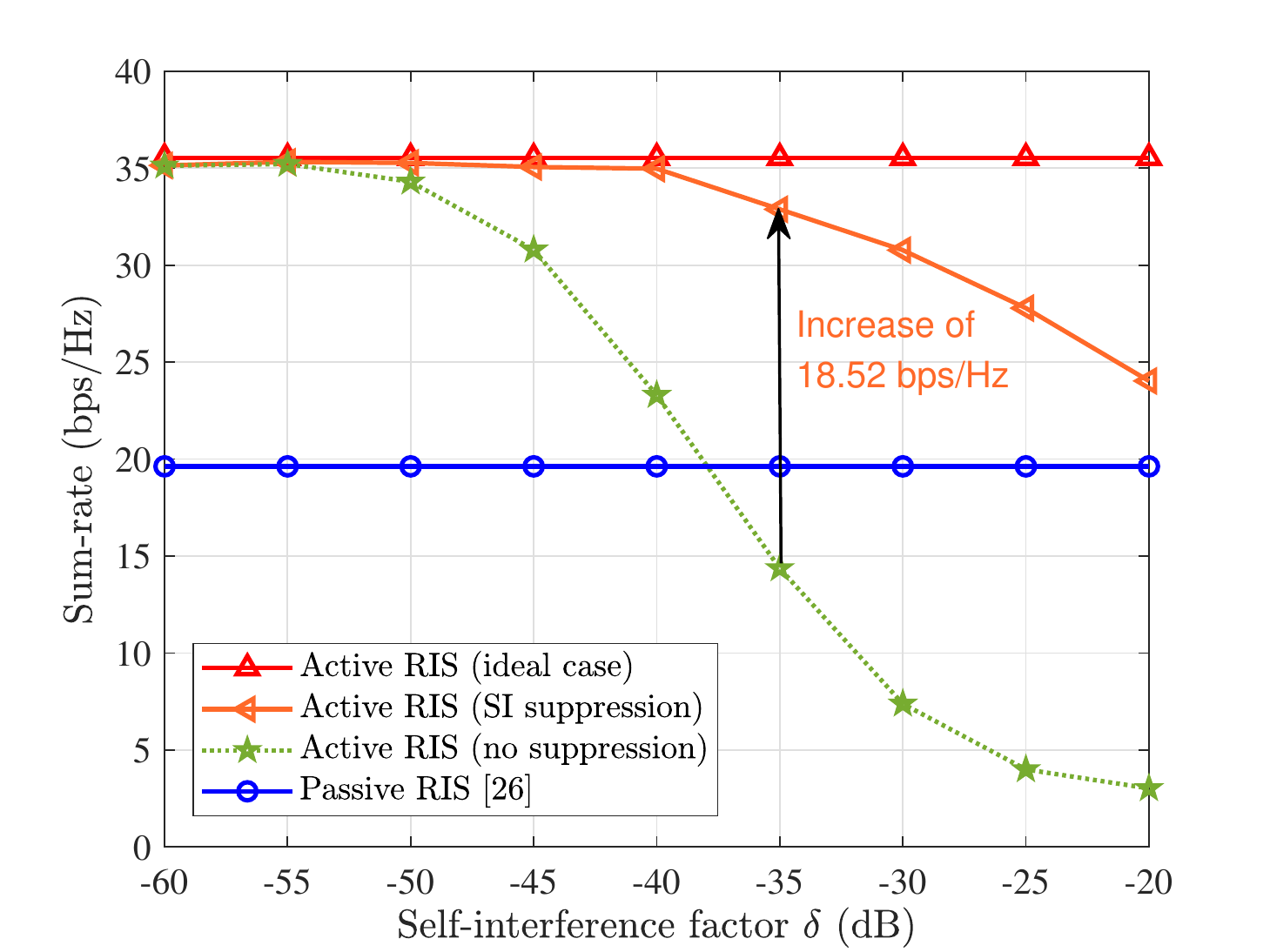}}
	%\vspace*{-0.5em}
	\caption{Simulation results for the sum-rate versus the self-interference factor $\delta$ of the active RIS.}	
	\label{img:val:SI}
	%\vspace{-1em}
\end{figure*}		
\subsubsection{Sum-rate versus total power consumption $P^{\max}$}
To evaluate the averaged performance in the coverage of active/passive RIS, we assume that all users are randomly distributed in a large circle with a radius of 50 m from the center (300 m, 0). We show the users' sum-rate versus the total power consumption $P^{\max}$ in Fig. \ref{img:val:sumratevsP}. From Fig. \ref{img:val:sumratevsP} we observe that the passive RIS achieves visible performance gains in scenario 1 where the direct link is weak, while the passive RIS only achieves limited sum-rate gains in scenario 2 where the direct link is strong. By contrast, in both scenarios, the active RIS realizes a high performance gain. Particularly, to achieve the same performance as the passive RIS aided system, the required power consumption for the active RIS aided system is much lower. For example, when the total power consumption of the passive RIS aided system is $P^{\max}=30$ dBm, to achieve the same sum-rate, the active RIS aided system only requires 7 dBm in scenario 1 and 12 dBm in scenario 2, which correspond to power savings of 23 dB and 18 dB, respectively. The reason for this result is that, for the passive RIS, the total power is only allocated to BS. Thus, all transmit power is affected by the large path loss of the full BS-RIS-user link. However, for the active RIS, part of the transmit power is allocated to the active RIS, and this part of the power is only affected by the path loss of the RIS-user link. Thus, the active RIS is promising for reducing the power consumption of communication systems. %This clearly explains why, to achieve the same performance, the total power required by the active RIS aided system is lower than that required by the passive RIS aided system. Thus, the active RIS is promising for reducing the power consumption of communication systems.
\subsubsection{Sum-rate versus number of RIS elements $N$}
	For the same setup as in Fig. \ref{img:val:sumratevsP}, we plot the users' sum-rate versus the number of RIS elements $N$ in Fig. \ref{img:val:sumratevsN}. We observe that, as the number of RIS elements $N$ increases, both the passive RIS and the active RIS achieve higher sum-rate gains, while the performance improvement for the active RIS aided system is much larger than that for the passive RIS aided system. For example, when $N$ increases from 100 to 900, the sum-rate of the passive RIS aided system increases from 5.23 bps/Hz to 12.44 bps/Hz in scenario 1 (increase of 7.21 bps/Hz) and from 17.57 bps/Hz to 20.85 bps/Hz in scenario 2 (increase of 3.28 bps/Hz), respectively. By contrast, the sum-rate of the active RIS aided system increases from 19.59 bps/Hz to 34.91 bps/Hz in scenario 1 (increase of 15.32 bps/Hz) and from 23.81 bps/Hz to 38.59 bps/Hz in scenario 2 (increase of 14.78 bps/Hz), respectively. These results show that the sum-rate increase of the active RIS aided system is much higher than that of the passive RIS aided system. This indicates that, as long as the number of RIS elements $N$ is not exceedingly large (such as millions of elements), compared with the passive RIS, increasing the number of elements of the active RIS is much more efficient for improving the communication performance, which is in agreement with the performance analysis in Section \ref{sec:PA}.	

\subsection{Simulation Results for Self-Interference Suppression}\label{sub:vr:3}	

In this subsection, we present simulation results to verify the effectiveness of the proposed self-interference suppression scheme for active RISs.  
\subsubsection{Simulation setup}
To avoid the impact of other factors, we adopt the same setup in Subsection \ref{sub:vr:2}, which is used in Fig. \ref{img:val:sumratevsP} and Fig. \ref{img:val:sumratevsN}. Without loss of generality, we assume that each element in self-interference matrix $\bf H$ is distributed as $\sim \mathcal{C} \mathcal{N}\left({0}, \delta^2\right)$ \cite{suraweera2014low,lioliou2010self,xing2016self}, where we name $\delta$ as the self-interference factor, which is inversely proportional to the inter-element isolation of practical arrays \cite{Mikko'15,Mikko'16,FDrelay_SI}. To evaluate the impact of self-interference on sum-rate, we add two new benchmarks for simulations:
\begin{itemize}
	\item {\bf Active RIS (SI suppression):} In a non-ideal active RIS-aided MU-MISO system with self-interference, {\bf Algorithm 1} is employed to optimize the BS beamforming and the active RIS precoding, and {\bf Algorithm 2} is employed to suppress the self-interference. Then, the performance is evaluated under the condition of self-interference.
	
	\item {\bf Active RIS (no suppression):} In a non-ideal active RIS-aided MU-MISO system with self-interference, only {\bf Algorithm 1} is employed to design the BS beamforming and the active RIS precoding and the self-interference is ignored. Then, the performance is evaluated under the condition of self-interference.
\end{itemize}

\subsubsection{Impact of self-interference on sum-rate}
We plot the users' sum-rate versus the self-interference factor $\delta$ in Fig. \ref{img:val:SI}. We observe that, when $\delta<-50$ dB, the self-interference has almost no impact on the sum-rate. However, as the self-interference strengthens, the active RIS aided system without self-interference suppression suffers an increasingly high performance loss. Particularly, when $\delta = -35$ dB, the active RIS without self-interference suppression does not even perform as well as the passive RIS in scenario 2. The reason is that, the existence of self-interference matrix $\bf H$ makes the reflected signals unable to focus on the users, or even worse, cancel the desired signals of the direct link. Fortunately, thanks to our proposed {\bf Algorithm 2}, the active RIS aided system with self-interference suppression can still hold a considerable performance. For example, when $\delta = -35$ dB, compared with the active RIS aided system without self-interference suppression, the system with self-interference suppression can compensate for the sum-rate loss of 14.72 bps/Hz in scenario 1 and that of 18.52 bps/Hz in scenario 2. 

% Besides, it is worth noting that, many real-world FD arrays \cite{Mikko'15,Mikko'16,FDrelay_SI} can achieve high inter-element isolation  to ensure $\delta<-50$ dB. For example, by integrating wave-trap circuits, \cite{FDrelay_SI} has reported a multi-antenna FD relay with the inter-element isolation of 110 dB, which can almost eliminate the negative impact of self-interference on the users' sum-rate. 

\section{Conclusions and Future Works}\label{sec:con}

In this paper, we have proposed the concept of active RISs to overcome the fundamental limitation of the “multiplicative fading” effect. Specifically, we have developed and verified a signal model for active RISs by a fabricated active RIS element through experimental measurements. Based on the verified signal model, we have analyzed the asymptotic performance of active RISs and then formulated an optimization problem to maximize the sum-rate in an active RIS aided MU-MISO system. Subsequently, we have proposed a joint beamforming and precoding scheme to solve this problem. Finally, experimental and simulation results have shown that, compared with the benchmark scheme without RIS, the passive RIS can realize only a limited sum-rate gain of about 22\% in a typical application scenario, while the proposed active RIS can achieve a substantial sum-rate gain of about 130\%, thus indeed overcoming the fundamental limitation of the “multiplicative fading” effect. In the future, many research directions for active RISs are worth pursuing, including hardware design \cite{Loncar'19}, prototype development \cite{LinglongDai}, channel estimation \cite{Huchen}, and energy efficiency analysis \cite{Huang'18'2}.

%In this paper, we tackled the fundamental limitation of the “multiplicative fading” effect of passive RISs by proposing the new concept of active RISs. The developed signal model for active RISs characterized the amplification of incident signals and incorporates the non-negligible thermal noise introduced by the active elements, which was validated by experimental measurements. Asymptotic analysis revealed the high \ac{snr} gain enabled by active RISs, and a joint beamforming and precoding algorithm was proposed to maximize the sum-rate in an active RIS aided \ac{mimo} system. Simulations results have shown that, the proposed active RISs achieves a 67\% sum-rate gain in a typical scenario thus it overcomes the “multiplicative fading” effect. 

\appendices

\section{Proof of Lemma 2}
For notational simplicity, we rewrite some matrices and vectors in (\ref{eqn:signal}) as ${\bf g}=\left[ {{g_1}, \cdots ,{g_N}} \right]^{\rm T}$, ${\bf f}=\left[ {{f_1}, \cdots ,{f_N}} \right]^{\rm T}$, and ${\bf w}_k:={w}$. Thus, the downlink transmission model in (\ref{eqn:signal}) can be rewritten as
	\begin{equation}\label{eqn:transmission_model_siso}
		\begin{aligned}
			r = \underbrace {p{{\bf{f}}^{\rm H}}{\bf{\Theta g}}}_{{\text{Reflected link}}}ws +\!\!\!\!  \underbrace {p{{\bf{f}}^{\rm H}}{\bf{\Theta v}}}_{{\text{Noise introduced by active RIS}}} +\!\! \underbrace z_{{\text{Noise introduced at user}}},
		\end{aligned}
	\end{equation}		
	where $r\in{\mathbb C}$ is the signal received by the user. Based on the transmission model in (\ref{eqn:transmission_model_siso}), the maximization of the user's \ac{snr} $\gamma$, subject to the power constraints at the \ac{bs} and the active RIS, can be formulated as follows:
	\begin{align}
		&\!\!\!\!\!\!\mathop {\max }\limits_{{w},\,{p},\,{\bf{\Theta }}} \,\,\, \gamma  = \frac{{{{\left| { { p{{\bf{f}}}^{\rm H}{\bf \Theta} {\bf{g}}}w} \right|}^2}}}{{{p^2}{{\left\| {{{\bf{f}}^{\rm H}}{\bf \Theta} } \right\|}^2}\sigma _{v}^2 + {\sigma ^2}}}, \notag \\
		\label{eqn:problem_pa}
		&{\rm s.t.}\,\,\,\,\,\, {\rm C_1}: {{{\left| {w} \right|}^2}} \le {P^{\max}_{\text{BS}}},\\
		&\quad \,\,\,\,\,\,\,\, {\rm C_2}: {p^2{{\left\| {{{\bf{\Theta }}}{\bf{g}}{{{w}}}} \right\|}^2}  +  p^2N\sigma _{v}^2} \le {P^{\max}_{\text{A}}}, \notag
	\end{align}
	where ${P^{\max}_{\text{BS}}}$ and ${P^{\max}_{\text{A}}}$ denote the maximum transmit power and the maximum reflect power at the \ac{bs} and the active \ac{ris}, respectively. Then, the optimal solution of problem (\ref{eqn:problem_pa}) can be obtained by the Lagrange multiplier method as follows:
	\begin{subequations}\label{eqn:optimal_so}
		\begin{align}
			&{w^{\rm opt}} = \sqrt{P^{\max}_{\text{BS}}}, \\& {\theta_n ^{\rm opt}} = \angle {f_n} - \angle {g_n},	{~~\forall n\in\{1,\cdots,N\}},											\\
			& {p^{\rm opt}} = \sqrt {\frac{{P_{\text{A}}^{{\rm{max}}}}}{{P_{{\text{BS}}}^{{\rm{max}}}\sum\nolimits_{n = 1}^N {{{\left| {{g_n}} \right|}^2} + N\sigma _v^2} }}} .
		\end{align}
	\end{subequations}
	By substituting (\ref{eqn:optimal_so}) into (\ref{eqn:problem_pa}), the user's maximum achievable \ac{snr} for active RISs can be obtained as 
	\begin{align}\label{eqn:snr_active}
		\gamma_{\text{active}}  \!=\! \frac{{P_{{\text{BS}}}^{{\max}}P_{\text{A}}^{{\max}}{{\left| {\sum\nolimits_{n = 1}^N \!{\left| {{f_n}} \right|\left| {{g_n}} \right|} } \right|}^2}}}{{P_{\text{A}}^{{\max}}\sigma _v^2\sum\nolimits_{n = 1}^N\! {{{\left| {{f_n}} \right|}^2}}  \!+\! {\sigma ^2}\left( {P_{{\text{BS}}}^{{\max}}\sum\nolimits_{n = 1}^N\! {{{\left| {{g_n}} \right|}^2} \!+\! N\sigma _v^2} } \right)}}.
	\end{align}	
	
	Note that we assume ${\bf{f}} \sim \mathcal{C} \mathcal{N}(\mathbf{0}_N, \varrho_{f}^{2} {\bf{I}}_N)$ and ${\bf{g}} \sim \mathcal{C} \mathcal{N}\left(\mathbf{0}_N, \varrho_{g}^{2} {\bf{I}}_N\right)$. Thus, by letting $N\to\infty$ in (\ref{eqn:snr_active}), according to the law of large numbers, we have $\sum\nolimits_{n = 1}^N {\left| {{f_n}} \right|\left| {{g_n}} \right|}  \to N\frac{{\pi {\varrho _f}{\varrho _g}}}{4}$, $\sum\nolimits_{n = 1}^N {{{\left| {{g_n}} \right|}^2}}  \to N\varrho _g^2$, and $\sum\nolimits_{n = 1}^N {{{\left| {{f_n}} \right|}^2}}  \to N\varrho _f^2$. After substituting these asymptotic equations into (\ref{eqn:snr_active}), we obtain for the asymptotic \ac{snr} for active RISs the expression in (\ref{eqn:ag_active}), which completes the proof.
	\section{Proof of Lemma 3}
	According to the related analysis in \cite{Wu'19} and Appendix A, the user's achievable \ac{snr} for an SU-SISO system aided by a passive RIS and that aided by an active RIS can be respectively written as follows:
	\begin{subequations}\label{eqn:double_snr}
	\begin{align}
		{\gamma _{{\text{passive}}}} &\!=\! \frac{{P_{{\text{BS-P}}}^{\max }{{\left| {\sum\nolimits_{n = 1}^N {\left| {{f_n}} \right|\left| {{g_n}} \right|} } \right|}^2}}}{{{\sigma ^2}}},\\
		\!\gamma_{\text{active}}  &\!=\! \frac{{P_{{\text{BS-A}}}^{{\max}}P_{\text{A}}^{{\max}}{{\left| {\sum\nolimits_{n = 1}^N\! {\left| {{f_n}} \right|\left| {{g_n}} \right|} } \right|}^2}}}{{P_{\text{A}}^{{\max}}\sigma _v^2\sum\nolimits_{n = 1}^N\! {{{\left| {{f_n}} \right|}^2}}  \!+\! {\sigma ^2}\!\left( {P_{{\text{BS-A}}}^{{\max}}\sum\nolimits_{n = 1}^N\! {{{\left| {{g_n}} \right|}^2} \!+\! N\sigma _v^2} } \right)}},
	\end{align}
	\end{subequations}
	where ${P^{{\max}}_{\text{BS-A}}}$ denotes the maximum BS transmit power for the active RIS aided system and ${P^{{\max}}_{\text{BS-P}}}$ denotes that for the passive RIS aided system. By solving ${\gamma _{{\text{passive}}}}  \ge  \gamma_{\text{active}}$ according to (\ref{eqn:double_snr}), we have
\begin{align}
	&N \ge \notag \\
	&\frac{{P_{{\text{BS-A}}}^{\max }}}{{P_{{\text{BS-P}}}^{\max }}}\frac{{P_{\text{A}}^{\max }{\sigma ^2}}}{{P_{\text{A}}^{\max }\sigma _v^2\frac{1}{N}\sum\nolimits_{n = 1}^N\! {{{\left| {{f_n}} \right|}^2}}  \!+\! {\sigma ^2}\left( {P_{{\text{BS}}}^{\max }\frac{1}{N}\sum\nolimits_{n = 1}^N\! {{\left| {{g_n}} \right|}^2} \!+\! \sigma _v^2 } \right)}} \notag \\
	&\approx \frac{{P_{{\text{BS-A}}}^{{\max}}}}{{P_{{\text{BS-P}}}^{{\max}}}}\frac{{P_{\text{A}}^{{\max}}{\sigma ^2}}}{{\left( {P_{\text{A}}^{{\max}}\sigma _v^2\varrho _f^2 + P_{{\text{BS-A}}}^{{\max}}{\sigma ^2}\varrho _g^2 + {\sigma ^2}\sigma _v^2} \right)}}, \label{eqn:N>}
\end{align}
where we assume again that ${\bf{f}} \sim \mathcal{C} \mathcal{N}(\mathbf{0}_N, \varrho_{f}^{2} {\bf{I}}_N)$ and ${\bf{g}} \sim \mathcal{C} \mathcal{N}\left(\mathbf{0}_N, \varrho_{g}^{2} {\bf{I}}_N\right)$. Since the number of RIS elements $N$ is usually large, the components $\frac{1}{N}\sum\nolimits_{n = 1}^N {{{\left| {{f_n}} \right|}^2}}$ and $\frac{1}{N}\sum\nolimits_{n = 1}^N {{\left| {{g_n}} \right|}^2}$ in (\ref{eqn:N>}) were approximated by $\varrho _f^2$ and $\varrho _g^2$. This completes the proof.

\footnotesize
%\balance 
\bibliographystyle{IEEEtran}

%\begin{spacing}{1.8}
%\bibliography{IEEEabrv,reference}
%\end{spacing}

\bibliography{IEEEabrv,reference}

\begin{IEEEbiography}[{\includegraphics[width=1in,height=1.25in,clip,keepaspectratio]{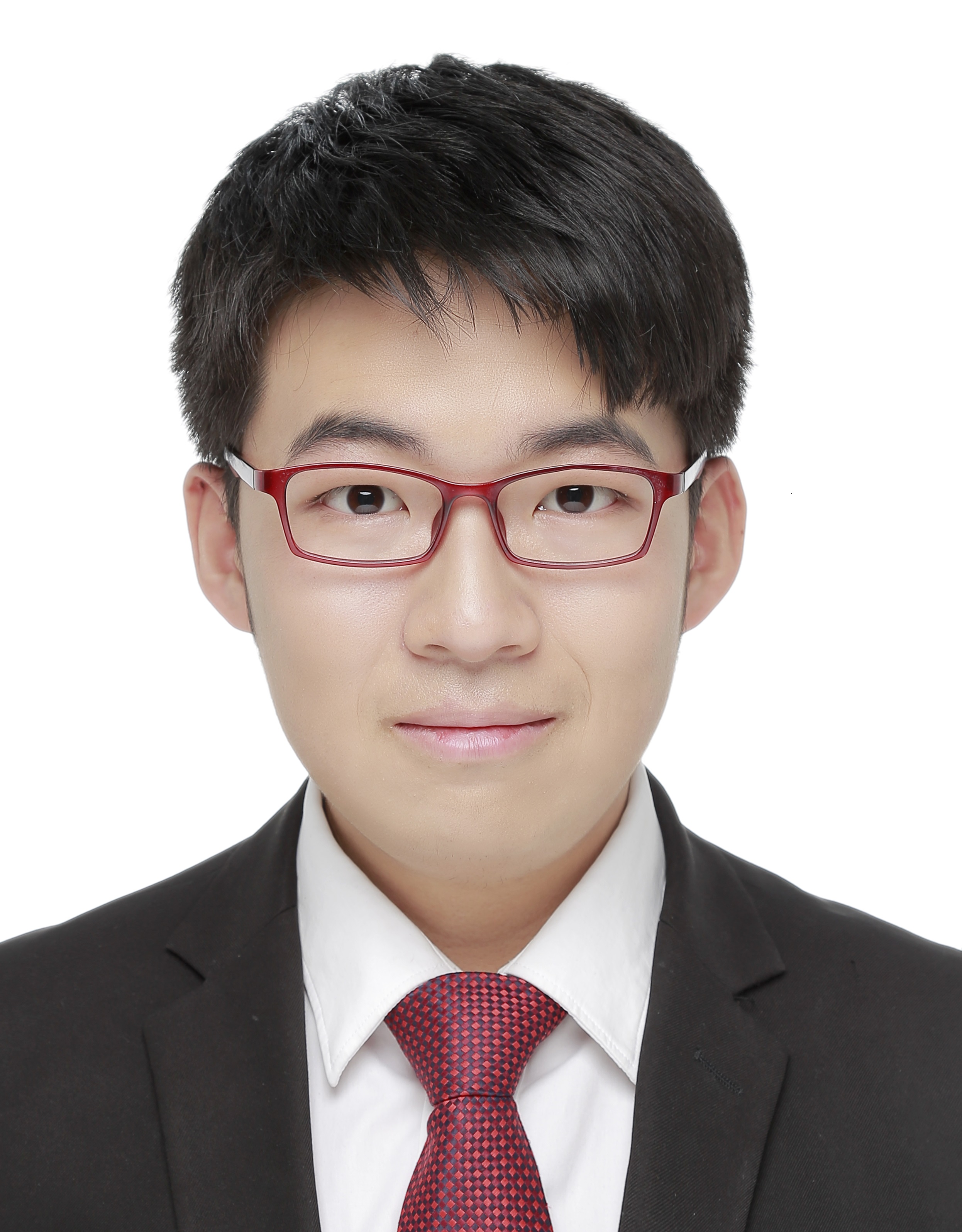}}]{Zijian Zhang.jpg}
(Student Member, IEEE) received the B.E. degree in electronic engineering from Tsinghua University, Beijing, China, in 2020. He is currently working toward the Ph.D. degree in electronic engineering from Tsinghua University, Beijing, China.
His research interests include physical-layer algorithms for massive MIMO and reconfigurable intelligent surfaces (RIS). He has received the National Scholarship in 2019 and the Excellent Thesis Award of Tsinghua University in 2020.
\end{IEEEbiography}

\begin{IEEEbiography}[{\includegraphics[width=1in,height=1.25in,clip,keepaspectratio]{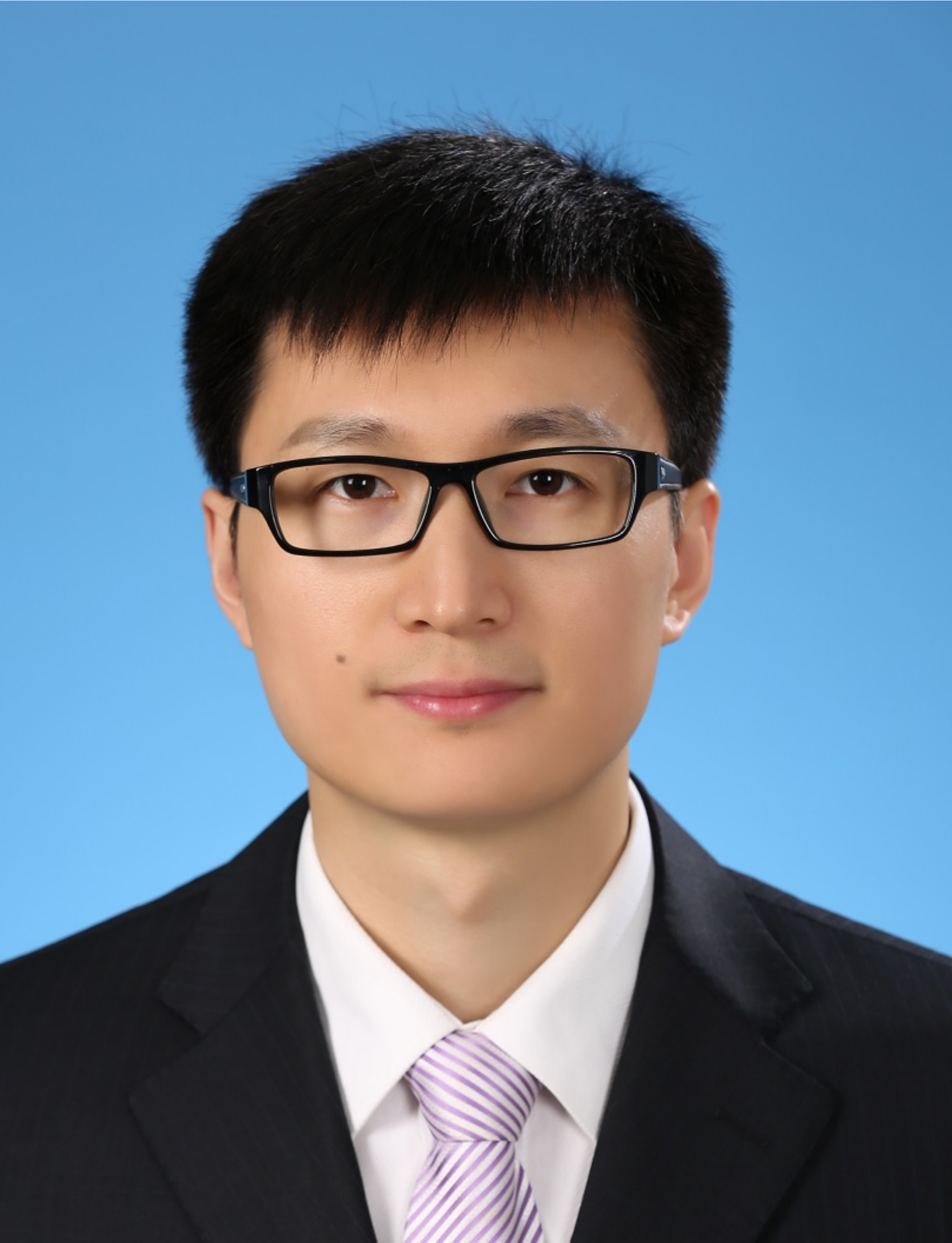}}]{Linglong Dai} (Fellow, IEEE) received the B.S. degree from Zhejiang University, Hangzhou, China, in 2003, the M.S. degree from the China Academy of Telecommunications Technology, Beijing, China, in 2006, and the Ph.D. degree from Tsinghua University, Beijing, in 2011. From 2011 to 2013, he was a Post-Doctoral Researcher with the Department of Electronic Engineering, Tsinghua University, where he was an Assistant Professor from 2013 to 2016, an Associate Professor from 2016 to 2022, and has been a Professor since 2022. His current research interests include massive MIMO, reconfigurable intelligent surface (RIS), millimeter-wave and Terahertz communications, wireless AI, and electromagnetic information theory. He has received the National Natural Science Foundation of China for Outstanding Young Scholars in 2017, the IEEE ComSoc Leonard G. Abraham Prize in 2020, and the IEEE ComSoc Stephen O. Rice Prize in 2022. He was elevated as an IEEE Fellow in 2021.
\end{IEEEbiography}

\begin{IEEEbiography}[{\includegraphics[width=1in,height=1.25in,clip,keepaspectratio]{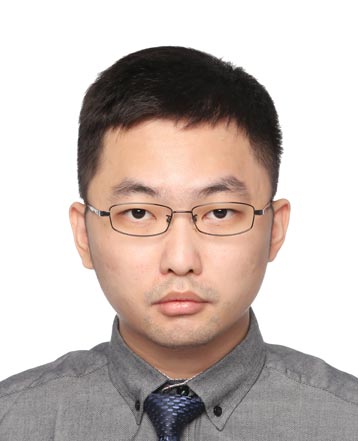}}]{Xibi Chen} (Student Member, IEEE) received the B.S. and M.S. degree from Tsinghua University, Beijing, China, in 2017 and 2020, respectively. Xibi Chen is currently a Ph.D. student at the Department of Electrical Engineering and Computer Science (EECS), Massachusetts Institute of Technology (MIT), Cambridge, MA. From 2015 to 2017, he was a Research Assistant with the Microwave and Antenna Institute, Department of Electronic Engineering, Tsinghua University. He later became a Graduate Student Researcher in the same institute from 2017 to 2019. In 2020, he joined EECS, MIT as a Ph.D. student. His current research interests include terahertz (THz) integrated electronic system, THz imaging/sensing, and CMOS electromagnetics/optics. He was the recipient of Analog Devices Outstanding Student Designer Award.	
\end{IEEEbiography}
\vspace*{-2em}

\begin{IEEEbiography}[{\includegraphics[width=1in,height=1.25in,clip,keepaspectratio]{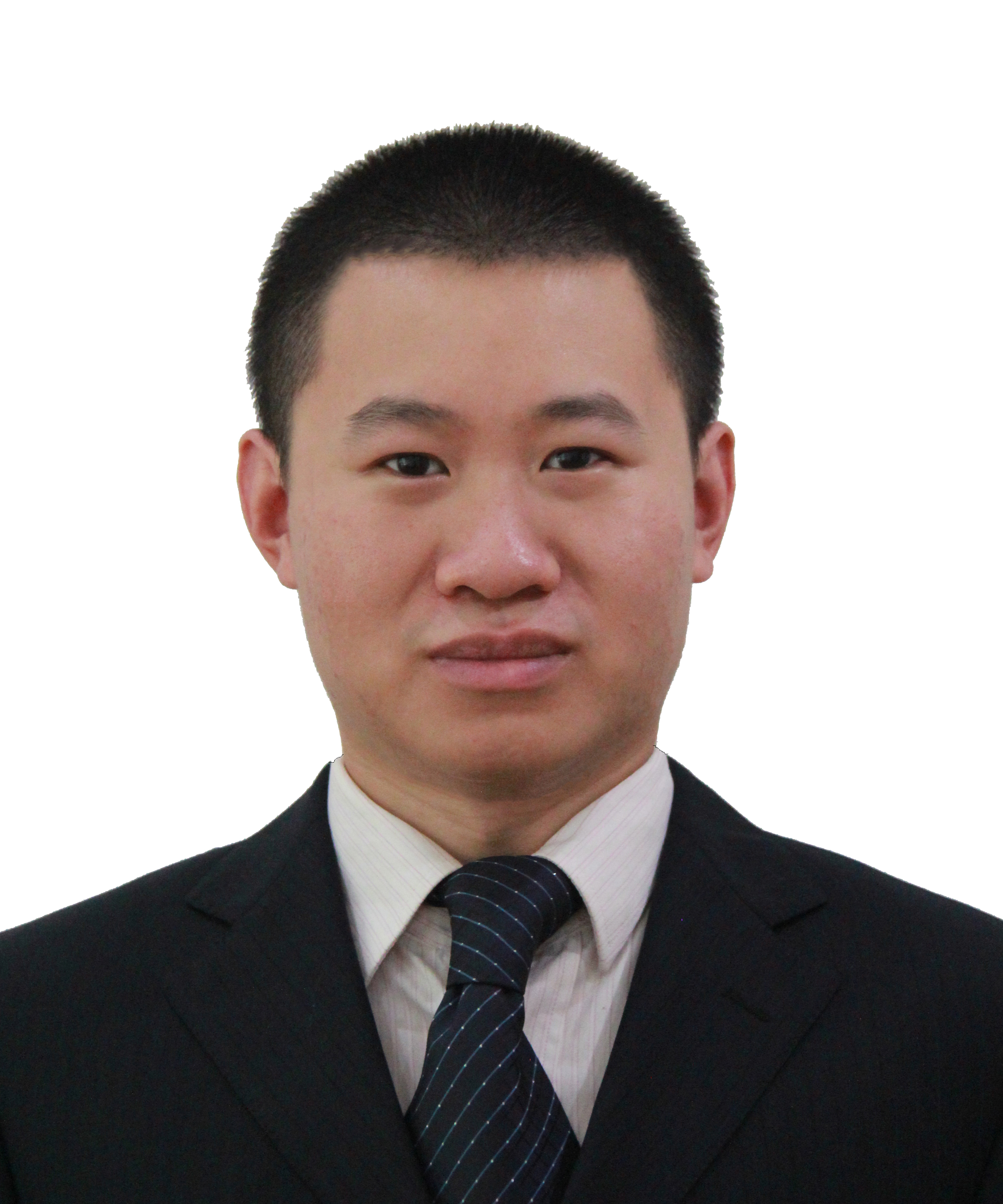}}]{Changhao Liu} (Graduate Student Member, IEEE) received the B.S. degree in electronic engineering from Tsinghua University, Beijing, China, in 2021. He is currently pursuing the Ph.D. degree in electronic engineering at Tsinghua University, Beijing, China.
	
From 2019 to 2021, he was a Research Assistant with the Microwave and Antenna Institute, Department of Electronic Engineering, Tsinghua University. His current research interests include reconfigurable metasurfaces, surface electromagnetics, reflectarray antennas, transmitarray antennas, terahertz metasurfaces, and reconfigurable intelligent surfaces.
\end{IEEEbiography}
\vspace*{-2em}

\begin{IEEEbiography}[{\includegraphics[width=1in,height=1.25in,clip,keepaspectratio]{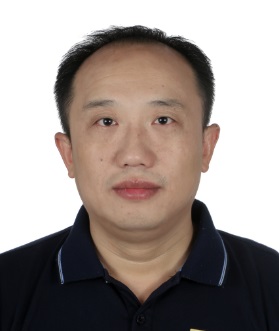}}]{Fan Yang} (Fellow, IEEE) received the B.S. and M.S. degrees from Tsinghua University, Beijing, China, in 1997 and 1999, respectively, and the Ph.D. degree from the University of California at Los Angeles (UCLA), in 2002.
	
From 1994 to 1999, he was a Research Assistant at the State Key Laboratory of Microwave and Digital Communications, Tsinghua University. From 1999 to 2002, he was a Graduate Student Researcher at the Antenna Laboratory, UCLA. From 2002 to 2004, he was a Post-Doctoral Research Engineer and Instructor at the Electrical Engineering Department, UCLA. In 2004, he joined the Electrical Engineering Department, The University of Mississippi as an Assistant Professor, and was promoted to an Associate Professor in 2009. In 2011, he joined the Electronic Engineering Department, Tsinghua University as a Professor, and served as the Director of the Microwave and Antenna Institute until 2020. 

Dr. Yang’s research interests include antennas, surface electromagnetics, computational electromagnetics, and applied electromagnetic systems. He has published over 500 journal articles and conference papers, eight book chapters, and six books entitled {\it Surface Electromagnetics} (Cambridge Univ. Press, 2019),  {\it Reflectarray Antennas: Theory, Designs, and Applications} (IEEE-Wiley, 2018),  {\it Analysis and Design of Transmitarray Antennas} (Morgan \& Claypool, 2017),  {\it Scattering Analysis of Periodic Structures Using Finite-Difference Time-Domain Method} (Morgan \& Claypool, 2012),  {\it Electromagnetic Band Gap Structures in Antenna Engineering} (Cambridge Univ. Press, 2009), and  {\it Electromagnetics and Antenna Optimization Using Taguchi’s Method} (Morgan \& Claypool, 2007). 

Dr. Yang served as an Associate Editor of the \textsc{IEEE Transactions on Antennas and Propagation} (2010-2013) and an Associate Editor-in-Chief of \textsc{Applied Computational Electromagnetics Society (ACES) Journal} (2008-2014). He was the Technical Program Committee (TPC) Chair of \textsc{2014 IEEE International Symposium on Antennas and Propagation and USNC-URSI Radio Science Meeting}. Dr. Yang has been the recipient of several prestigious awards and recognitions, including the Young Scientist Award of the 2005 URSI General Assembly and of the 2007 International Symposium on Electromagnetic Theory, the 2008 Junior Faculty Research Award of the University of Mississippi, the 2009 inaugural IEEE Donald G. Dudley Jr. Undergraduate Teaching Award, and the 2011 Recipient of Global Experts Program of China. He is an ACES Fellow and IEEE Fellow, as well as an IEEE APS Distinguished Lecturer for 2018-2021.
\end{IEEEbiography}

\begin{IEEEbiography}[{\includegraphics[width=1in,height=1.25in,clip,keepaspectratio]{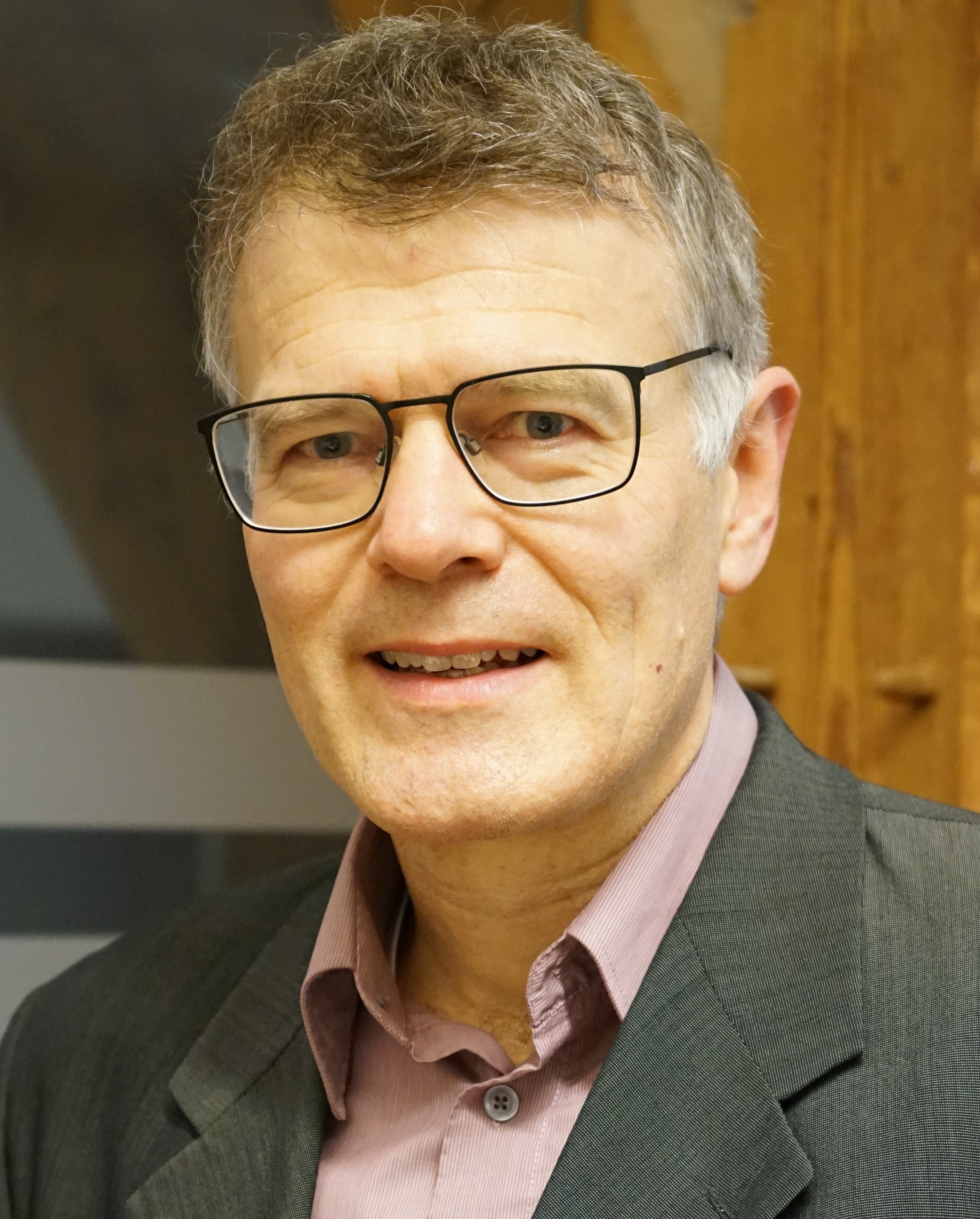}}]{Robert Schober} (Fellow, IEEE) received the Diplom (Univ.) and the Ph.D. degrees in electrical engineering from Friedrich-Alexander University of Erlangen-Nuremberg (FAU), Germany, in 1997 and 2000, respectively. From 2002 to 2011, he was a Professor and Canada Research Chair at the University of British Columbia (UBC), Vancouver, Canada. Since January 2012 he is an Alexander von Humboldt Professor and the Chair for Digital Communication at FAU. His research interests fall into the broad areas of Communication Theory, Wireless and Molecular Communications, and Statistical Signal Processing.

Robert received several awards for his work including the 2002 Heinz Maier­ Leibnitz Award of the German Science Foundation (DFG), the 2004 Innovations Award of the Vodafone Foundation for Research in Mobile Communications, a 2006 UBC Killam Research Prize, a 2007 Wilhelm Friedrich Bessel Research Award of the Alexander von Humboldt Foundation, the 2008 Charles McDowell Award for Excellence in Research from UBC, a 2011 Alexander von Humboldt Professorship, a 2012 NSERC E.W.R. Stacie Fellowship, a 2017 Wireless Communications Recognition Award by the IEEE Wireless Communications Technical Committee, and the 2022 IEEE Vehicular Technology Society Stuart F. Meyer Memorial Award. Furthermore, he received numerous Best Paper Awards for his work including the 2022 ComSoc Stephen O. Rice Prize. Since 2017, he has been listed as a Highly Cited Researcher by the Web of Science. Robert is a Fellow of the Canadian Academy of Engineering, a Fellow of the Engineering Institute of Canada, and a Member of the German National Academy of Science and Engineering.

He served as Editor-in-Chief of the \textsc{IEEE Transactions on Communications} from 2012 to 2015 and as VP Publications of the IEEE Communication Society (ComSoc) in 2020 and 2021. Currently, he serves as Member of the Editorial Board of the \textsc{Proceedings of the IEEE}, as Member at Large of the ComSoc Board of Governors, and as ComSoc Treasurer. He is the ComSoc President-Elect for 2023.
\end{IEEEbiography}

\begin{IEEEbiography}[{\includegraphics[width=1in,height=1.25in,clip,keepaspectratio]{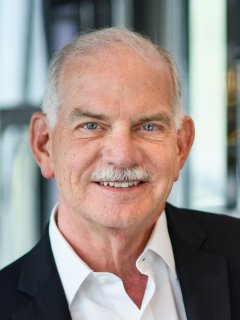}}]{H. Vincent Poor} (Life Fellow, IEEE) received the Ph.D. degree in EECS from Princeton University in 1977.  From 1977 until 1990, he was on the faculty of the University of Illinois at Urbana-Champaign. Since 1990 he has been on the faculty at Princeton, where he is currently the Michael Henry Strater University Professor. During 2006 to 2016, he served as the dean of Princeton’s School of Engineering and Applied Science. He has also held visiting appointments at several other universities, including most recently at Berkeley and Cambridge. His research interests are in the areas of information theory, machine learning and network science, and their applications in wireless networks, energy systems and related fields. Among his publications in these areas is the recent book {\it Machine Learning and Wireless Communications}.  (Cambridge University Press, 2022). Dr. Poor is a member of the National Academy of Engineering and the National Academy of Sciences and is a foreign member of the Chinese Academy of Sciences, the Royal Society, and other national and international academies. He received the IEEE Alexander Graham Bell Medal in 2017.
\end{IEEEbiography}
	
\end{document}